\documentclass[3p,preprint]{elsarticle}

\usepackage{graphicx}
\usepackage{physics}
\usepackage{amsmath}
\usepackage{amssymb}
\usepackage{float}
\usepackage{subcaption}
\usepackage{multirow}
\usepackage{xcolor}
\usepackage{hyperref}

\hypersetup{hidelinks}
\hypersetup{colorlinks=true,allcolors=blue}
\hypersetup{pdfstartview={XYZ null null 1.00}} 
\hypersetup{pdfpagemode=UseNone} 

\allowdisplaybreaks  

\begin{document}

\begin{frontmatter}
\title{High-Order Large-Eddy Simulations of a Wind Turbine in Ducted and Open-Rotor Configurations}

\author[mae]{Chi Ding\corref{cor1}}
\ead{chid@clarkson.edu}

\author[mae]{Bin Zhang}

\author[mae]{Chunlei Liang}

\author[mae]{Kenneth Visser}

\author[math]{Guangming Yao}

\cortext[cor1]{Corresponding author.}

\address[mae]{Department of Mechanical and Aerospace Engineering, Clarkson University, Potsdam, New York 13699}

\address[math]{Department of Mathematics, Clarkson University, Potsdam, New York 13699}

\begin{abstract}
High-order large-eddy simulations are performed to study the performance and flow field of a ducted wind turbine operating at different tip speed ratios. To evaluate the effects of the duct, simulations with the same tip speed ratios are also performed on the corresponding open-rotor turbine. It is found that the ducted turbine consistently obtains higher power outputs than the open-rotor counterpart, and the duct itself enhances flow turbulence and blade trailing-edge vortices but weakens tip and hub vortices. Flow bifurcation is observed at the largest tip speed ratio and is identified to be caused by blade blockage effects. Comparative simulations are also performed on both turbines under different yaw angles. It is noticed that the ducted configuration is insensitive to small yaw angles and maintains higher power outputs than the open-rotor configuration at all yaw angles. Moreover, it is observed that the wakes of both configurations recover more quickly as the yaw angle increases.
\end{abstract}

\end{frontmatter}


\section{Introduction}
Wind power is an important energy source and has a long history of being exploited \cite{shepherd-2017}. In the 1970s, the interest in developing wind power was boosted by the oil crisis. At that time, a lot of state-funded projects were launched to develop wind power. In 1973, the U.S. government approved about 200 million dollars to support studies on wind turbines \cite{hau-2013}. Between 1975 and 1987, the MOD series of wind turbines were erected in the U.S. \cite{burton-2011}. During the same period, Denmark built two large experimental turbines in the vicinity of Aalborg \cite{hau-2013}. In Sweden, two large wind turbines: WTS-75 and WTS-3, were installed in cooperation with a German company and a U.S. company, respectively \cite{hau-2013}.

One major design task is to extract more energy from wind via, for example, increasing its blade size. However, the increasing size leads to larger aerodynamic loads and makes turbine blades more vulnerable to aeroelastic issues like flutters \cite{ochieng-2018}. In addition, large wind turbines are more difficult to install in urban areas than small ones \cite{dighe-2020}. Another solution to improve turbine efficiency is to place the rotor into a diffuser, called a ducted wind turbine (DWT) or a diffuser augmented wind turbine (DAWT). By doing so, the mass flow rate across the rotor plane increases, producing more power output \cite{bontempo-2020}. The DWT design avoids the problems caused by increasing blade size and is more suitable for installation in urban areas. DWTs have even more advantages, including insensitivity to yawed flow and less tip loss \cite{bontempo-2020}. Therefore, DWTs are a promising way to efficiently harvest wind energy.

To study the performance of DWTs,  a lot of theoretical and experimental research has been carried out. In 1956, Lilley and Rainbird \cite{lilley-1956} performed a one-dimensional theoretical analysis on DWTs. They concluded that adding a duct could improve the power performance, and higher power outputs can be achieved via larger duct expansion ratios. However, a large expansion ratio may cause boundary-layer separation. During the 1970s and 1980s, a series of experiments verified the concept of DWT. Some experiments employing techniques like ring-shaped flaps and multi-slotted diffusers were also performed to prevent flow separations \cite{foreman-1978,gilbert-1979,igra-1981}.

Many computational studies on DWTs have also been reported. In 1981, Fletcher \cite{fletcher-1981} analyzed a DWT using the blade element momentum (BEM) theory, which took into account the effects of Reynolds number and wake rotation. Good agreement with experimental measures was achieved. Vaz and Wood \cite{vaz-2018}  improved this BEM method by including a high rotor thrust correction and a new formulation for the far-wake velocity. Koras and Georgalas \cite{koras-1988} modeled the rotor of a DWT by a lifting line and the duct by a combination of vortex rings and source rings. They used this potential flow method to study the influence of several geometrical parameters on the power output. However, their method was limited to DWTs with large tip clearance. Politis and Koras \cite{politis-1995} later made progress by using a lifting-surface approach for duct modeling, and their method was able to handle DWTs with any tip clearance.

The rapid development of computational fluid dynamics (CFD) technologies has also boosted the computational studies on DWTs. One of the most popular approaches is to combine a CFD solver with an actuator-disk (AD) model that represents a rotor to study DWTs. This approach is usually referred to as the CFD-AD approach and can provide more details of a DWT flow field at low computational costs. Phillips et al. \cite{phillips-2002} applied a CFD-AD approach to investigate the Vortec 7 turbine --- a full-scale DWT design. In the same way, Hansen et al. \cite{hansen-2000} analyzed the performance of a DWT and verified that adding a duct increases the mass flow rate. Abe and Ohya \cite{abe-2004} employed a CFD-AD approach to study a turbine with a flanged diffuser. Their focus was on how the loading coefficient and the diffuser's opening angle affect the turbine performance. Venters et al. \cite{venters-2018} used a CFD-AD approach to find an optimized design for a DWT. They employed two objective functions for the optimization: one based on rotor power coefficient and the other on total power coefficient. That study was continued by Sadeghi et al. \cite{bagheri-2018} using new optimization algorithms.

These computational studies gave a lot of helpful guidance on the design of DWTs. However, the aforementioned methods inevitably bear significant simplifications on the turbine geometries. Nowadays, it is viable to simulate fluid flows about real turbines using more advanced CFD techniques. Furthermore, the development of high-order methods made it possible to simulate a flow field at higher spatial accuracies than traditional finite volume methods. The most popular high-order methods include the discontinuous Galerkin (DG) method \cite{reed-1973, cockburn-2012}, the spectral element (SE) method \cite{patera-1984, karniadakis-2013}, the spectral difference (SD) method \cite{kopriva-1996a, kopriva-1996b, kopriva-1998, liu-2006}, and the flux reconstruction (FR) method \cite{huynh-2007, huynh-2009, wang-2009}. The SD and the FR methods are based on the differential-form governing equations and are two of the most efficient high-order methods. The FR method is a unified framework that can recover many existing high-order schemes (e.g., DG and SD schemes) and produce new schemes that were never reported before. To deal with rotating objects, Zhang and Liang \cite{zhang-2015a, zhang-2015b} introduced the curved dynamic mortar concept and applied it to developing high-order sliding-mesh SD and FR methods. These methods were later extended to sliding-deforming meshes \cite{zhang-2016a}, 3D geometries \cite{zhang-2016c}, and general nonuniform sliding interfaces \cite{zhang-2018}. Zhang et al. \cite{zhang-2021b} further introduced the transfinite mortar concept that has no geometric error and makes a sliding-mesh method arbitrarily high-order accurate in space and high-order in time. This method has been applied to simulate flows around rotating cylinders of different cross-sectional shapes \cite{zhang-2019b}, flapping wings for energy harvesting \cite{zhang-2019a}, and, more recently, the first high-order eddy-resolving simulation of flow over a marine propeller \cite{zhang-2021a}.

The authors of the present work also applied the above techniques to a preliminary study of a DWT (designed at Clarkson University by Dr. Kenneth Visser) at its design condition \cite{ding-2022a}. In this work, we further the study by comparing the DWT with the corresponding open-rotor wind turbine (OWT) at different working conditions. Given that DWTs can be installed in urban areas where the directions of winds may be affected by buildings, simulations for the DWT and OWT under yawed inflows are also performed.

The rest of this paper is organized as follows. Section \ref{sec:num} briefly introduces the numerical methods. Section \ref{sec:setup} gives details on the simulation setup and the numerical validation. In Section \ref{sec:axial}, computations for turbines under axial flows are carried out, and results on the aerodynamic loads, pressure, and velocities are presented and analyzed. Section \ref{sec:yaw} reports simulation results for the yawed cases. Finally, Section \ref{sec:concl} concludes this study.

\section{Numerical Methods}
\label{sec:num}

\subsection{The Governing Equations}
The three-dimensional unsteady Navier-Stokes equations in the following conservative form are numerically solved,
\begin{equation}
\pdv{\mathbf{Q}}{t}  + \pdv{\mathbf{F}}{x} + \pdv{\mathbf{G}}{y} + \pdv{\mathbf{H}}{z} = \mathbf{0},
\label{eq:physical}
\end{equation}
where $\mathbf{Q}$ is the vector of conservative variables, $\mathbf{F}$, $\mathbf{G}$, and $\mathbf{H}$ are the flux vectors in each coordinate direction. These terms have the following expressions,
\begin{align}
\mathbf{Q} & = [\rho ~ \rho u ~ \rho v ~ \rho w ~ E]^{\intercal},\label{eq:Q}\\[1mm]
\mathbf{F} & = \mathbf{F}_\text{inv} (\mathbf{Q}) +
\mathbf{F}_\text{vis} (\mathbf{Q},\grad{\mathbf{Q}}), \label{eq:F}\\[1mm]
\mathbf{G} & = \mathbf{G}_\text{inv} (\mathbf{Q}) +
\mathbf{G}_\text{vis} (\mathbf{Q},\grad{\mathbf{Q}}), \label{eq:G}\\[1mm]
\mathbf{H} & = \mathbf{H}_\text{inv} (\mathbf{Q}) +
\mathbf{H}_\text{vis} (\mathbf{Q},\grad{\mathbf{Q}}), \label{eq:H}
\end{align}
where $\rho$ is fluid density, $u$, $v$, and $w$ are the velocity components, $E$ is the total energy per unit volume defined as
$E=p/(\gamma-1)+\frac{1}{2}\rho(u^2+v^2+w^2)$, $p$ is pressure, and $\gamma$ is the ratio of specific heats which is set to 1.4 in this work. The fluxes have been split into inviscid and viscous parts. The inviscid fluxes are only functions of the conservative variables and have the following expressions,
\begin{align}
\mathbf{F}_\text{inv} &= [\rho u ~~ \rho u^2\! +\! p ~~ \rho uv ~~ \rho uw ~~ u(E\!+\!p)]^\intercal, \\[0.2em]
\mathbf{G}_\text{inv} &= [\rho v ~~ \rho uv ~~ \rho v^2\! +\! p ~~ \rho vw ~~ v(E\!+\!p)]^\intercal, \\[0.2em]
\mathbf{H}_\text{inv} &= [\rho w ~~ \rho uw ~~ \rho vw ~~ \rho w^2\! +\! p ~~ w(E\!+\!p)]^\intercal.
\label{eq:FGHinvx}
\end{align}
The viscous fluxes are functions of the conservative variables and the gradients. Their expressions are
\begin{align}
\mathbf{F}_\text{vis} &= -[0 ~~ \tau_{xx} ~~ \tau_{yx} ~~ \tau_{zx} ~~ u\tau_{xx}\! +\! v\tau_{yx}\! +\! w\tau_{zx}\! +\! \kappa T_{x}]^\intercal,\\[0.3em]
\mathbf{G}_\text{vis} &= -[0 ~~ \tau_{xy} ~~ \tau_{yy} ~~ \tau_{zy} ~~ u\tau_{xy}\! +\! v\tau_{yy}\! +\! w\tau_{zy}\! +\! \kappa T_{y}]^\intercal,\\[0.3em]
\mathbf{H}_\text{vis} &= -[0 ~~ \tau_{xz} ~~ \tau_{yz} ~~ \tau_{zz} ~~ u\tau_{xz}\! +\! v\tau_{yz}\! +\! w\tau_{zz}\! +\! \kappa T_{z}]^\intercal,
\label{eq:FGHvis}
\end{align}
where $\tau_{ij}$ is viscous stress tensor which is related to velocity gradients as $\tau_{ij} = \mu(u_{i,j}+u_{j,i})+ \lambda \delta_{ij}u_{k,k}$, $\mu$ is the dynamic viscosity, $\lambda=-\frac{2}{3}\mu$ based on Stokes' hypothesis, $\delta_{ij}$ is the Kronecker delta, $\kappa$ is the thermal conductivity, $T$ is temperature which is related to density and pressure through the ideal gas law $p=\rho R T$, and $R$ is the gas constant.

\subsection{The Computational Equations}
Each mesh element in the physical space is mapped to a standard element in a computational space. Assume the mapping is: $t=\tau$, $x=x(\tau,\xi,\eta,\zeta)$, $y=y(\tau,\xi,\eta,\zeta)$, and $z=z(\tau,\xi,\eta,\zeta)$, where $(\tau,\xi,\eta,\zeta)$ are the computational time and coordinates. It can be shown that the governing equations will take the following conservative form in the computational space,
\begin{equation}
\pdv{\widetilde{\mathbf{Q}}}{t} + \pdv{\widetilde{\mathbf{F}}}{\xi} + \pdv{\widetilde{\mathbf{G}}}{\eta} + \pdv{\widetilde{\mathbf{H}}}{\zeta} = \mathbf{0}.
\label{eq:computational}
\end{equation}
The computational variables and fluxes are related to the physical ones as
\begin{equation}
\begin{bmatrix}
\widetilde{\mathbf{Q}} \\[1mm]
\widetilde{\mathbf{F}} \\[1mm]
\widetilde{\mathbf{G}} \\[1mm]
\widetilde{\mathbf{H}}
\end{bmatrix}
=
\abs{\mathcal{J}}\mathcal{J}^{-1}
\begin{bmatrix}
{\mathbf{Q}} \vphantom{\widetilde{\mathbf{Q}}}\\[1mm]
{\mathbf{F}} \vphantom{\widetilde{\mathbf{F}}}\\[1mm]
{\mathbf{G}} \vphantom{\widetilde{\mathbf{G}}}\\[1mm]
{\mathbf{H}} \vphantom{\widetilde{\mathbf{H}}}
\end{bmatrix}
\label{eq:jacobi},
\end{equation}
where $\mathcal{J}$ is the Jacobian matrix, $\abs{\mathcal{J}}$ is the determinant, and $\mathcal{J}^{-1}$ is the inverse Jacobian matrix. Their expressions are
\begin{eqnarray}
\mathcal{J} =
\pdv{(t,x,y,z)}{(\tau,\xi,\eta,\zeta)} =
\begin{bmatrix}
1           & 0          & 0           & 0            \\
x_{\tau}    & x_{\xi}    & x_{\eta}    & x_{\zeta}    \\
y_{\tau}    & y_{\xi}    & y_{\eta}    & y_{\zeta}    \\
z_{\tau}    & z_{\xi}    & z_{\eta}    & z_{\zeta}
\end{bmatrix},\\[0.5em]
\mathcal{J}^{-1} =
\pdv{(\tau,\xi,\eta,\zeta)}{(t,x,y,z)} =
\begin{bmatrix}
1          & 0          & 0           & 0            \\
\xi_{t}     & \xi_{x}    & \xi_{y}     & \xi_{z}      \\
\eta_{t}    & \eta_{x}   & \eta_{y}    & \eta_{z}     \\
\zeta_{t}   & \zeta_{x}  & \zeta_{y}   & \zeta_{z}
\end{bmatrix}.
\end{eqnarray}
Besides the flow equations, the geometric conservation law (GCL) \cite{thomas-1979} also needs to be numerically satisfied to ensure free-stream preservation on moving grids. The GCL equations and the steps for solving them  are described in, e.g., \cite{zhang-2016a}.

\subsection{The Flux Reconstruction Method}
Only hexahedral elements are used in this work, and each element is mapped to a unit computational element $0\leq\xi,\eta,\zeta\leq 1$ via the following iso-parametric mapping,
\begin{equation}
\begin{bmatrix}
x \\[1mm]
y \\[1mm]
z
\end{bmatrix}
= \sum^{K}_{i=1} M_i(\xi,\eta,\zeta)
\begin{bmatrix}
x_i(t)  \\[1mm]
y_i(t)  \\[1mm]
z_i(t)
\end{bmatrix},
\end{equation}
where $K$ is the number of nodes that define a physical element, $M_i$ is the shape function (detailed expressions can be found in, e.g., \cite{bathe-2006}), and $(x_i,y_i,z_i)$ are the coordinates, of the $i$-th node.

\begin{figure}[H]
\centering
\includegraphics[width=2.0in]{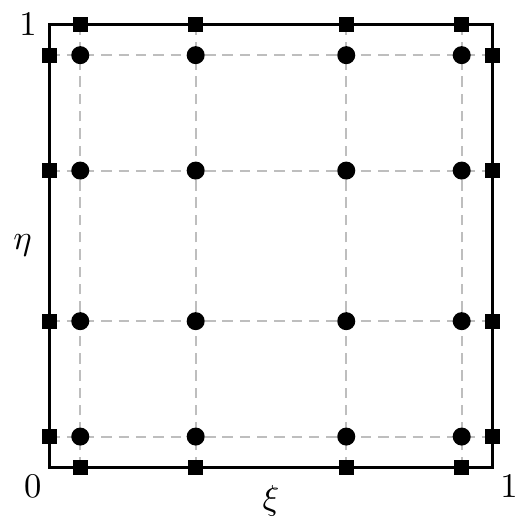}
\caption{Schematic of SPs (round dots) and FPs (square dots) in the $\xi$-$\eta$ plane for a fourth-order FR scheme.}
\label{fig:FP_SP}
\end{figure}

Solution points (SPs, denoted by $X_s$) are defined inside each computational element, and flux points (FPs, denoted by $X_f$) are defined on the boundaries. Figure \ref{fig:FP_SP} shows a schematic of the distribution of the SPs and FPs in the $\xi$-$\eta$ plane for a fourth-order FR scheme. For an $N$-th order FR scheme, there are $N$ SPs and $N$ FPs in each direction. The SPs and FPs are chosen as the Legendre points in this study.
At the SPs, the following Lagrange interpolation bases can be defined (where $X_i$ is the coordinate of the $i$-th SP),
\begin{equation}
h_i(X)= \prod^{N}_{s=1,s\neq i}(\frac{X-X_s}{X_i-X_s}), ~~ i=1,2,\cdots,N.
\end{equation}
The above polynomials also form a basis for polynomials of degrees less than or equal to $N-1$, i.e., $\boldsymbol{\mathsf{P}}_{N-1}$. The solution and fluxes within each element can be approximated via tensor products of the interpolation bases, e.g.,
\begin{align}
\widetilde{\mathbf{Q}}(\xi,\eta,\zeta) &= \sum_{i=1}^N \sum_{j=1}^N \sum_{k=1}^N \widetilde{\mathbf{Q}}_{ijk} h_i(\xi) h_j(\eta) h_k(\zeta),\label{eq:Qt}\\[1mm]
\widetilde{\mathbf{F}}(\xi,\eta,\zeta) &= \sum_{i=1}^N \sum_{j=1}^N \sum_{k=1}^N \widetilde{\mathbf{F}}_{ijk} h_i(\xi) h_j(\eta) h_k(\zeta),  \label{eq:Ft}
\end{align}
where the subscript $ijk$ denotes the discrete value at the $(i,j,k)$-th SP. Obviously, the solution and flux polynomials are in $\boldsymbol{\mathsf{P}}_{N-1,N-1,N-1}$ and are continuous within each element but discontinuous across cell boundaries. Therefore, common values need to be defined at cell boundaries. In this work, the common solution is calculated as the average of the discontinuous values from the two sides of a boundary; the common inviscid fluxes are computed using the Rusanov solver \cite{rusanov-1961}; the common viscous fluxes are computed from the common solutions and common gradients.

After taking the spatial derivatives in Eq. (\ref{eq:computational}), the three flux terms are reduced to elements of $\boldsymbol{\mathsf{P}}_{N-2,N-1,N-1}$, $\boldsymbol{\mathsf{P}}_{N-1,N-2,N-1}$, and $\boldsymbol{\mathsf{P}}_{N-1,N-1,N-2}$, respectively, which are inconsistent with the solution term. To fix this issue, the degrees of the original flux polynomial need to be raised, which can be achieved using higher-degree correction functions \cite{huynh-2007}. For example, the corrected/reconstructed flux in the $\xi$ direction is
\begin{equation}
\widehat{\mathbf{F}} =  \widetilde{\mathbf{F}}(\xi,\eta,\zeta)
+ \left[\widetilde{\mathbf{F}}^{\text{com}}(0,\eta,\zeta) -  \widetilde{\mathbf{F}}(0,\eta,\zeta)\right]\cdot g_\text{\tiny L}(\xi)
+ \left[\widetilde{\mathbf{F}}^{\text{com}}(1,\eta,\zeta) -  \widetilde{\mathbf{F}}(1,\eta,\zeta)\right]\cdot g_\text{\tiny R}(\xi),
\end{equation}
where $\widetilde{\mathbf{F}}$ is from (\ref{eq:Ft}), $\widetilde{\mathbf{F}}^{\text{com}}$ is the common flux on a cell boundary, and $g_L$ and $g_R$ are the left and right correction functions that are required to satisfy
\begin{equation}
\begin{alignedat}{2}
g_{\text{\tiny L}}(0) = 1,\quad g_{\text{\tiny L}}(1) = 0,\\
g_{\text{\tiny R}}(0) = 0,\quad g_{\text{\tiny R}}(1) = 1.
\end{alignedat}
\end{equation}
These conditions ensure that
\begin{equation}
\widehat{\mathbf{F}}(0,\eta,\zeta) = \widetilde{\mathbf{F}}^{\text{com}}(0,\eta,\zeta),\quad
\widehat{\mathbf{F}}(1,\eta,\zeta) = \widetilde{\mathbf{F}}^{\text{com}}(1,\eta,\zeta),
\end{equation}
i.e., the reconstructed flux still takes the common values on cell boundaries. In this work, the $g_\text{DG}$ function \cite{huynh-2007} is chosen as the correction function. The other two fluxes are reconstructed in the same way. Finally, the governing equations can be written in the following residual form,
\begin{equation}
\eval{\pdv{\widetilde{\mathbf{Q}}}{t}}_{ijk} = - \left[ \pdv{\widehat{\mathbf{F}}}{\xi} + \pdv{\widehat{\mathbf{G}}}{\eta} + \pdv{\widehat{\mathbf{H}}}{\zeta} \right]_{ijk} = \mathbf{R}_{ijk}, \quad i,j,k = 1,2,\cdots,N,
\end{equation}
where $\mathbf{R}_{ijk}$ is the residual at the $(i,j,k)$-th SP. This system can be time marched either explicitly or implicitly. In the present work, a four-stage third-order explicit Runge-Kutta method \cite{spiteri-2002,ruuth-2006} is employed.

\subsection{The Sliding-mesh SD/FR Method}
\begin{figure}[H]
\centering
\includegraphics[width=1.3in]{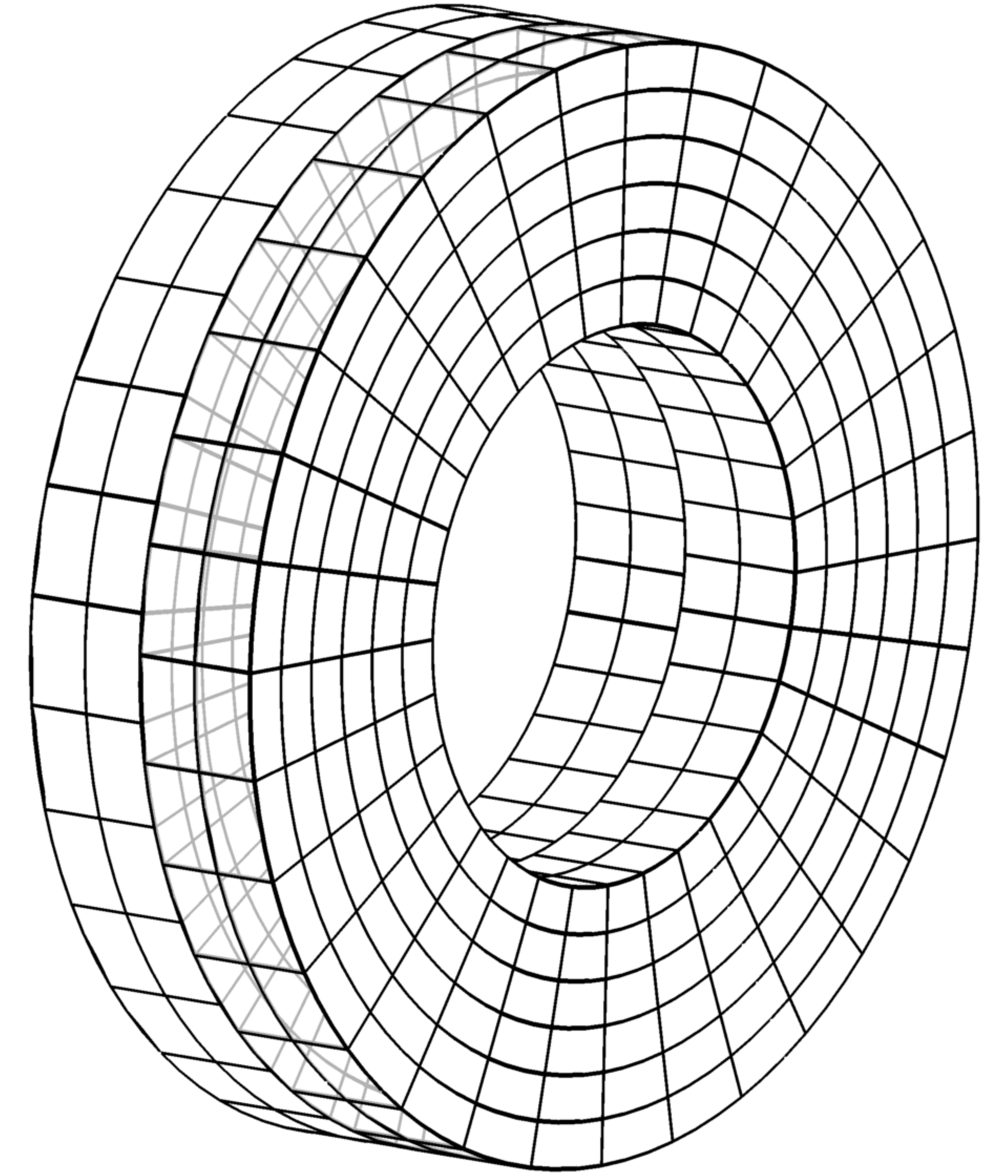} \hspace{1cm}
\includegraphics[width=1.3in]{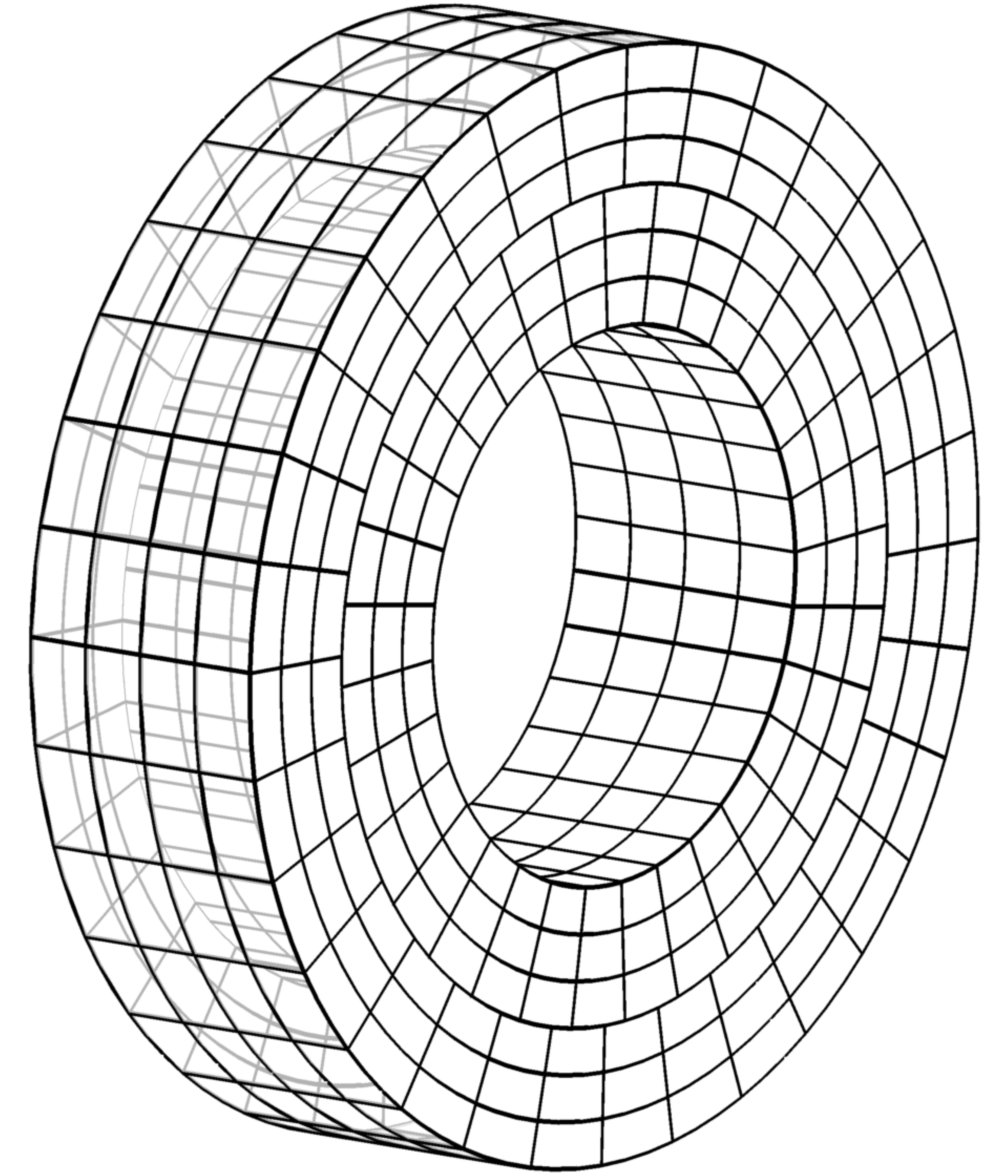}
\caption{Two types of sliding interfaces.}
\label{fig:sliding_mesh}
\end{figure}

There are two fundamental types of sliding interfaces in 3D as shown in Fig. \ref{fig:sliding_mesh}. For simplicity, assume that the mesh points do not match in the azimuthal direction but match in the other direction. Further assume that the azimuthal direction is uniformly meshed. We take the second type to briefly explain how the method works. More detailed explanation can be found in previous papers, e.g., \cite{zhang-2016c,zhang-2021b,zhang-2021a}.

\begin{figure}[H]
\centering
\includegraphics[width=4.0in]{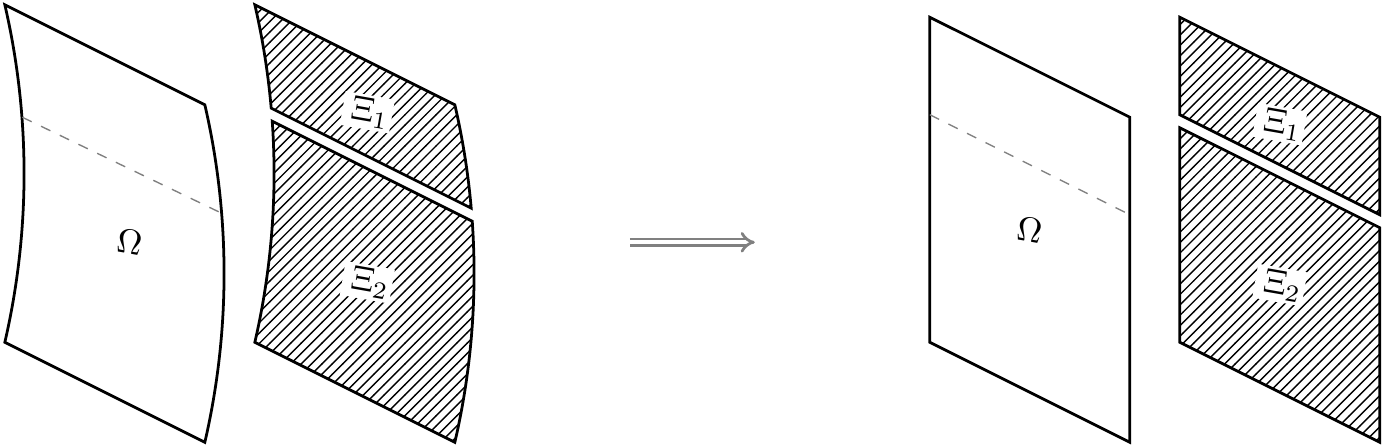}
\caption{Map curved cell face and mortar elements to straight ones.}
\label{fig:mortar_map}
\end{figure}

Curved dynamic mortar elements \cite{zhang-2015b} are employed to communicate between the two sides of a sliding interface. As shown in Fig. \ref{fig:mortar_map}, at each time instant, a cell face $\Omega$ is connected to two mortar elements $\Xi_1$ and $\Xi_2$. These curved geometries are mapped to straight ones first. For $\Omega$, assume that the azimuthal direction is mapped to $\xi$ and the other direction is mapped to $\eta$ in the computational space. Similarly, for each mortar, assume these two directions are mapped to $\xi'$ and $\eta'$, respectively, in the mortar space. Then these two spaces are related as: $\xi = o + s \cdot \xi^{\prime},~ \eta = \eta^{\prime}$, where $0\leq\xi,\eta,\xi',\eta'\leq 1$, and $o$ and $s$ are the offset and scaling of a mortar with respect to a cell face.

\begin{figure}[!htb]
\centering
\includegraphics[width=4.0in]{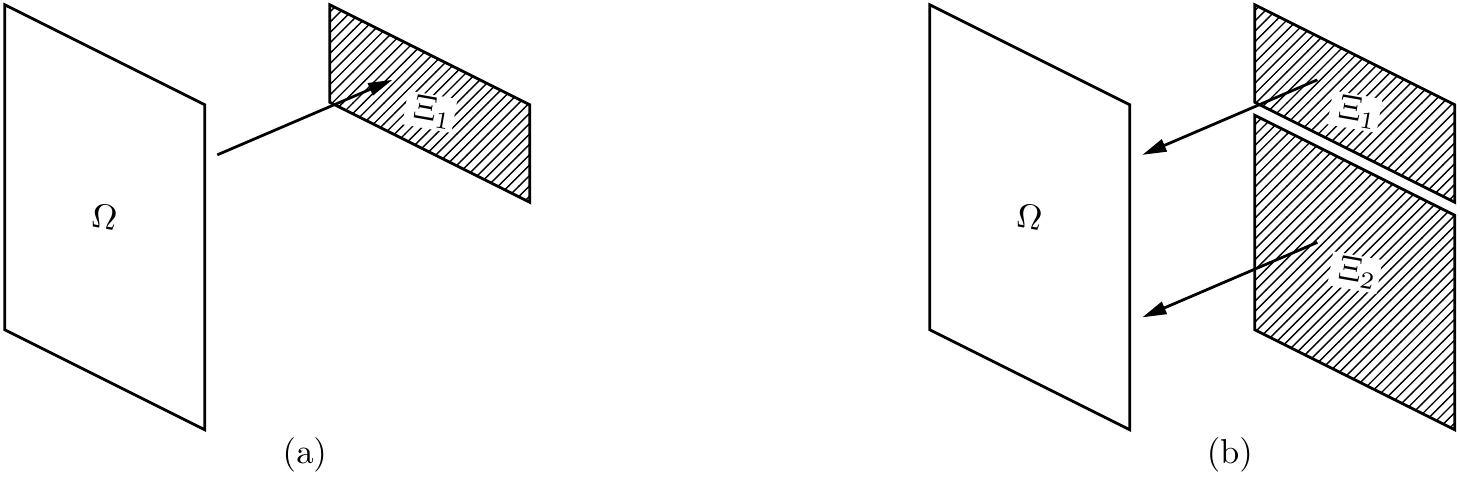}
\caption{Projections between cell face and mortars.}
\label{fig:prj}
\end{figure}

Any variable $\phi$ on a cell face $\Omega$ and on the left side of a mortar $\Xi$ can be represented as
\begin{align}
\phi^{\Omega}(\xi,\eta)   &= \sum_{i=1}^N \sum_{j=1}^N \phi^{\Omega}_{ij} h_i(\xi) h_j(\eta),\\[1mm]
\phi^{\Xi,L} (\xi',\eta') &= \sum_{i=1}^N \sum_{j=1}^N \phi^{\Xi,L}_{ij}  h_i(\xi')h_j(\eta'),
\end{align}
where the $(\phi^{\Xi,L}_{ij})^,$s are unknown and can be obtained through the following projection (refer to Fig. \ref{fig:prj}(a)),
\begin{equation}
\int_0^1 \int_0^1  (\phi^{\Xi,L}(\xi',\eta') -\phi^{\Omega}(\xi,\eta))
h_{\alpha} (\xi') h_{\beta}(\eta') \mathrm{d}\xi' \mathrm{d}\eta' = 0,
\quad \forall \alpha,\beta = 1,2,...,N.
\label{eq:project_1}
\end{equation}
The same process is repeated on the right side of the mortar. Following that, a common value, denoted as $\Phi$, are computed on mortars and then projected back to cell faces (see Fig. \ref{fig:prj}(b)) according to
\begin{equation}
\sum_{k=1}^{2} \int_{\xi=o_k}^{\xi=o_k+s_k} \int_{\eta=0}^{\eta=1} (\Phi^{\Omega}(\xi,\eta) - \Phi^{\Xi_k}(\xi',\eta')) h_\alpha(\xi) h_\beta(\eta) \dd{\xi} \dd{\eta}   = 0, ~~\forall\, \alpha,\beta=1,...,N.
\label{eq:prj2a}
\end{equation}
It can be shown that the two projections, (\ref{eq:project_1}) and (\ref{eq:prj2a}), are equivalent to a series of 1D projections \cite{zhang-2016b}, which make the process very efficient.

\section{Simulation Setup}
\label{sec:setup}

\subsection{Geometry}
\begin{figure}[!htb]
\centering
\begin{subfigure}[b]{0.30\textwidth}
\centering
\includegraphics[width=0.9\textwidth]{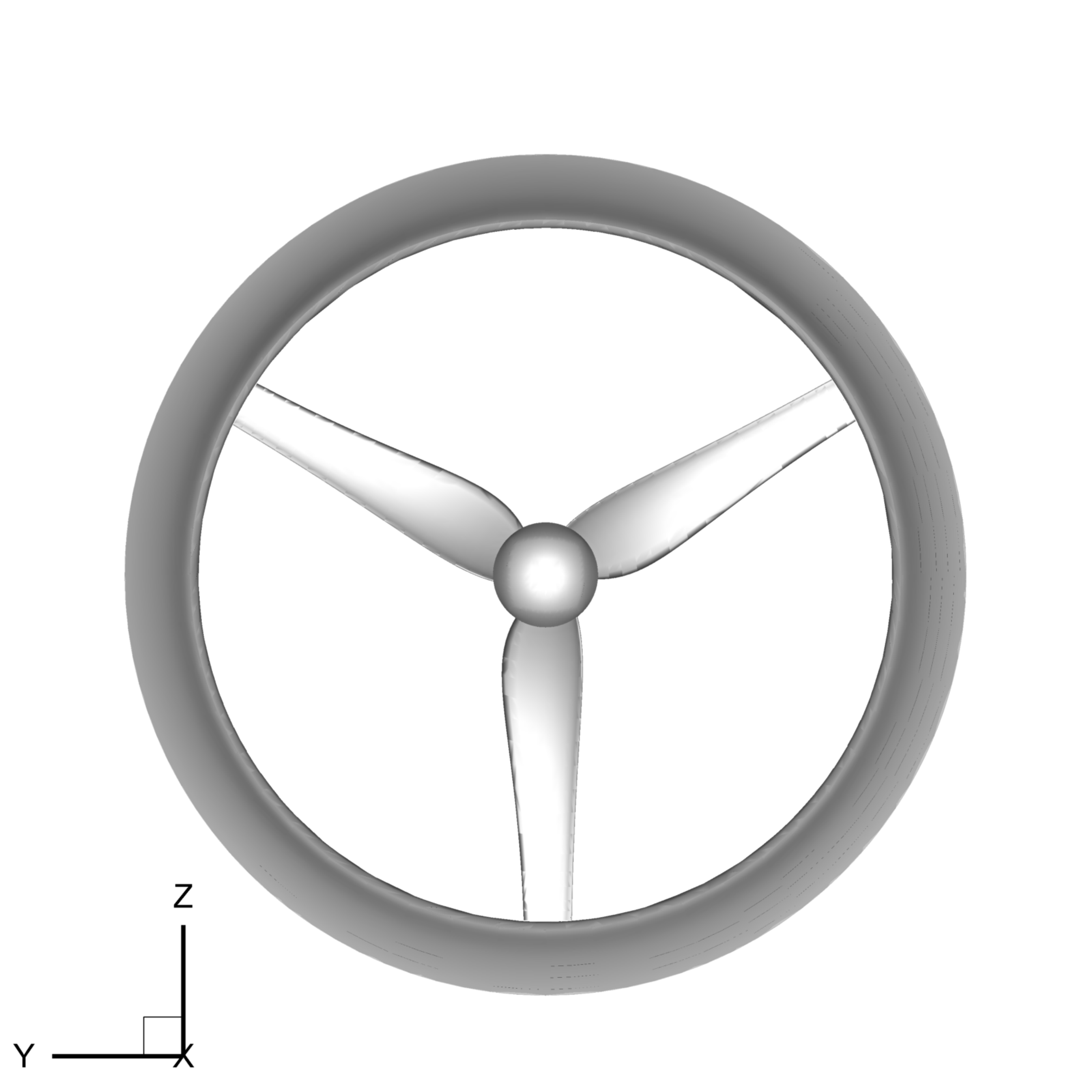}
\caption{}
\end{subfigure}
\begin{subfigure}[b]{0.30\textwidth}
\centering
\includegraphics[width=0.9\textwidth]{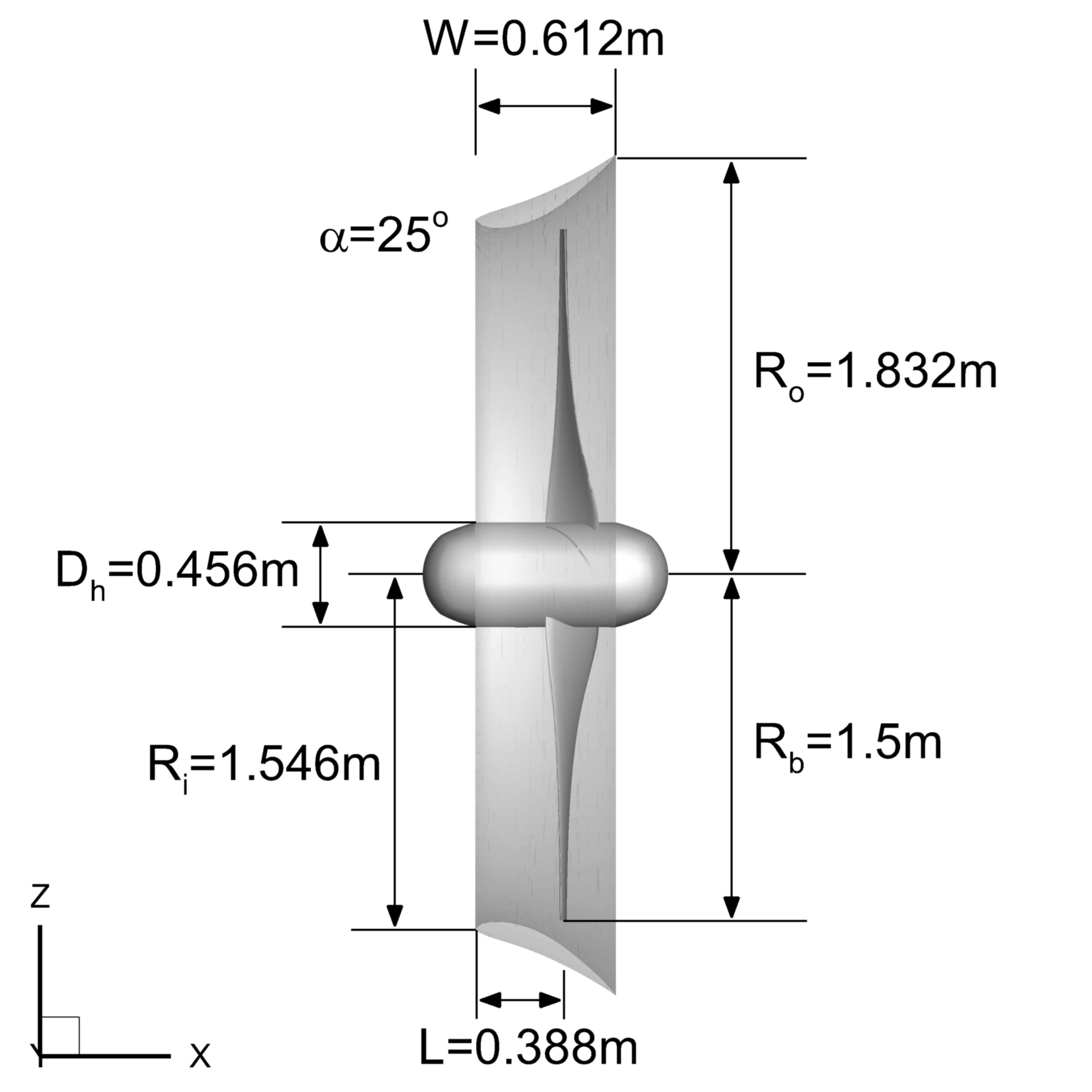}
\caption{}\label{fig:dwt_geo_side}
\end{subfigure}
\begin{subfigure}[b]{0.30\textwidth}
\centering
\includegraphics[width=0.8\textwidth]{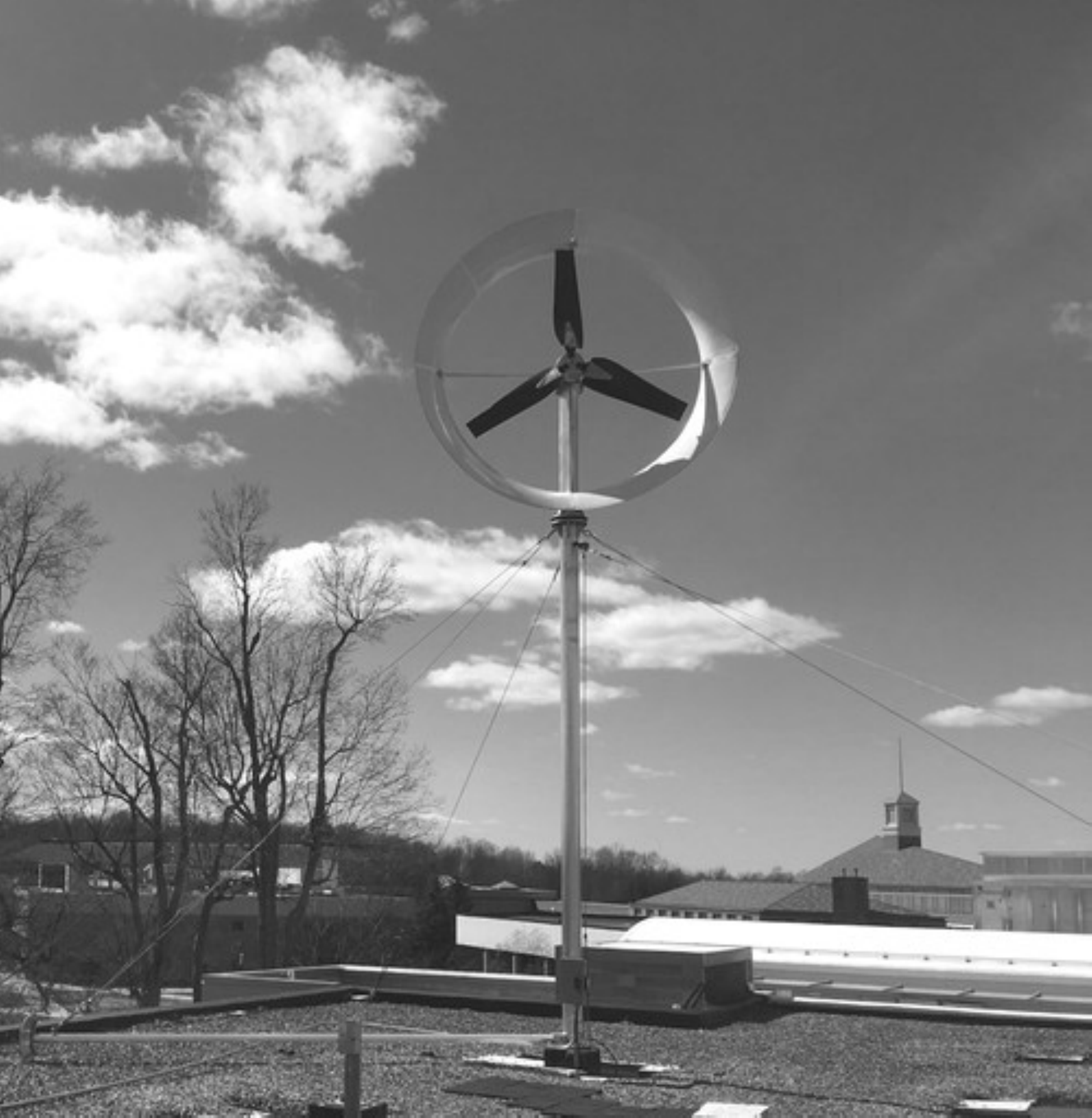}
\caption{}
\end{subfigure}
\caption{The ducted wind turbine geometry: (a) upstream view, (b) lateral view, (c) photo of the real product.}
\label{fig:dwt_geo}
\end{figure}

The DWT considered here was designed at Clarkson University by Dr. Kenneth Visser based on his previous experimental study \cite{kanya-2018}. Figure \ref{fig:dwt_geo} shows two views of the geometry and a photo of the real product. The diffuser duct has a profile of the E423 high lift airfoil, with an inlet radius of $R_i=1.546$m, an exit radius of $R_o=1.832$m, a width of $W=0.612$m, and an angle of attack of $\alpha=25 ^\circ$. The rotor has three 1.5m-long blades (i.e., $R_b=1.5$m) that are $0.388$m downstream from the inlet. More information about the blade shape can be found in \cite{kanya-2018}. The cylindrical hub has a diameter of $D_h=0.456$m and is closed by two hemispherical ends of the same diameter. This configuration results in a small gap of approximately 0.113m between the blade tips and the duct's inner surface.

\subsection{Meshes and Boundary Conditions}
\begin{figure}[!htb]
\centering
\begin{subfigure}[b]{0.3\textwidth}
\centering
\includegraphics[width=0.75\textwidth]{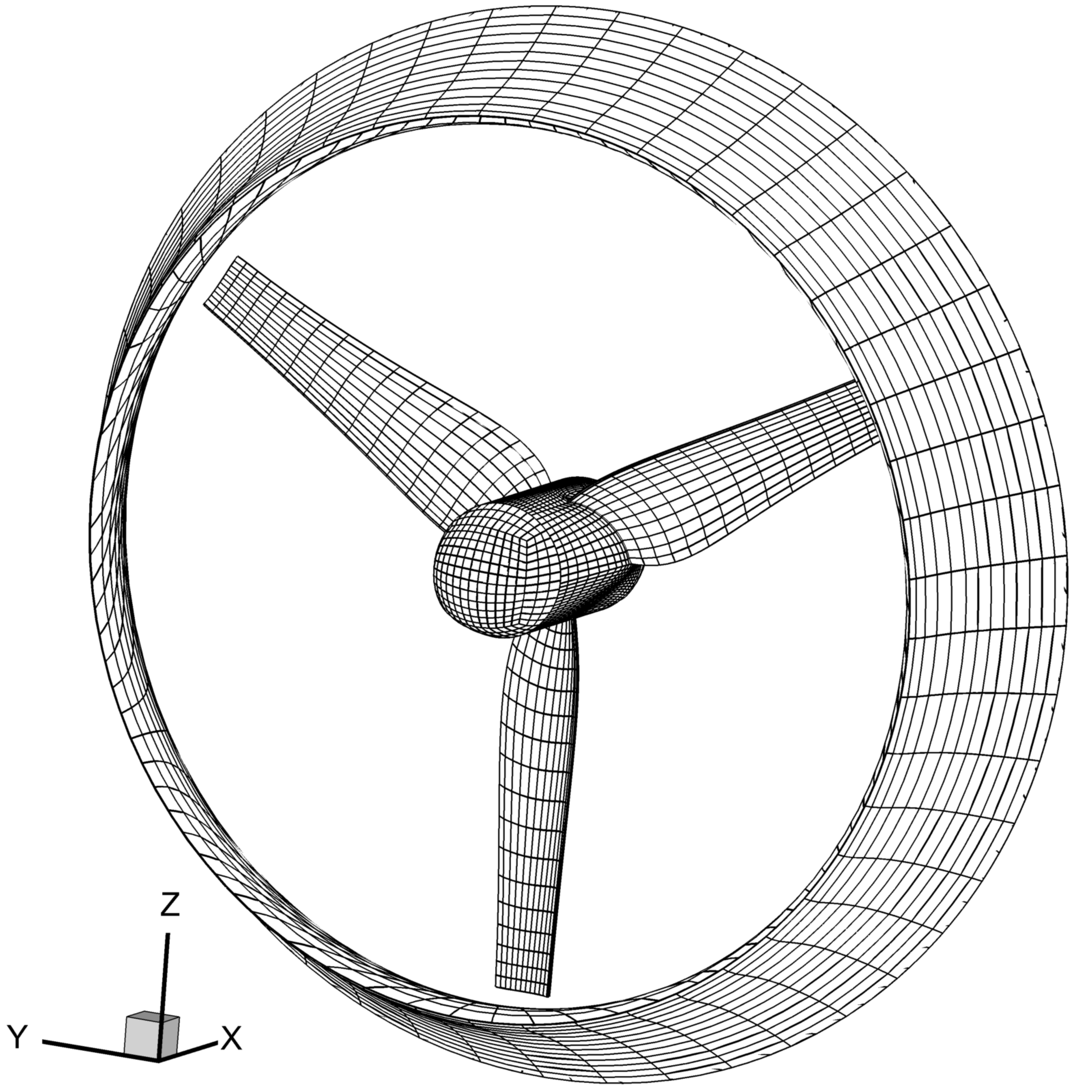}
\caption{}\label{fig:dwt_surface}
\end{subfigure}
\begin{subfigure}[b]{0.3\textwidth}
\centering
\includegraphics[width=0.75\textwidth]{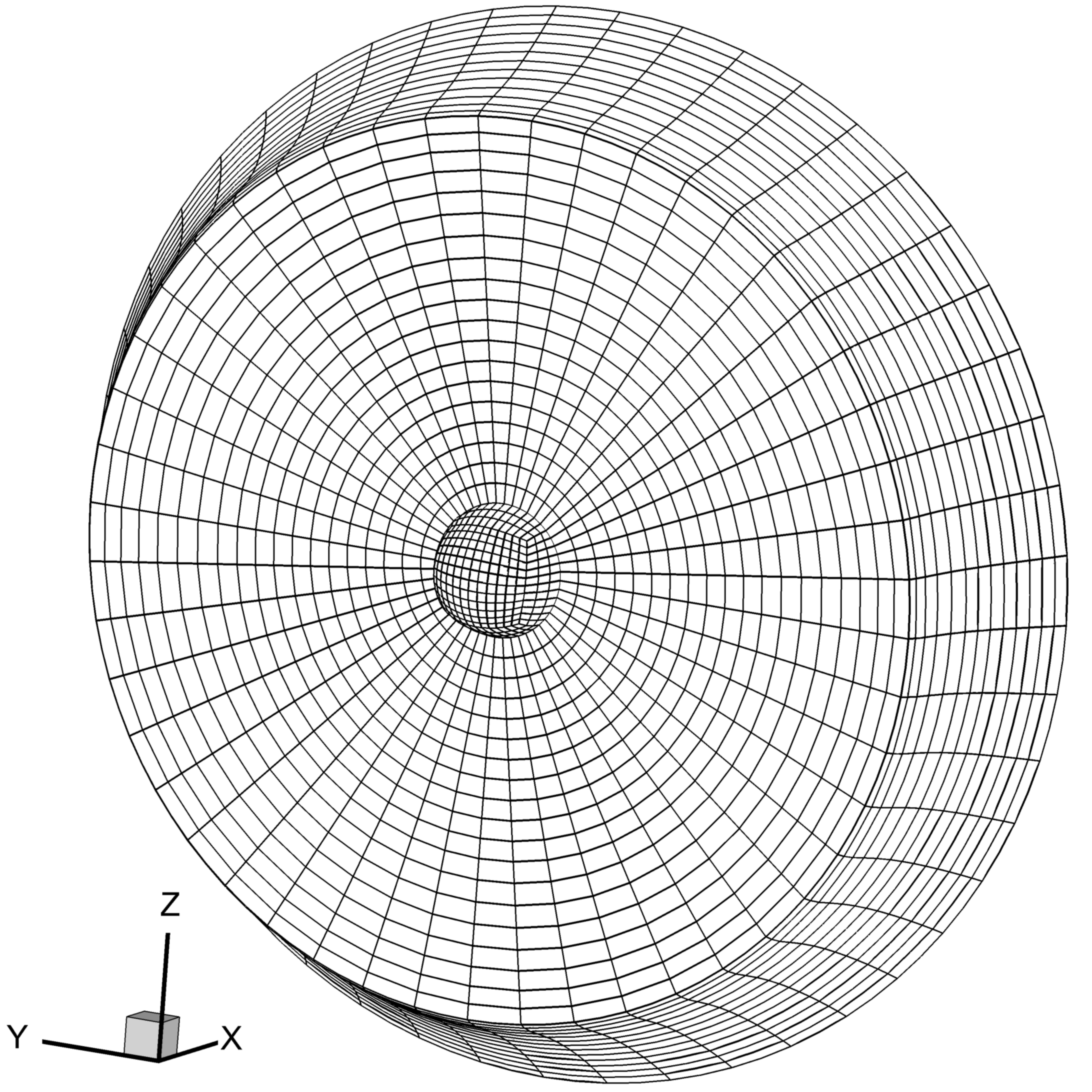}
\caption{}\label{fig:dwt_sliding}
\end{subfigure}
\begin{subfigure}[b]{0.3\textwidth}
\centering
\includegraphics[width=0.75\textwidth]{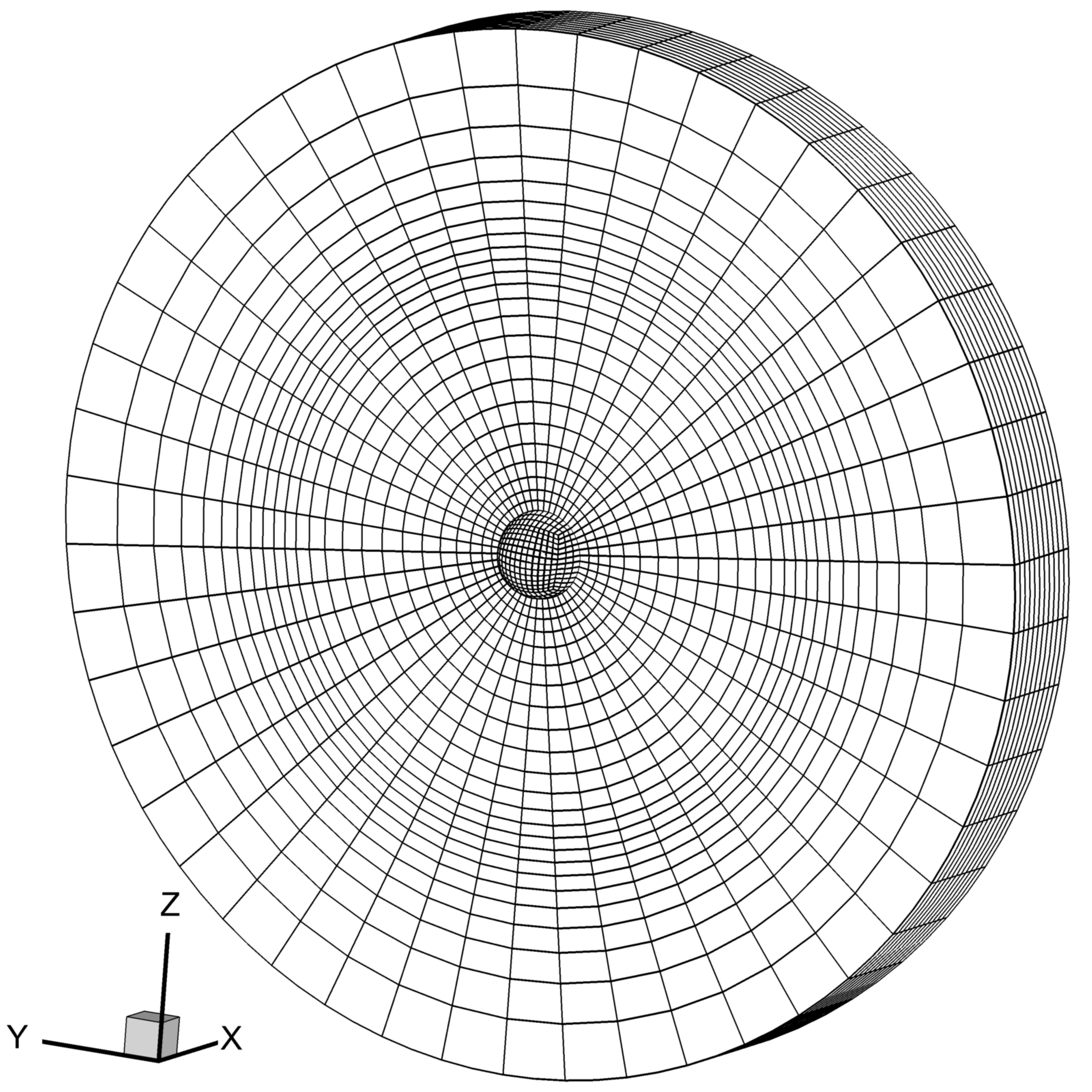}
\caption{}\label{fig:owt_sliding}
\end{subfigure}
\caption{Surface meshes: (a) for DWT, (b) for DWT's sliding interfaces, (c) for OWT's sliding interfaces.}
\label{fig:surface_mesh}
\end{figure}

Figure \ref{fig:dwt_surface} shows the surface meshes of the DWT's wall boundaries. Along these boundaries, the first layer of volume mesh has a height of approximately $3\times10^{-3}D_o$, and the first layer of solution points is about $1.4\times10^{-4}D_o$ off the walls (for a fifth-order scheme), where $D_o=2R_o$ is the diameter of the duct's exit. Figure \ref{fig:dwt_sliding} shows the meshes on the upstream sliding interface and the duct's inner surface. In fact, there are two sliding interfaces, and the other one is downstream at the duct's exit. Because the gap between the blade tips and the duct's inner surface is too small to allow a third sliding interface there, the duct's inner surface is set to rotate with the two sliding interfaces. The speed on this inner surface is then overwritten to zero. The surface mesh of the OWT's rotor is identical to that of the DWT. The OWT allows three sliding interfaces that form a disk region, as shown in Fig. \ref{fig:owt_sliding}. This region has a thickness of $W$ and diameter of $1.5D_o$ and encloses the blades. For both turbines, the hubs and the hemispherical ends rotate at the same angular speed as the blades.

\begin{figure}[!htb]
\centering
\begin{subfigure}[b]{0.42\textwidth}
\centering
\includegraphics[width=0.8\textwidth]{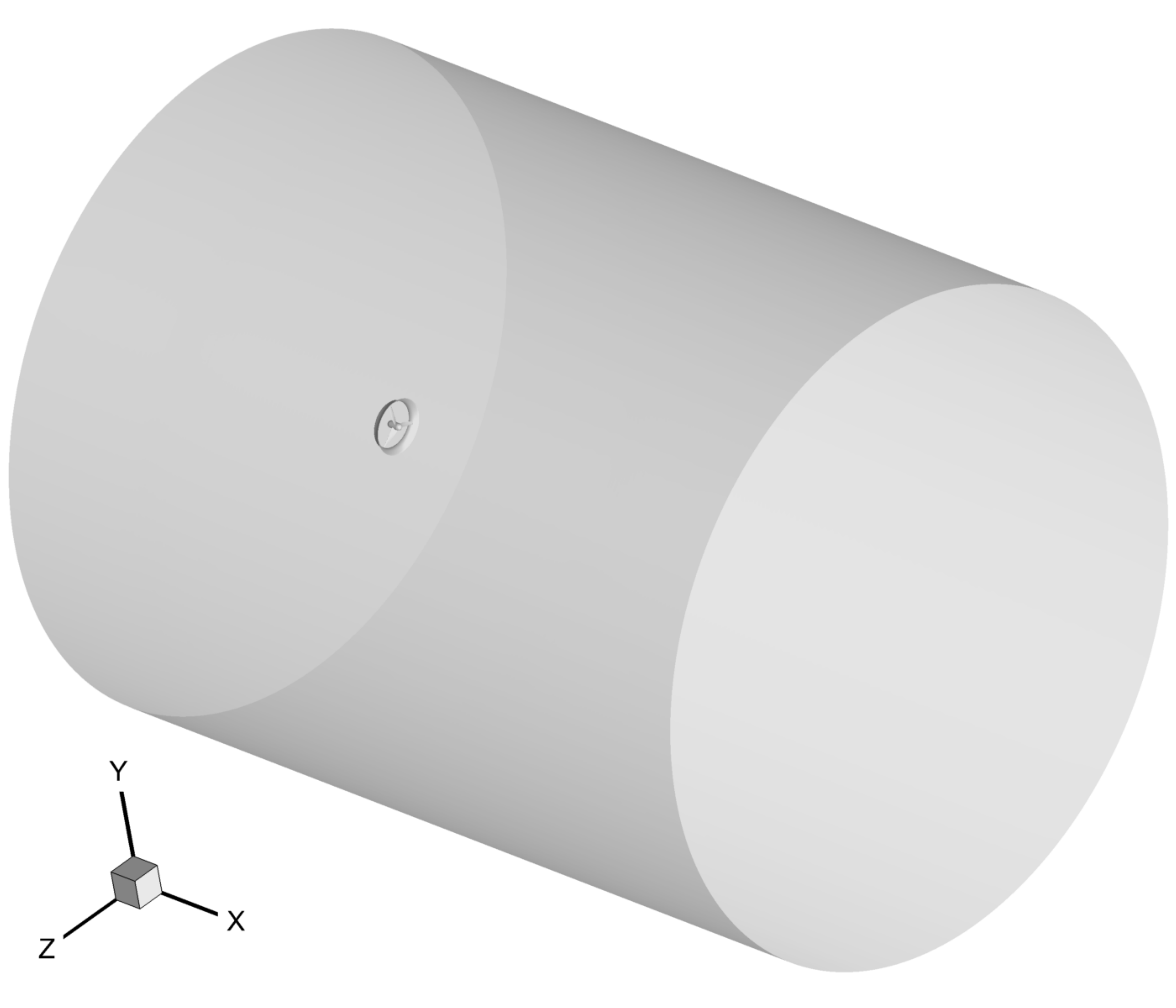}
\caption{}\label{fig:geo_overall}
\end{subfigure}
\begin{subfigure}[b]{0.42\textwidth}
\centering
\includegraphics[width=0.8\textwidth]{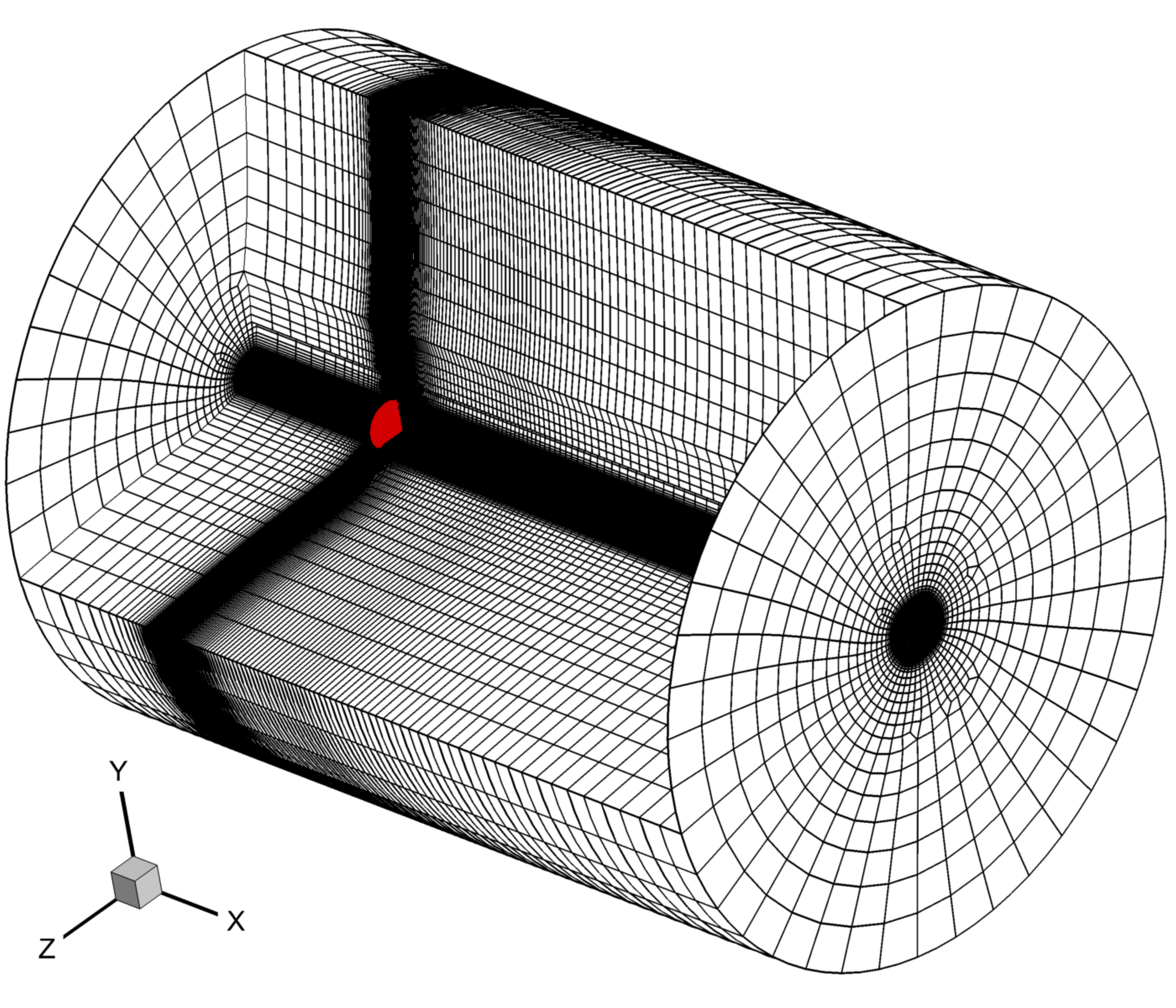}
\caption{}\label{fig:msh_overall}
\end{subfigure}
\caption{Overall computational domain and mesh: (a) domain, (b) mesh.}
\label{fig:dwt_mesh}
\end{figure}

The overall computational domain for the DWT has a cylindrical shape, as shown in Fig. \ref{fig:geo_overall}. The length and diameter of the domain are both $12D_o$. The domain for the OWT has the same shape and sizes. The resulting blockage ratio is $0.69\%$ for the DWT and $0.47\%$ for the OWT. The overall mesh with a 1/4 cutout is shown in Fig. \ref{fig:msh_overall}, where the red area represents a sliding subdomain. For the DWT, the sliding subdomain has 29,469 elements, and the outer subdomain has 266,969 elements, resulting in 296,438 total elements (or 37.1 million degrees of freedom (DOFs) for a fifth-order scheme). For the OWT, the sliding region has 54,432 elements, and the outer subdomain has 240,905 elements, adding up to 295,337 elements in total (or 37.0 DOFs for a fifth-order scheme). The inlet is treated as a Dirichlet boundary. The outlet and outer cylindrical surface are treated as characteristic farfields. All solid walls are treated as no-slip adiabatic walls. The Mach number of the incoming freestream flow is set to $Ma_{\infty}=0.08$ to ensure negligible compressibility effect.

\subsection{Nondimensional Parameters}

\begin{table}[H]
\centering
\begin{tabular}{cc}
\hline
Parameter                         &  Value					          \\
\hline
Freestream velocity ($U_\infty$)  &  10 m/s       		      	\\
Kinematic viscosity ($\nu$)       &  $1.4616\times10^{-5}$ m\textsuperscript{2}/s  	\\
Rotation axis                     &  $x$-direction            	\\
Rotation speed ($\omega$)         &  $-26.18$ rad/s  	     		\\
\hline
\end{tabular}
\caption{Physical operating conditions.}
\label{tab.para}
\end{table}

The most common physical operating conditions for the present DWT are listed in Table \ref{tab.para}. Once the geometry of the turbine is given, there are two nondimensional parameters that govern the flow. One is the Reynolds number $Re$, and the other is the tip speed ratio $\lambda$. In this study, we take the following definitions
\begin{equation}
Re=\frac{U_\infty D_o}{\nu} \quad\text{and}\quad \lambda=\frac{\omega R_b}{U_\infty}.
\end{equation}
The values in Table \ref{tab.para} give $Re = 2.5\times10^6$ and $\lambda=3.93$. Hereinafter, the Reynolds number is kept constant for all the cases. Two more values, $\lambda=3.11$ and $\lambda=4.75$, are studied for some cases to examine the effects of the tip speed ratio.

In what follows, all the simulations use a small nondimensional time step size of $\Delta t^* = \Delta t U_{\infty}/D_o = 1.25\times 10^{-5}$ for stability consideration. This time step size corresponds to a blade rotation between 0.0054 degree (for $\lambda=3.11$) and 0.0083 degree (for $\lambda=4.75$), which is small enough to provide sufficient sampling resolution. Each simulation was ran for 25 nondimensional time units, and data sets from the last 12.5 time units were used to calculate the statistics.

\subsection{Resolution Verification}
For the method used, the total number of DOFs is $N_\text{DOF} = N_\text{elem}\cdot N^3$, where $N_\text{elem}$ is the total number of mesh elements, and $N$ is the scheme order. Once a mesh is generated, we can vary the scheme order $N$ to identify the resolution requirement for a simulation. The Reynolds number in this study is extremely high, which makes it prohibitively expensive to get fully resolved solutions (i.e., to resolve turbulent eddies of all sizes). Thus, we need to identify a proper scheme order that gives a good balance between computational cost and solution qualities.

To do this, we employ the thrust coefficient $C_T$ and power coefficient $C_P$ as two measurement criteria. These two terms are defined as below
\begin{gather}
C_T = \frac{T}{\frac{1}{2}\rho U_{\infty}^2 A_\text{rot}}, \\[1ex]
C_P = \frac{Q\omega}{\frac{1}{2}\rho U_{\infty}^3 A_\text{rot}},
\end{gather}
where $T$ and $Q$, respectively, are the thrust (in fact drag for a turbine) and the torque (only the $x$-component) acted on turbine blades, and $A_\text{rot}=\pi R_b^2$ is the swept area of rotor blades.

Three scheme orders, $N=4$, $5$, and $6$, are applied to the OWT at the operating conditions listed in Table \ref{tab.para}. The time-averaged results are summarized in Table \ref{tab.validation}. It is observed that the $\overline{C}_T$'s from $N=4$ and $5$ are $16\%$ and $5\%$, respectively, smaller than that from $N=6$. Meanwhile, the $\overline{C}_P$'s from $N=4$ and $5$ are $32\%$ and $7\%$, respectively, smaller than that from $N=6$. The data obviously shows a converging trend as $N$ increases. Going from $N=5$ to $N=6$ increases the computational cost by $73\%$, but the changes in the coefficients are relatively small. Thus $N=5$ is chosen as the scheme order to perform the simulations.
\begin{table}[H]
\centering
\begin{tabular}{cccc}
\hline
&   $N=4$ & $N=5$ & $N=6$\\
\hline
$\overline{C}_T$ & 5.31E-01  & 6.02E-01  & 6.34E-01 \\
$\overline{C}_P$ & 2.74E-01  & 3.74E-01  & 4.02E-01 \\
\hline
\end{tabular}
\caption{Results from different scheme orders.}
\label{tab.validation}
\end{table}

\section{Simulation of Turbines in Axial Flows}
\label{sec:axial}

In this section, the results and analysis for the axial flow cases are presented, in which the free stream moves along the turbines' axial direction. Besides the designed tip speed ratio $\lambda = 3.93$ ($\omega^*=\omega D_o/U_\infty = -9.60$), two other tip speed ratios: $\lambda = 3.11$ ($\omega^*=-7.60$) and $\lambda = 4.75$ ($\omega^*=-11.60$), are considered, and the loads, vortex fields and velocity fields are analyzed.

\subsection{Load Analysis}
\begin{figure}[!htb]
\centering
\begin{subfigure}[b]{0.495\textwidth}
\includegraphics[width=0.99\textwidth]{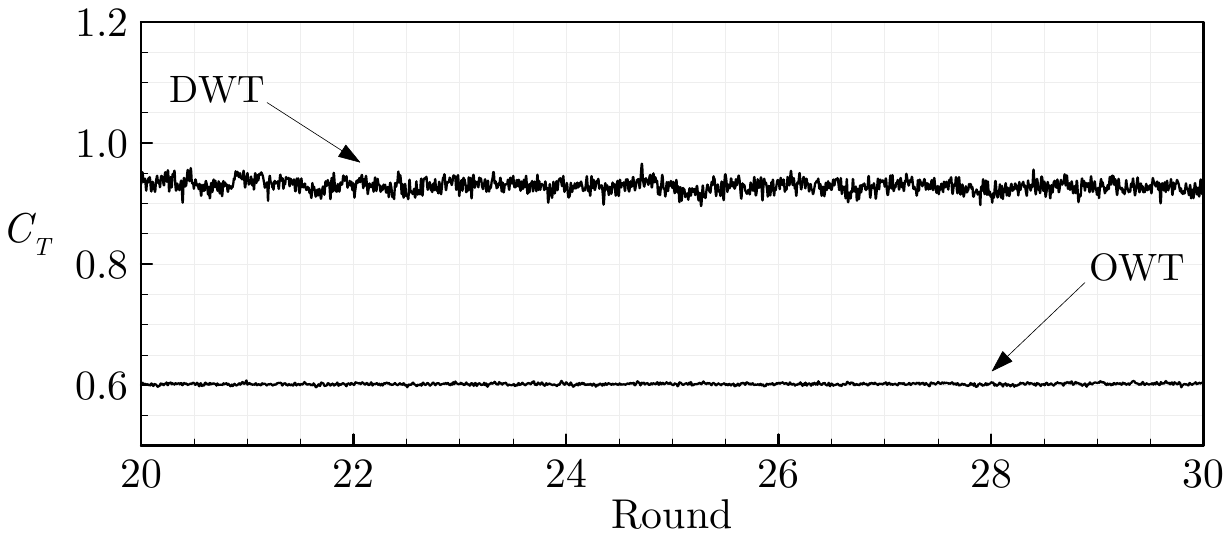}
\caption{Thrust}
\label{fig:dwt_history}
\end{subfigure}
\begin{subfigure}[b]{0.495\textwidth}
\includegraphics[width=0.99\textwidth]{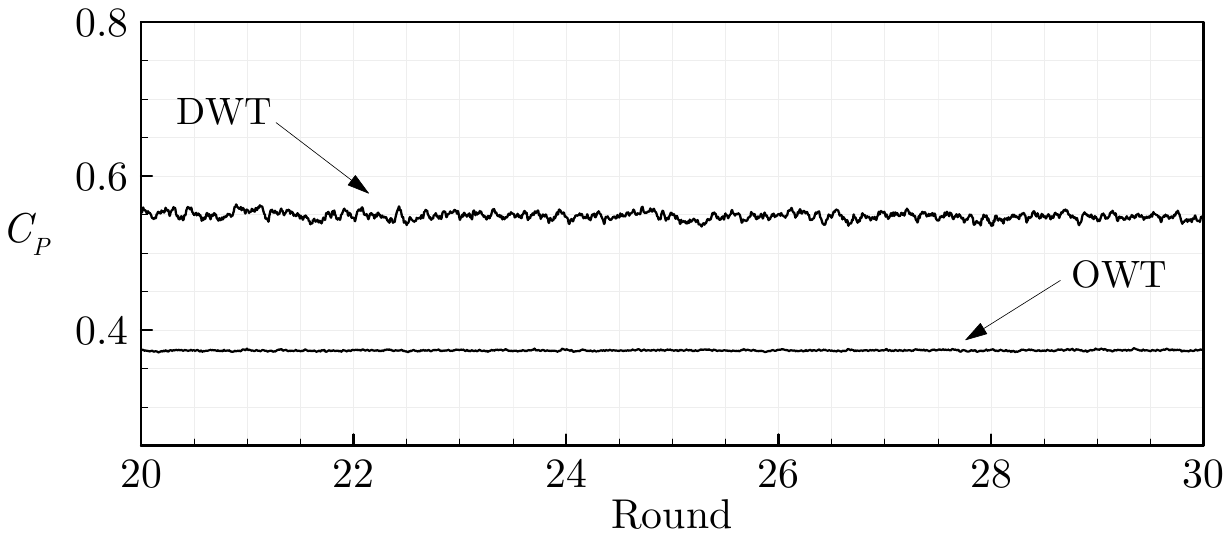}
\caption{Power}
\label{fig:owt_history}
\end{subfigure}
\caption{Time histories of the loads for the two configurations at $\lambda=3.93$.}
\label{fig:history}
\end{figure}

Figure \ref{fig:history} demonstrates the time histories of $C_T$ and $C_P$ for the two configurations at the designed tip speed ratio. It is seen that the thrust and power outputs for both configurations are well converged, and the DWT's loads have larger fluctuations than those of the OWT, which indicates that the flow over the DWT is more turbulent.

\begin{table}[!htb]
\centering
\begin{tabular}{cccccccc}
\hline
&  & $C_T$ & $C_{T,p}$ & $C_{T,v}$ &  $C_{P}$ & $C_{P,p}$ & $C_{P,v}$\\
\hline
\multirow{ 2}{*}{DWT}     & mean  &  9.28E-01 & 9.28E-01 & 1.97E-04 & 5.47E-01 & 5.49E-01 &-1.20E-03\\
& r.m.s &  9.79E-03 & 9.79E-03 & 5.79E-07 & 4.87E-03 & 4.87E-03 & 8.27E-06 \\
\hline
\multirow{ 2}{*}{OWT} & mean  & 6.02E-01 & 6.01E-01 & 1.91E-04 & 3.74E-01 & 3.75E-01 & -1.44E-03 \\
& r.m.s & 1.55E-03 & 1.55E-03 & 5.41E-07 & 7.38E-04 & 7.39E-04 & 1.45E-06 \\
\hline
\end{tabular}
\caption{Loads and their components of the two turbines at $\lambda=3.93$.}
\label{tab.load}
\end{table}

The mean (time-averaged) values, r.m.s (root-mean-square) deviations, and the corresponding contributions from pressure (with subscript `$p$') and viscosity (with subscript `$v$') of the loads are listed in Table \ref{tab.load}. Overall, the r.m.s values are about two orders of magnitude smaller than the mean values, suggesting that the loads are relatively steady. Meanwhile, viscosity contributions are over three orders of magnitude smaller than pressure contributions for $C_T$, and about two orders of magnitude smaller for $C_P$, revealing that pressure plays a much dominant role on the loads. Moreover, viscosity has positive contributions to the mean thrust but negative contributions to the power output. Most interestingly, the DWT's mean thrust coefficient is about $54\%$ higher, and the mean power coefficient is about $46\%$ higher than the OWT's.

\begin{figure}[!htb]
\centering
\begin{subfigure}[b]{0.45\textwidth}
\centering
\includegraphics[width=2.6in]{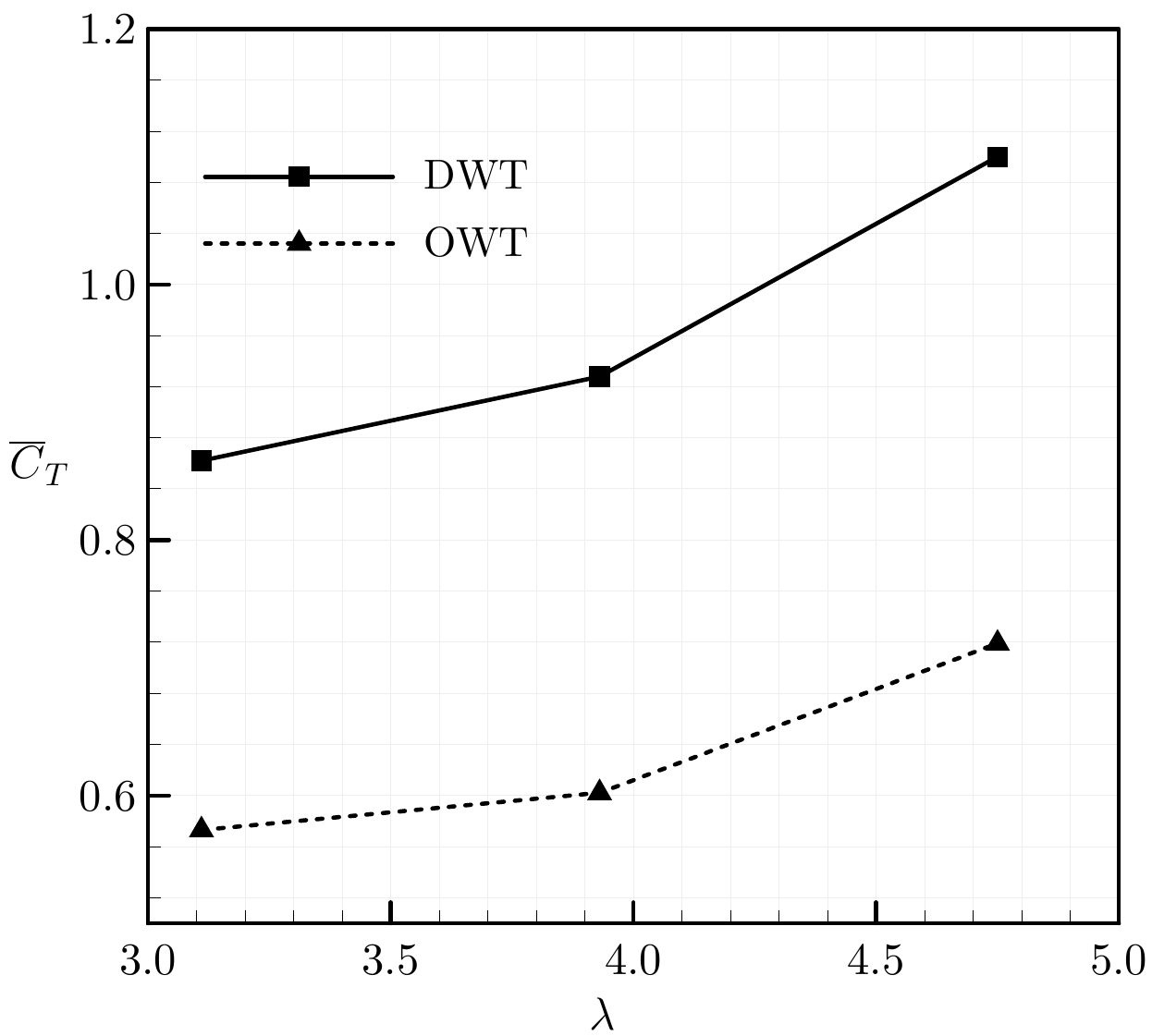}
\caption{Thrust}
\label{fig:dwt_perform}
\end{subfigure}
\begin{subfigure}[b]{0.45\textwidth}
\centering
\includegraphics[width=2.6in]{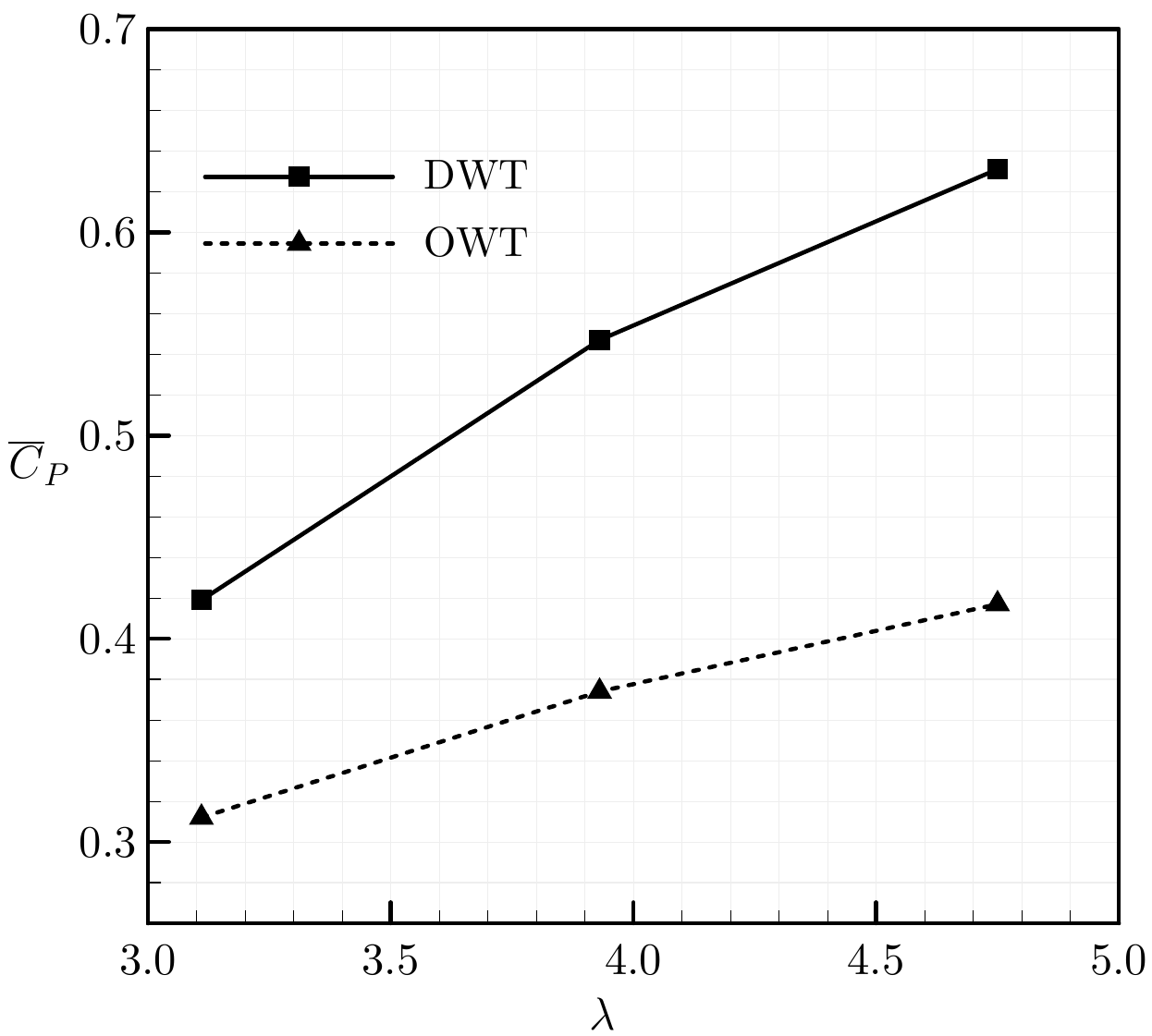}
\caption{Power}
\label{fig:owt_perform}
\end{subfigure}
\caption{Mean loads for different tip-speed ratios.}
\label{fig:perform}
\end{figure}

The time-averaged load coefficients (denoted by $\overline{C}_T$ and $\overline{C}_P$) at the three tip speed ratios are shown in Fig. \ref {fig:perform}. For both turbines, $\overline{C}_T$ and $\overline{C}_P$ increase with $\lambda$ in this range. The $\overline{C}_T$'s of the two turbines grow at almost the same pace (with a difference of approximately 0.3 between the two curves), while the $\overline{C}_P$ of the DWT grows faster than that of the OWT.

\begin{figure}[!htb]
\centering
\begin{subfigure}[b]{0.28\textwidth}
\centering
\includegraphics[width=1.8in]{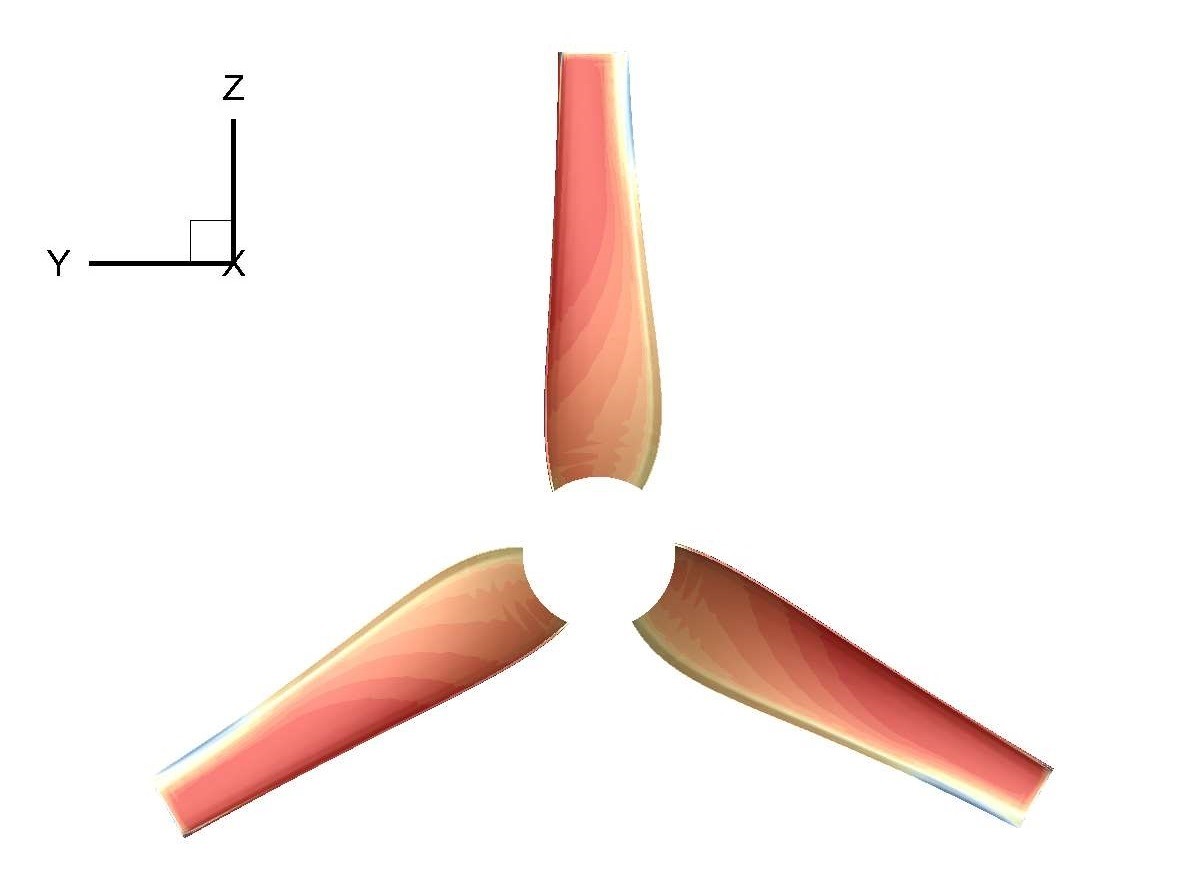}
\caption{$\lambda=3.11$}
\label{fig:dwt_l_3.7_p_up}
\end{subfigure}
\begin{subfigure}[b]{0.28\textwidth}
\centering
\includegraphics[width=1.8in]{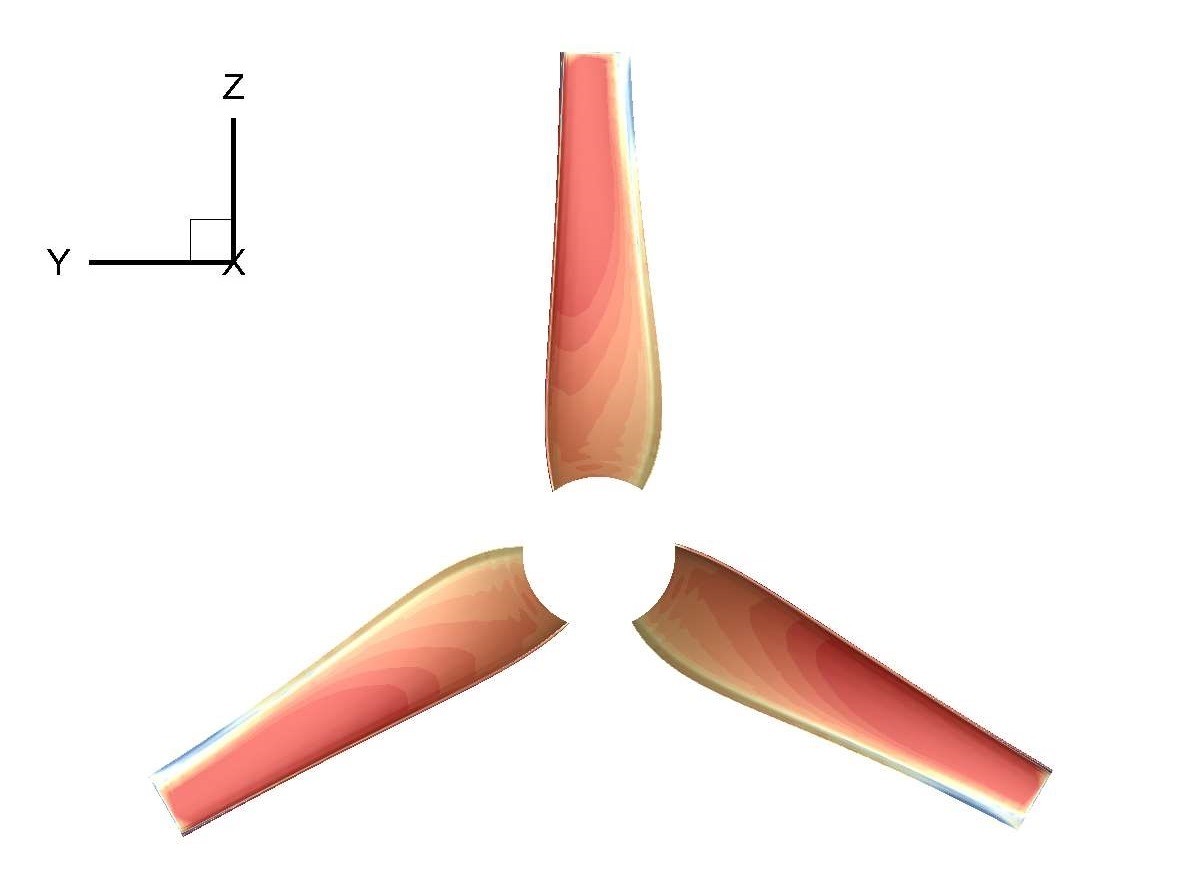}
\caption{$\lambda=3.93$}
\label{fig:dwt_l_4.7_p_up}
\end{subfigure}
\begin{subfigure}[b]{0.3\textwidth}
\centering
\includegraphics[width=2.0in]{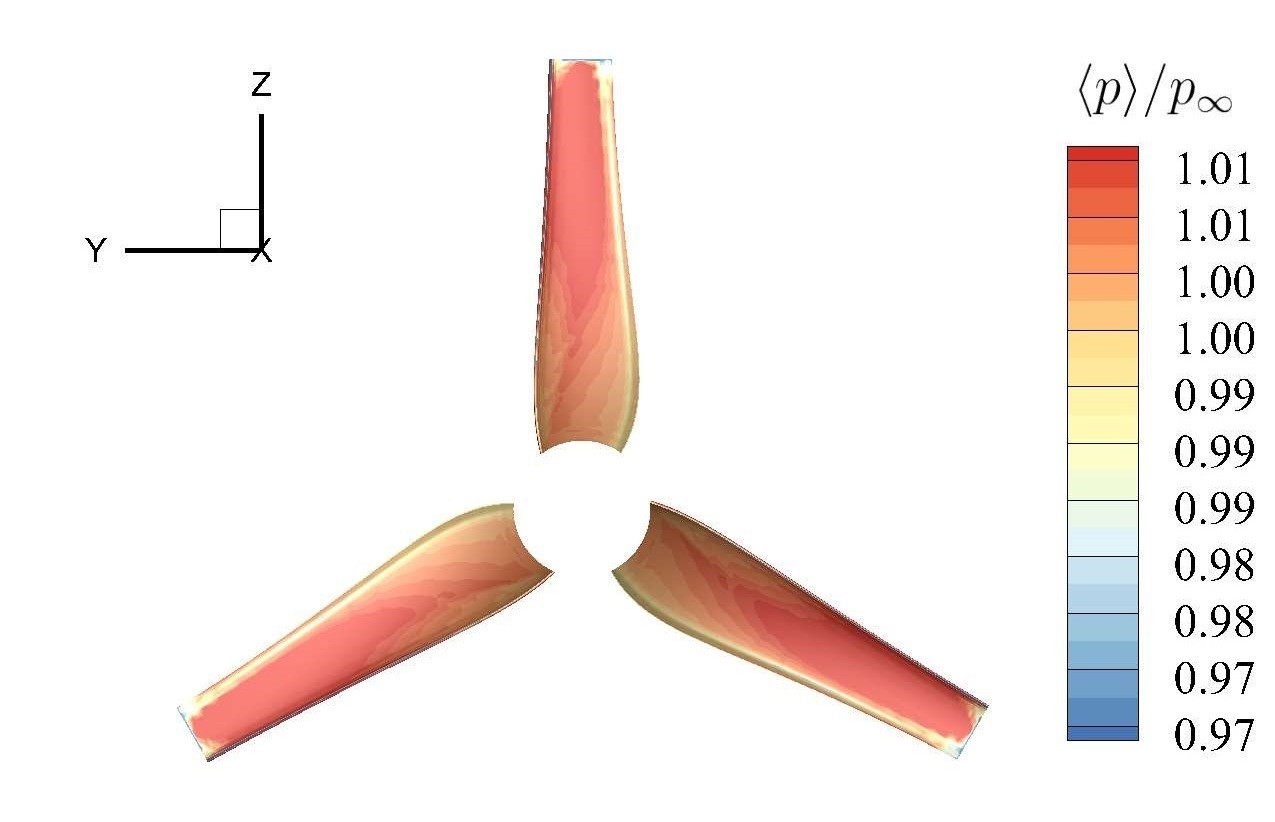}
\caption{$\lambda=4.75$}
\label{fig:dwt_l_5.7_p_up}
\end{subfigure}
\caption{The DWT's phase-averaged upstream surface pressure.}
\label{fig:dwt_p_up}
\end{figure}

\begin{figure}[!htb]
\centering
\begin{subfigure}[b]{0.28\textwidth}
\centering
\includegraphics[width=1.8in]{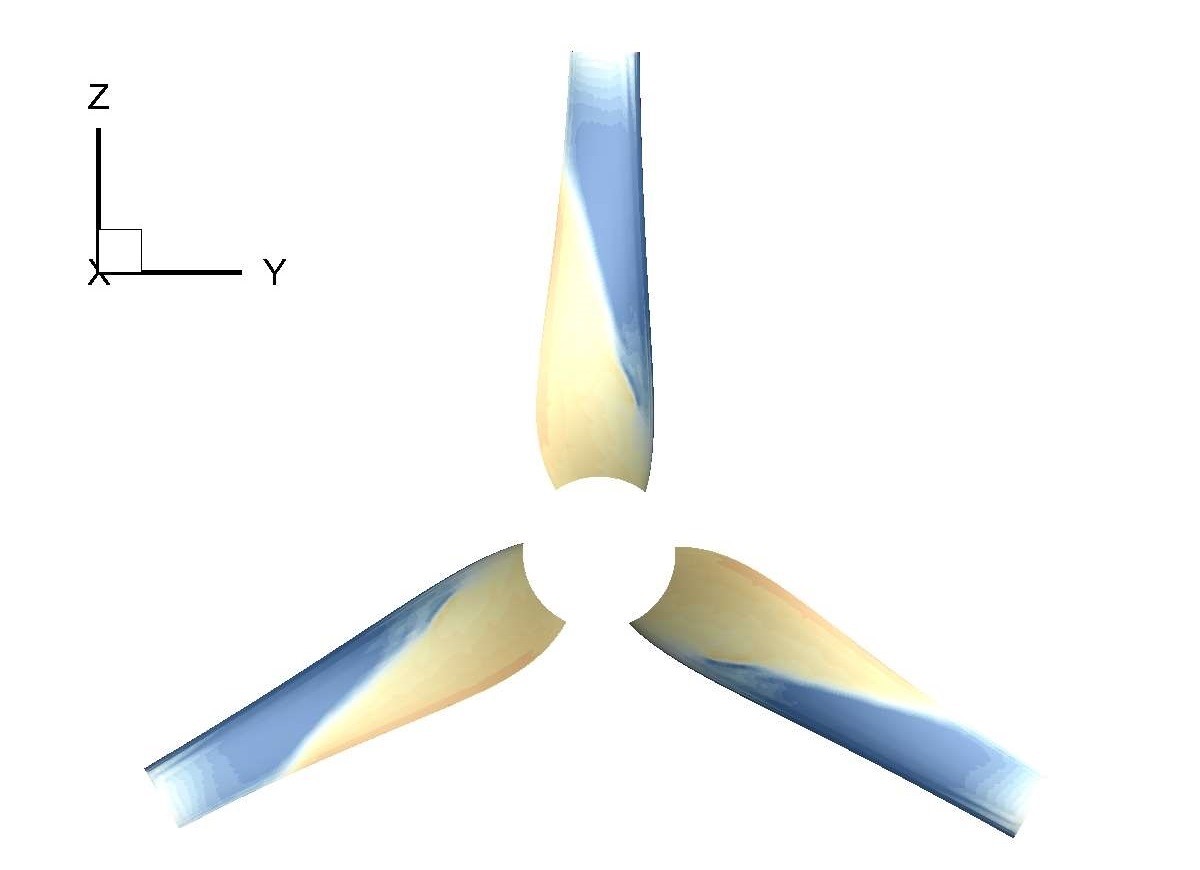}
\caption{$\lambda=3.11$}
\label{fig:dwt_l_3.7_p_down}
\end{subfigure}
\begin{subfigure}[b]{0.28\textwidth}
\centering
\includegraphics[width=1.8in]{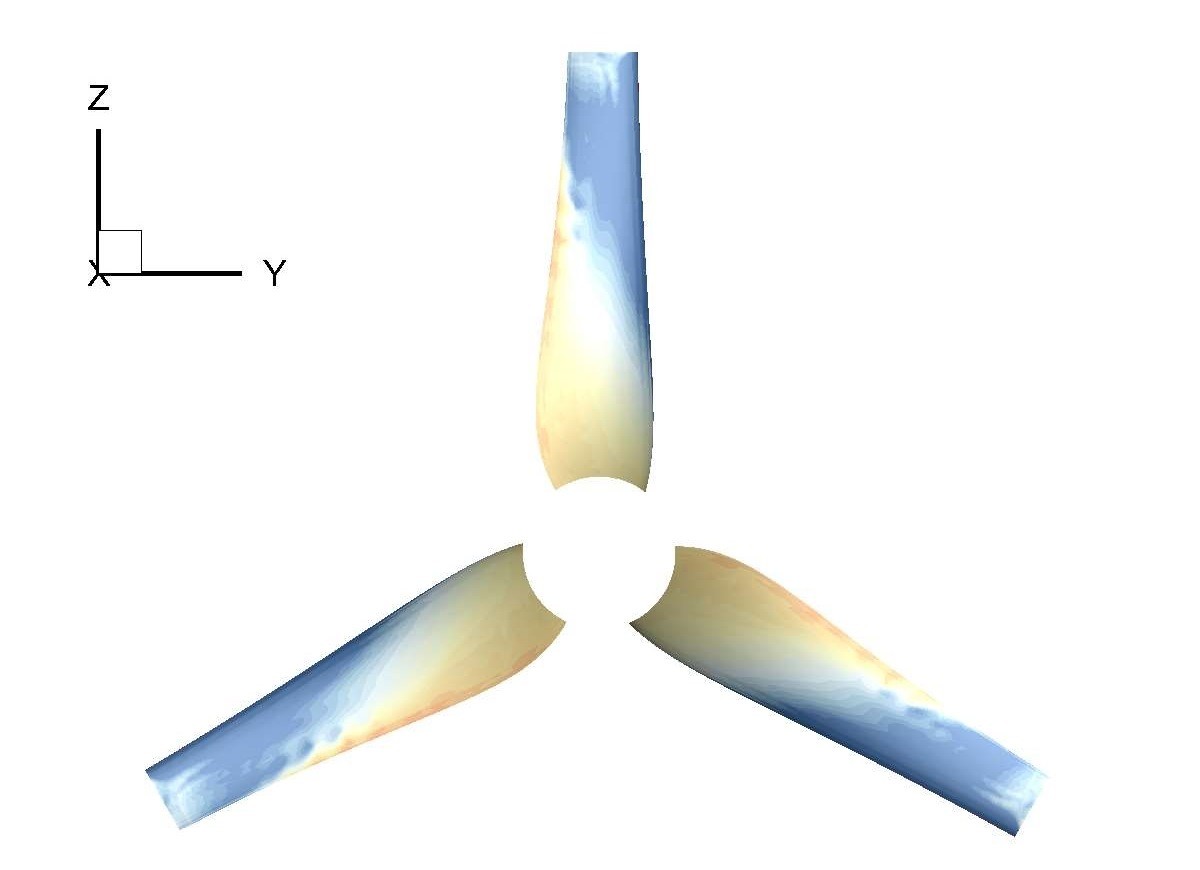}
\caption{$\lambda=3.93$}
\label{fig:dwt_l_4.7_p_down}
\end{subfigure}
\begin{subfigure}[b]{0.3\textwidth}
\centering
\includegraphics[width=2.0in]{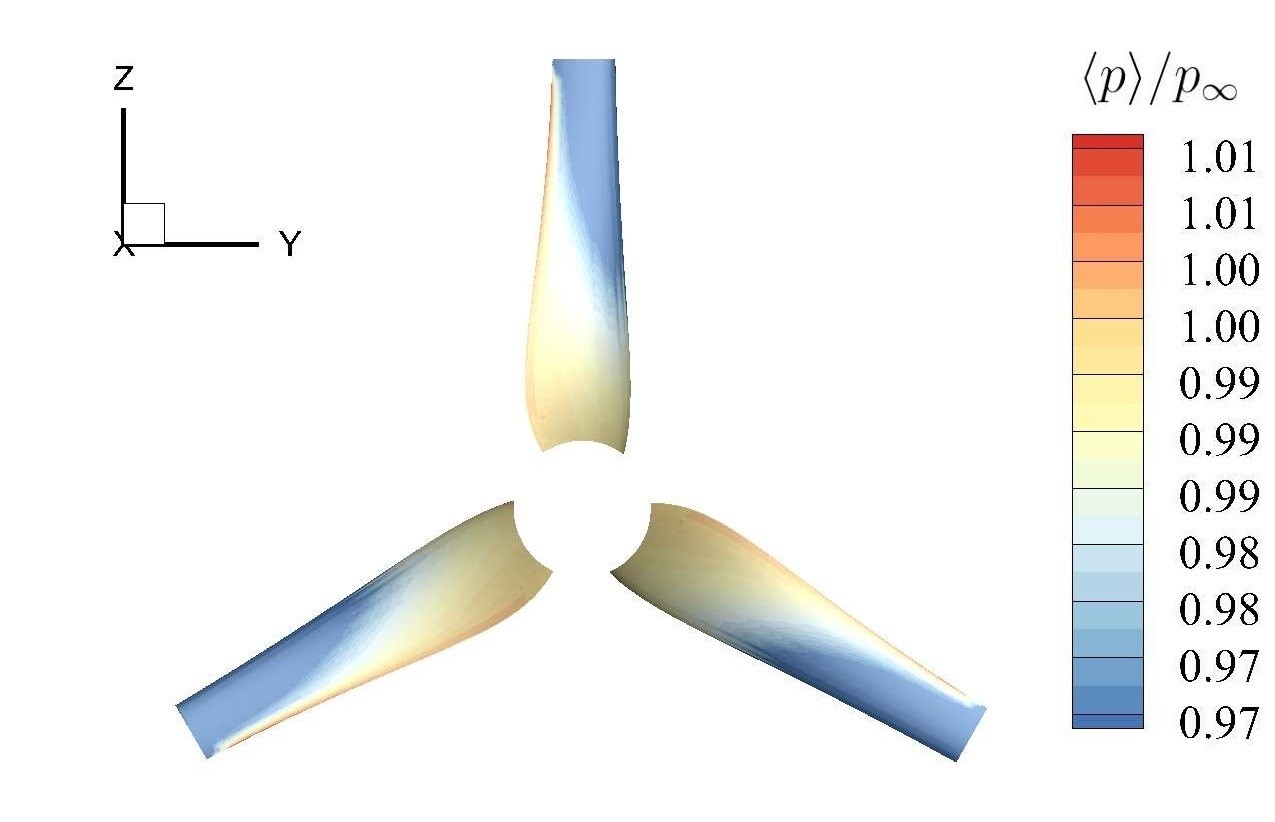}
\caption{$\lambda=4.75$}
\label{fig:dwt_l_5.7_p_down}
\end{subfigure}
\caption{The DWT's phase-averaged downstream surface pressure.}
\label{fig:dwt_p_down}
\end{figure}

\begin{figure}[!htb]
\centering
\begin{subfigure}[b]{0.28\textwidth}
\centering
\includegraphics[width=1.8in]{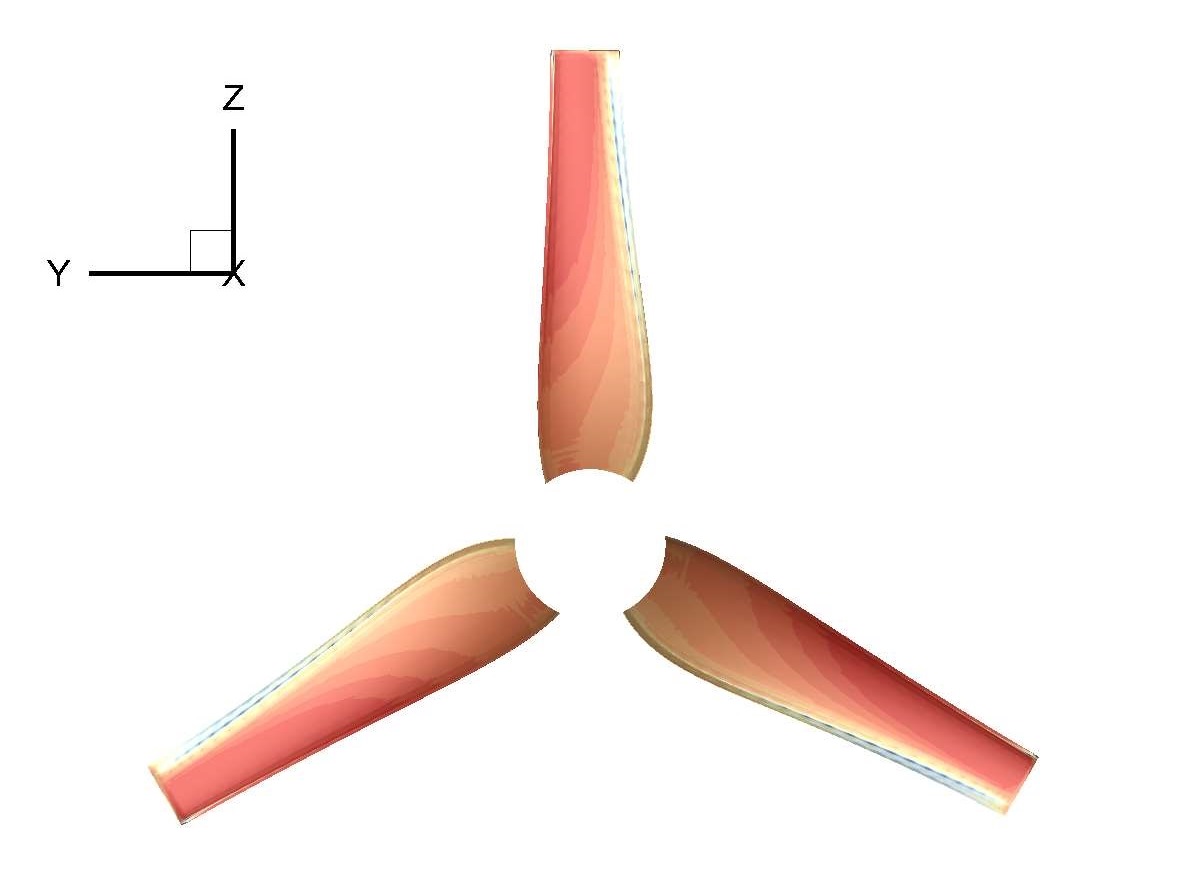}
\caption{$\lambda=3.11$}
\label{fig:owt_l_3.7_p_up}
\end{subfigure}
\begin{subfigure}[b]{0.28\textwidth}
\centering
\includegraphics[width=1.8in]{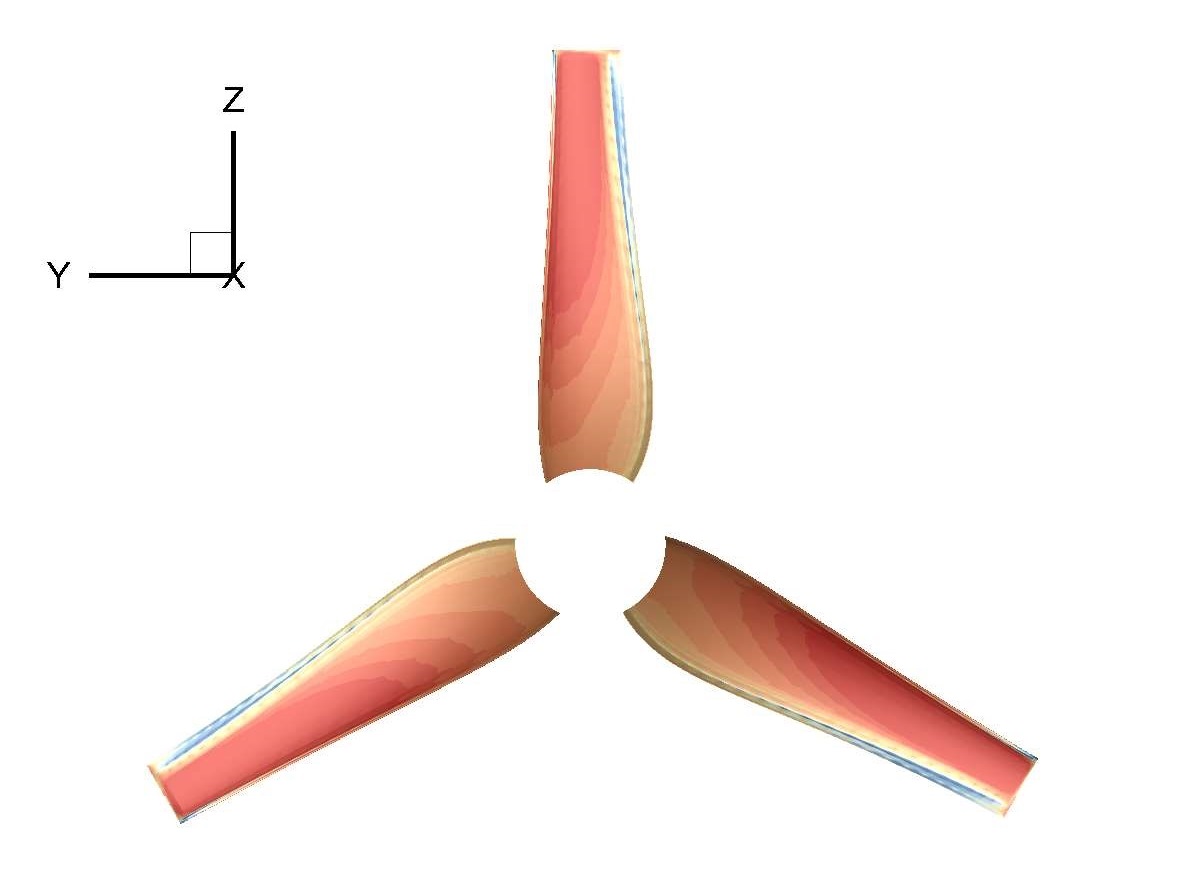}
\caption{$\lambda=3.93$}
\label{fig:owt_l_4.7_p_up}
\end{subfigure}
\begin{subfigure}[b]{0.3\textwidth}
\centering
\includegraphics[width=2.0in]{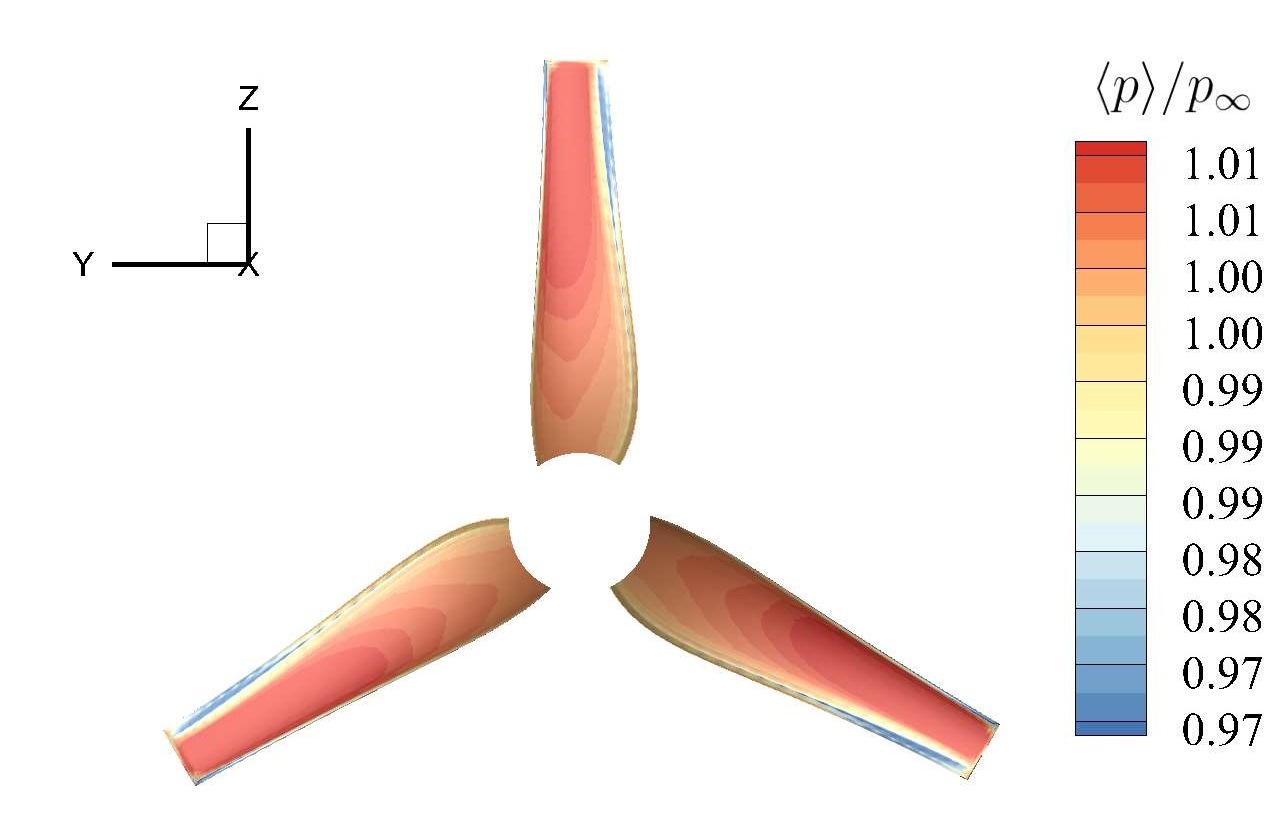}
\caption{$\lambda=4.75$}
\label{fig:owt_l_5.7_p_up}
\end{subfigure}
\caption{The OWT's phase-averaged upstream surface pressure.}
\label{fig:owt_p_up}
\end{figure}

\begin{figure}[!htb]
\centering
\begin{subfigure}[b]{0.28\textwidth}
\centering
\includegraphics[width=1.8in]{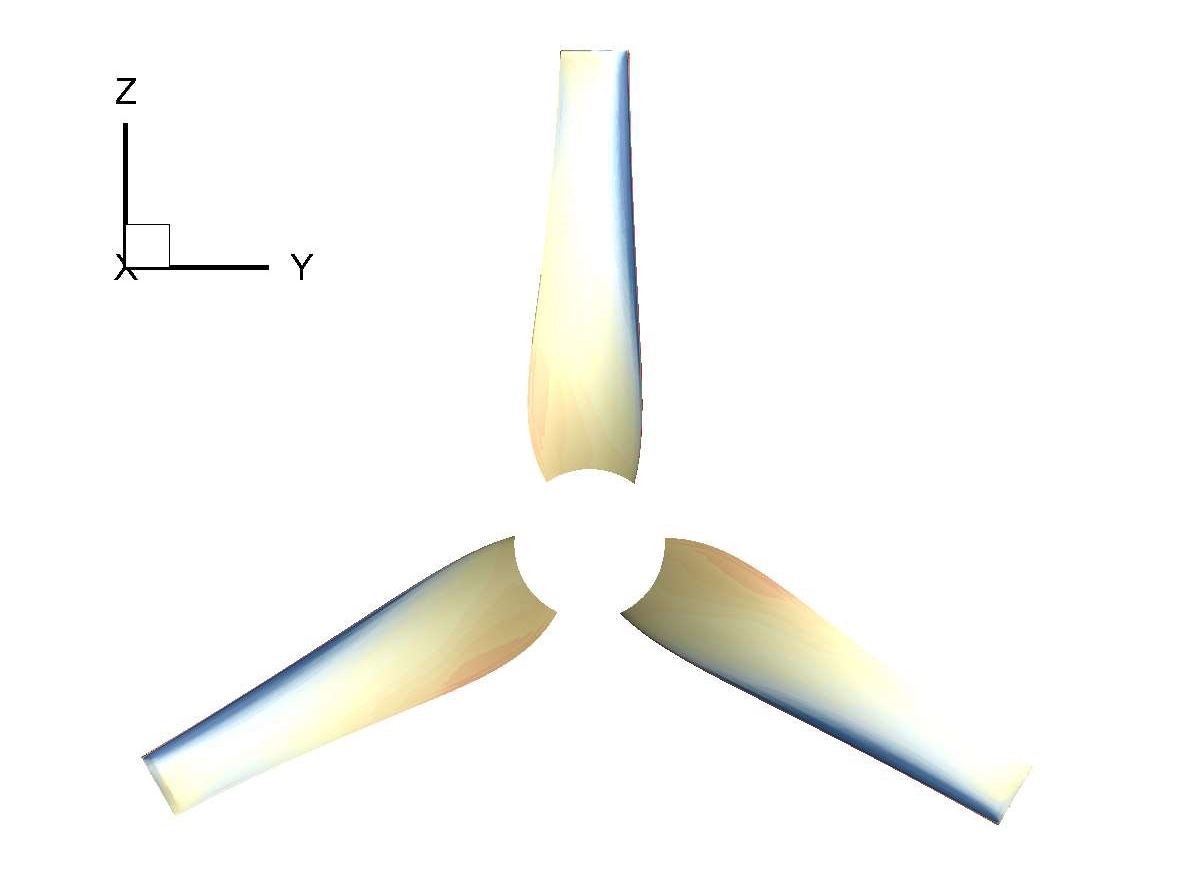}
\caption{$\lambda=3.11$}
\label{fig:owt_l_3.7_p_down}
\end{subfigure}
\begin{subfigure}[b]{0.28\textwidth}
\centering
\includegraphics[width=1.8in]{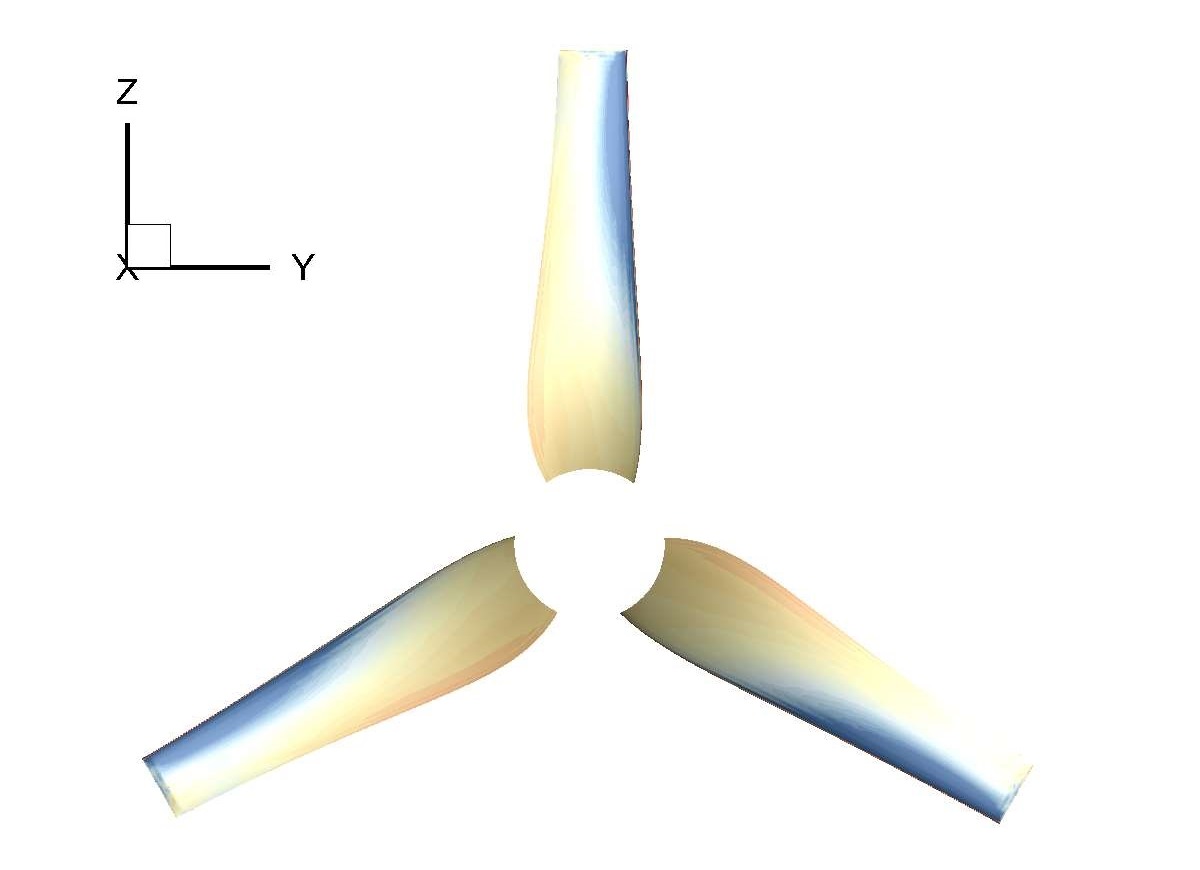}
\caption{$\lambda=3.93$}
\label{fig:owt_l_4.7_p_down}
\end{subfigure}
\begin{subfigure}[b]{0.3\textwidth}
\centering
\includegraphics[width=2.0in]{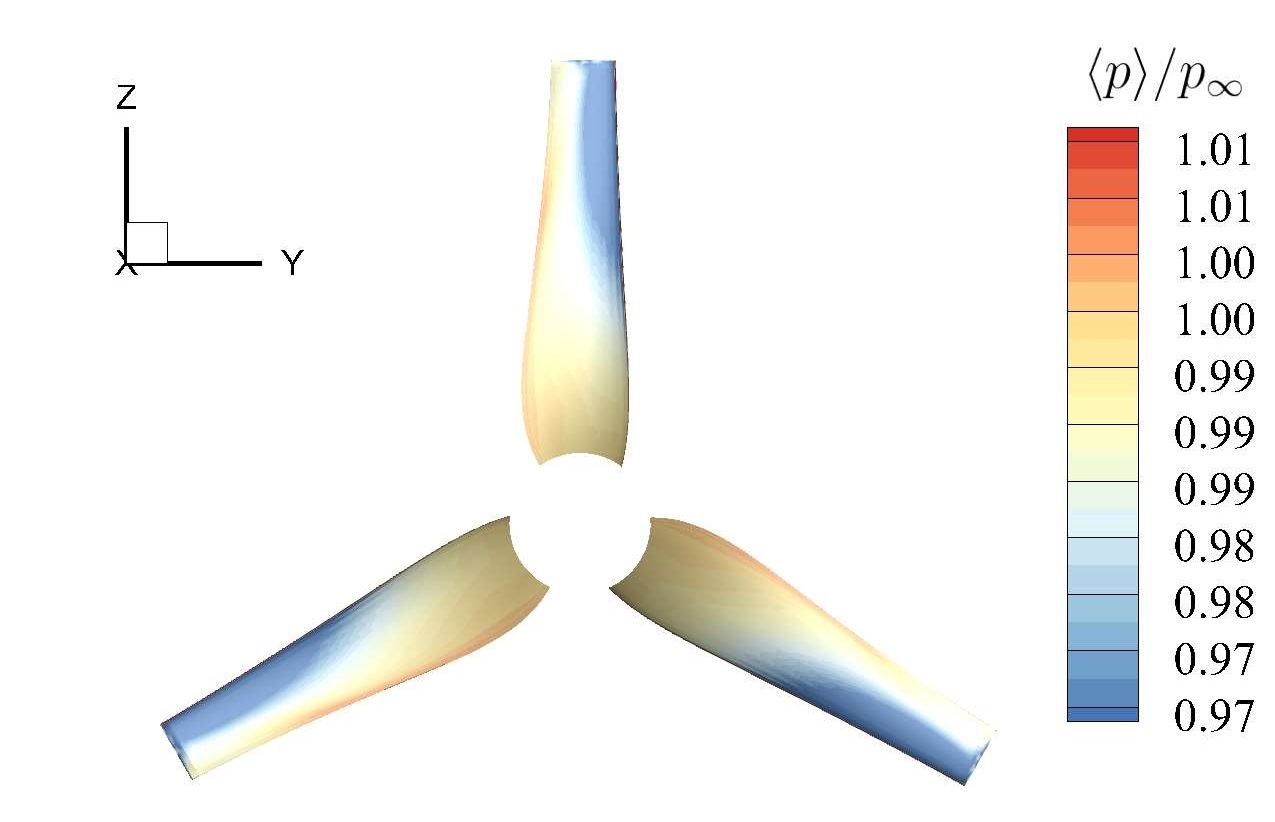}
\caption{$\lambda=4.75$}
\label{fig:owt_l_5.7_p_down}
\end{subfigure}
\caption{The OWT's phase-averaged downstream surface pressure.}
\label{fig:owt_p_down}
\end{figure}

The phase-averaged pressure on the blade surfaces is plotted to show more details of the load distributions on the blades. As shown in Figs. \ref{fig:dwt_p_up} and \ref{fig:dwt_p_down}, for the DWT, the upstream surface experiences a higher pressure than the downstream surface. Thus the rotor experiences a drag force pointed towards the x-direction. The outboard part of the blades, especially the area around the leading edge, has a higher pressure difference, which means the load strength is higher in this region. The OWT's phase-averaged pressure contours are given in Figs. \ref{fig:owt_p_up} and \ref{fig:owt_p_down}. A similar pressure distribution as that of the DWT is observed. One major difference is that the downstream surface has a smaller area of low pressure than that of the DWT, which is responsible for the smaller loads than the DWT's.

\begin{figure}[!htb]
\centering
\begin{subfigure}[b]{0.28\textwidth}
\centering
\includegraphics[width=1.8in]{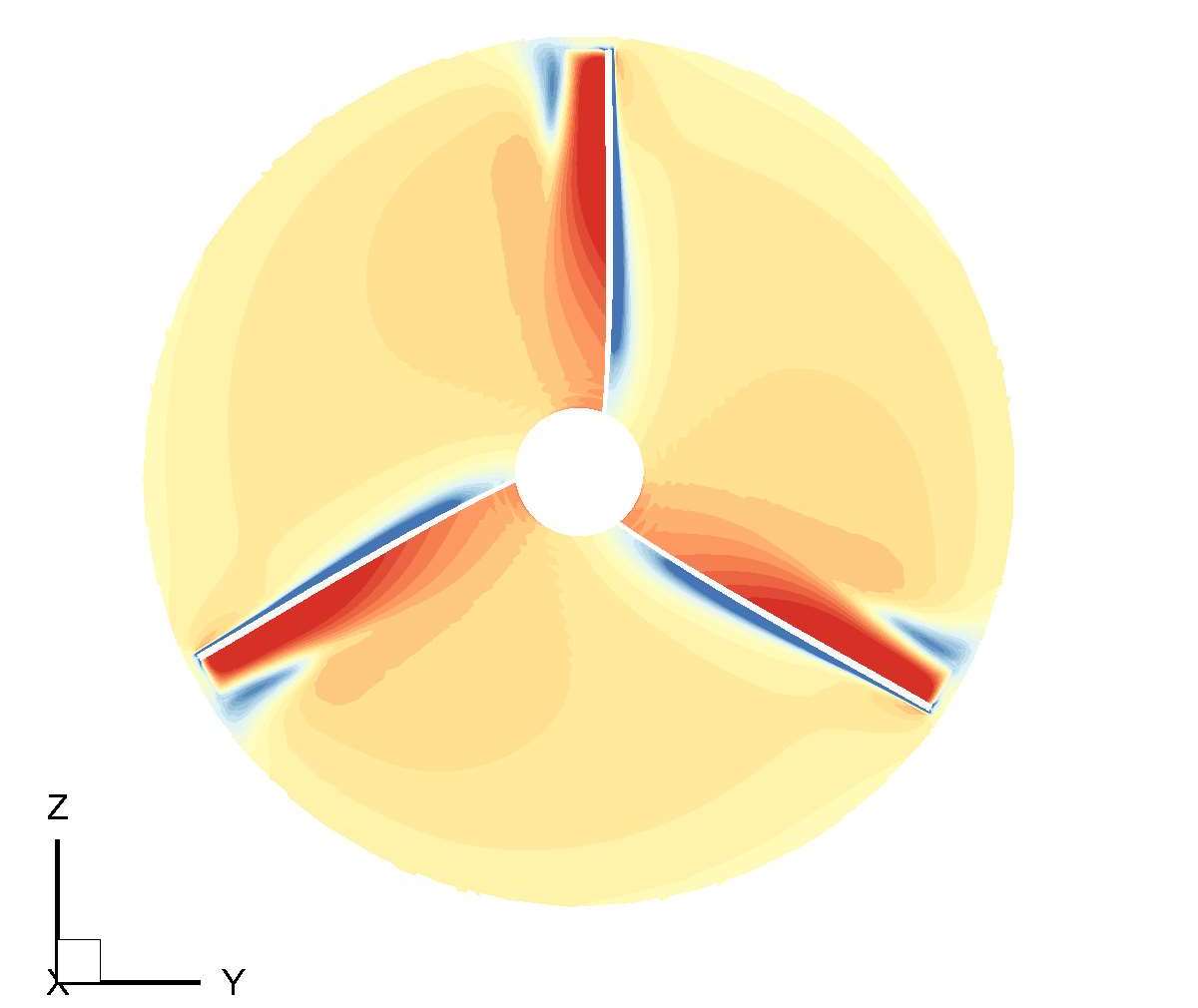}
\caption{$\lambda=3.11$}
\label{fig:dwt_l_3.7}
\end{subfigure}
\begin{subfigure}[b]{0.28\textwidth}
\centering
\includegraphics[width=1.8in]{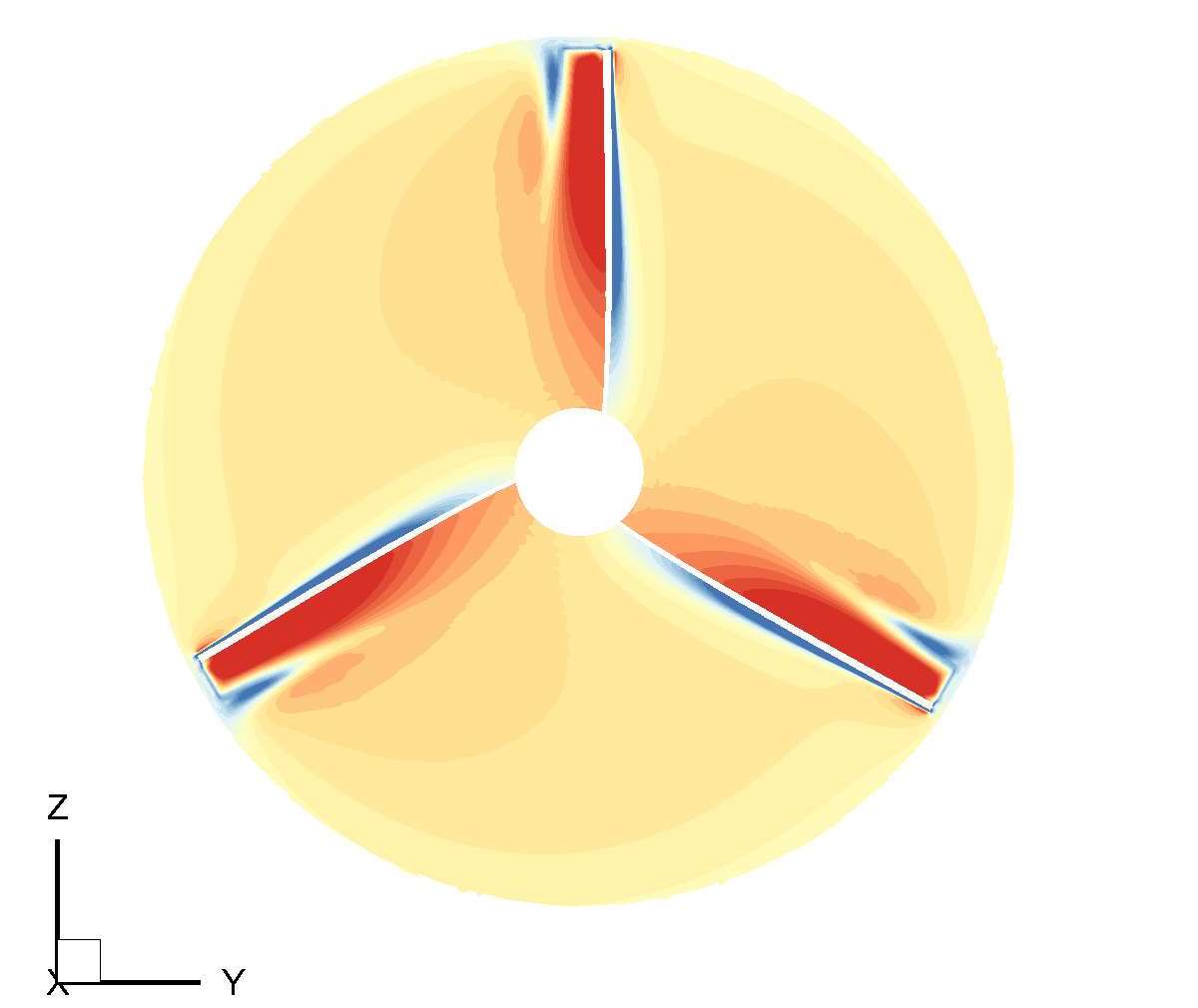}
\caption{$\lambda=3.93$}
\label{fig:dwt_l_4.7}
\end{subfigure}
\begin{subfigure}[b]{0.3\textwidth}
\centering
\includegraphics[width=2.0in]{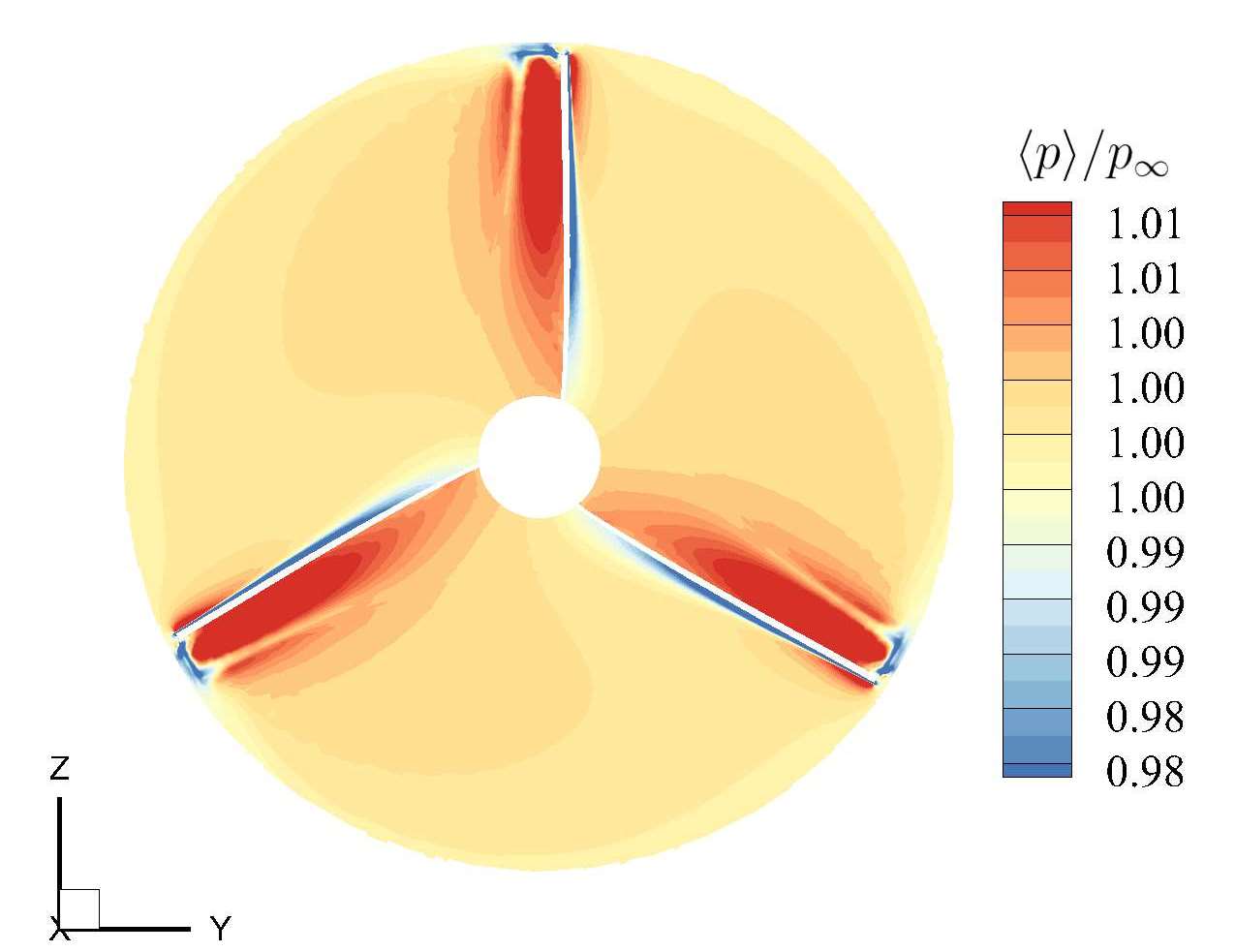}
\caption{$\lambda=4.75$}
\label{fig:dwt_l_5.7}
\end{subfigure}
\caption{The DWT's phase-averaged pressure at $x/D=0.02$.}
\label{fig:dwt_p_torque}
\end{figure}

\begin{figure}[!htb]
\centering
\begin{subfigure}[b]{0.28\textwidth}
\centering
\includegraphics[width=1.8in]{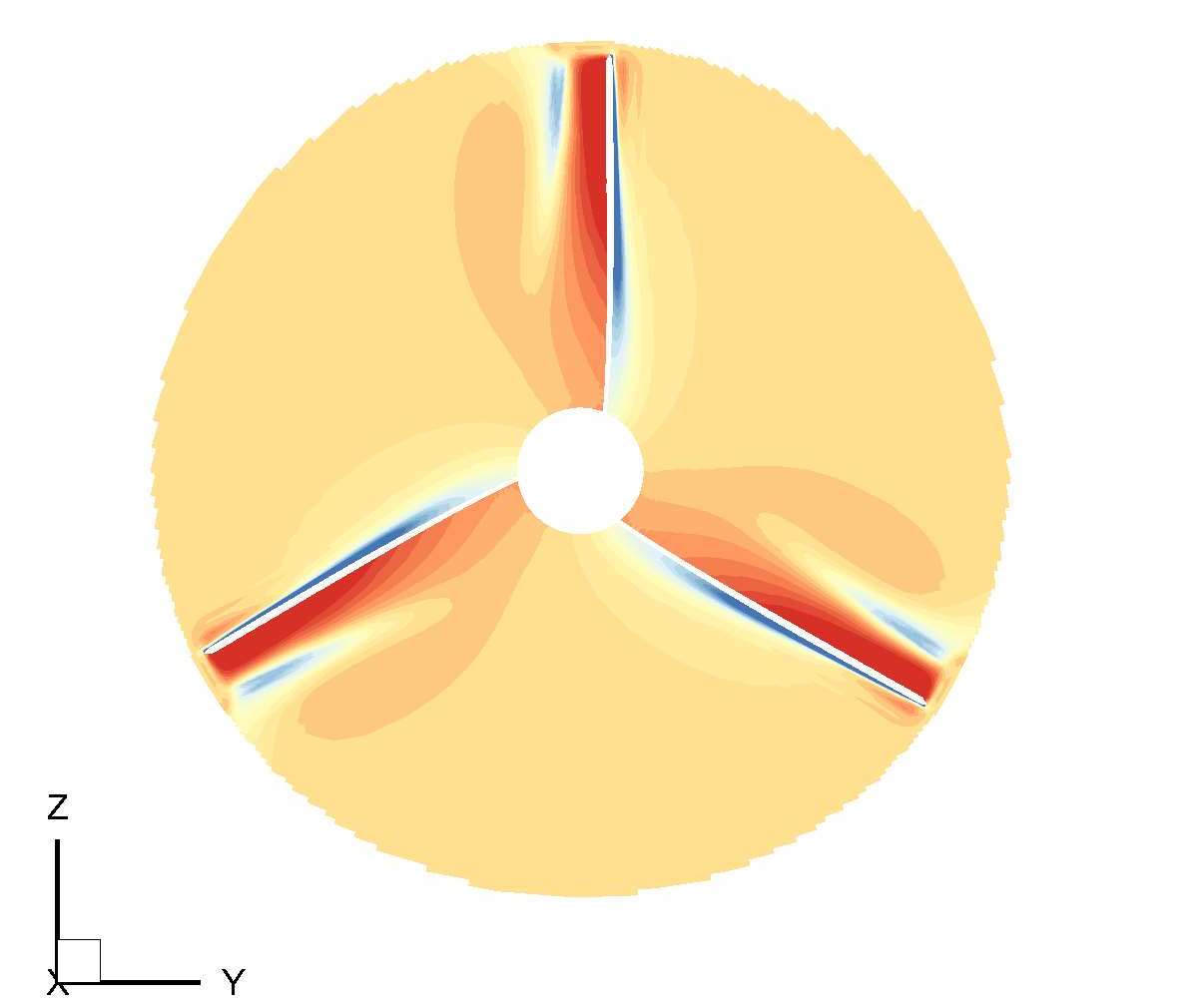}
\caption{$\lambda=3.11$}
\label{fig:owt_l_3.7}
\end{subfigure}
\begin{subfigure}[b]{0.28\textwidth}
\centering
\includegraphics[width=1.8in]{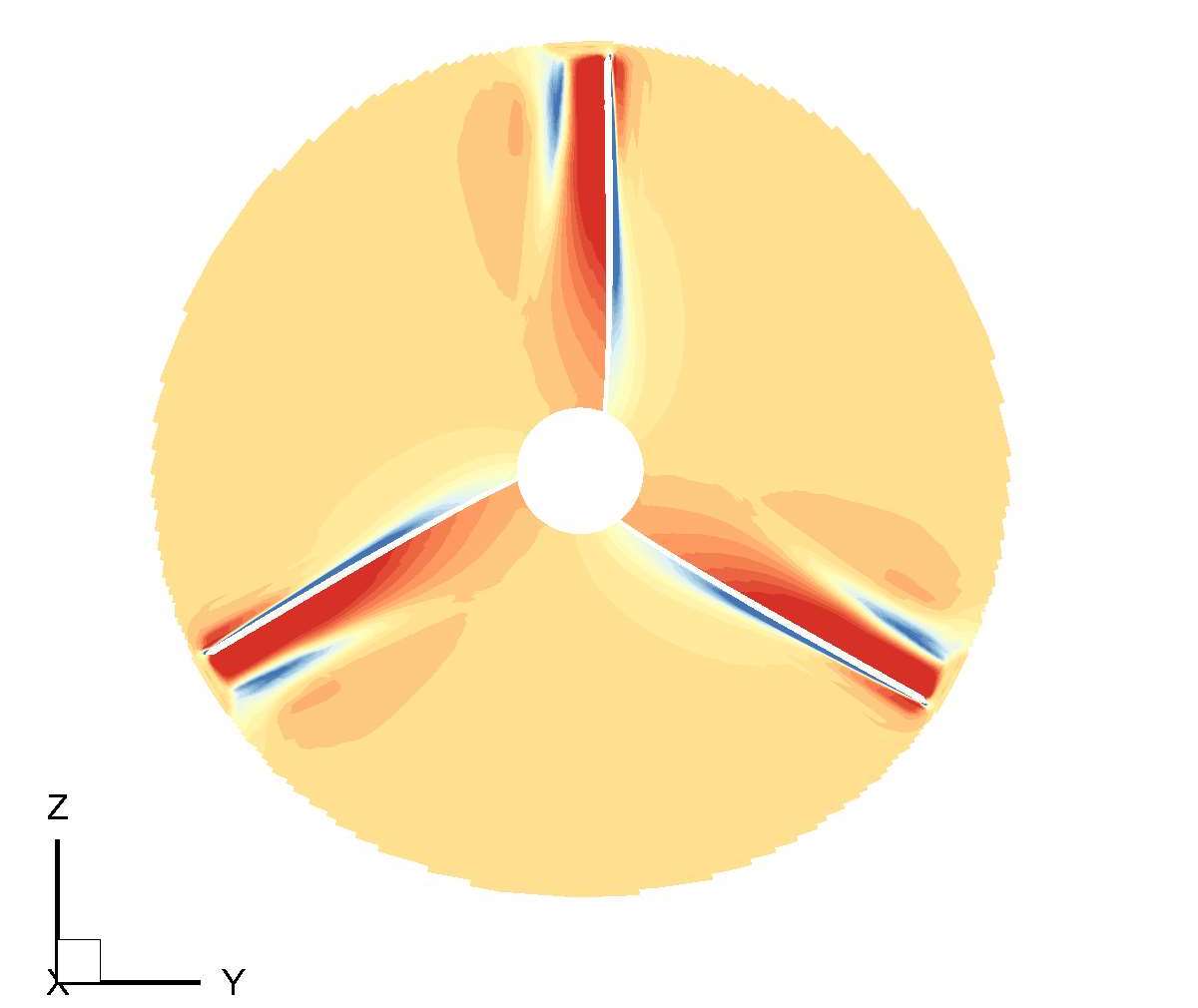}
\caption{$\lambda=3.93$}
\label{fig:owt_l_4.7}
\end{subfigure}
\begin{subfigure}[b]{0.3\textwidth}
\centering
\includegraphics[width=2.0in]{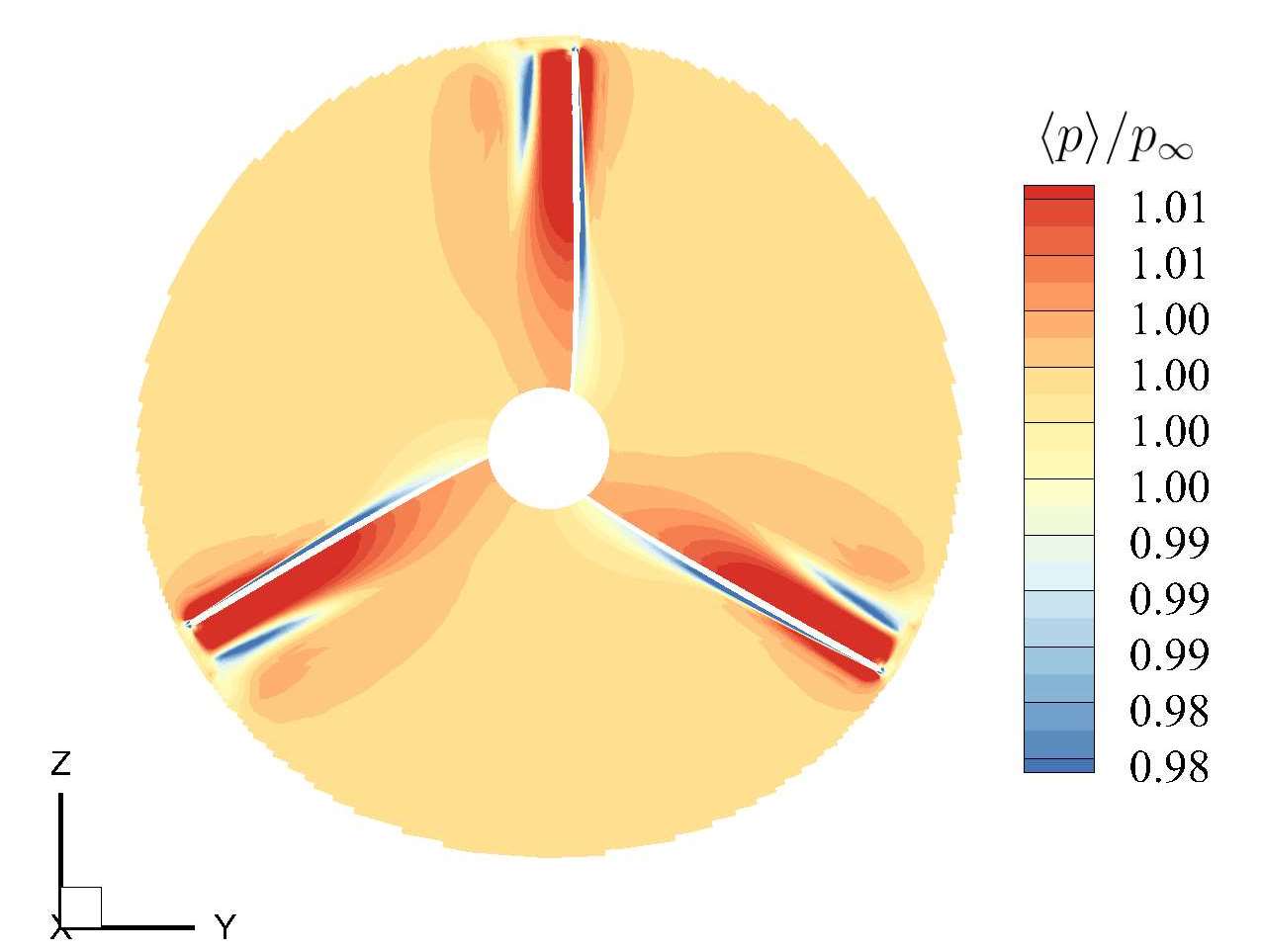}
\caption{$\lambda=4.75$}
\label{fig:owt_l_5.7}
\end{subfigure}
\caption{The OWT's phase-averaged pressure at $x/D=0.02$.}
\label{fig:owt_p_torque}
\end{figure}

For the DWT, the phase-averaged pressure at $x/D=0.02$ is shown in Fig. \ref{fig:dwt_p_torque}. Due to the pressure difference between the two sides of each blade, all three blades experience a torque pointed towards the negative x-axis. This torque direction is the same as the rotation direction, which means positive work is done on the rotor, and the wind energy is transferred to the rotor's mechanical energy. Fig. \ref{fig:owt_p_torque} plots the pressure contours for the OWT, and a similar pressure distribution is observed. The major difference between the two turbines is still that the DWT has larger low-pressure regions.

\subsection{Vortex Fields}
\begin{figure}[!htb]
\centering
\begin{subfigure}[b]{0.98\textwidth}
\centering
\includegraphics[width=3.1in]{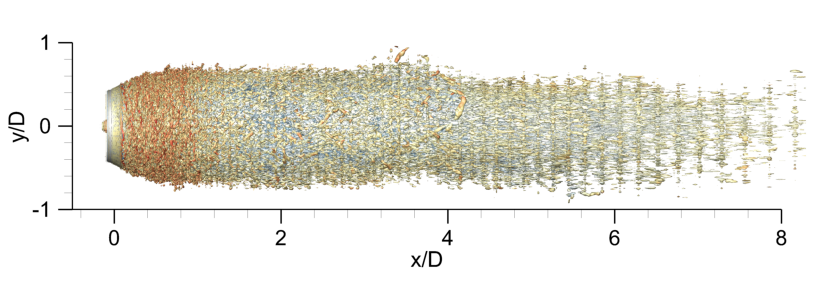}
\includegraphics[width=3.1in]{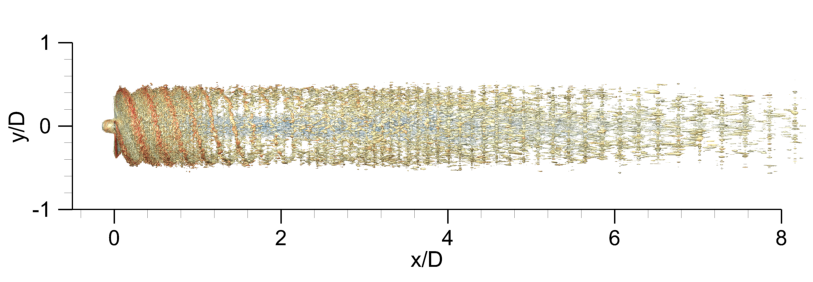}
\includegraphics[width=0.3in]{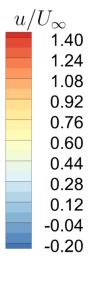}
\caption{$\lambda=3.11$}
\end{subfigure}
\begin{subfigure}[b]{0.98\textwidth}
\centering
\includegraphics[width=3.1in]{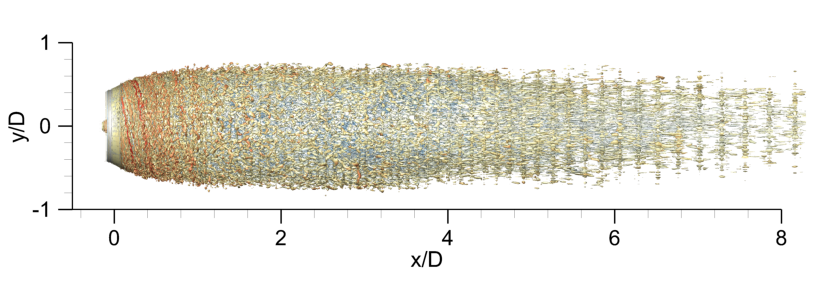}
\includegraphics[width=3.1in]{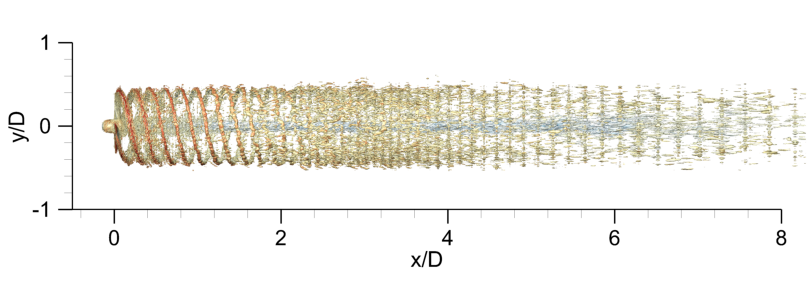}
\includegraphics[width=0.3in]{figures/Ch4/Vorticity/u_legend.png}
\caption{$\lambda=3.93$}
\end{subfigure}
\begin{subfigure}[b]{0.98\textwidth}
\centering
\includegraphics[width=3.1in]{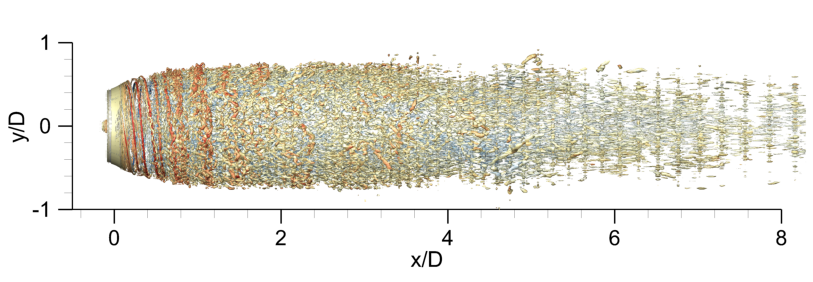}
\includegraphics[width=3.1in]{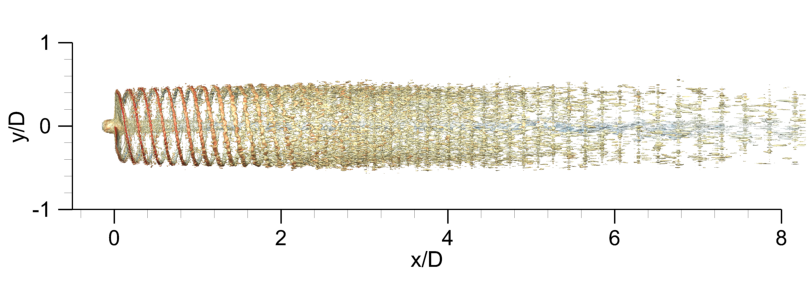}
\includegraphics[width=0.3in]{figures/Ch4/Vorticity/u_legend.png}
\caption{$\lambda=4.75$}
\end{subfigure}
\caption{Isosurface of instantaneous Q-criterion $Q_{cr}D^2/U_{\infty}^2=60$ in both configurations.}
\label{fig:Qcr_ins}
\end{figure}

\begin{figure}[!htb]
\centering
\centering
\includegraphics[width=4.5in]{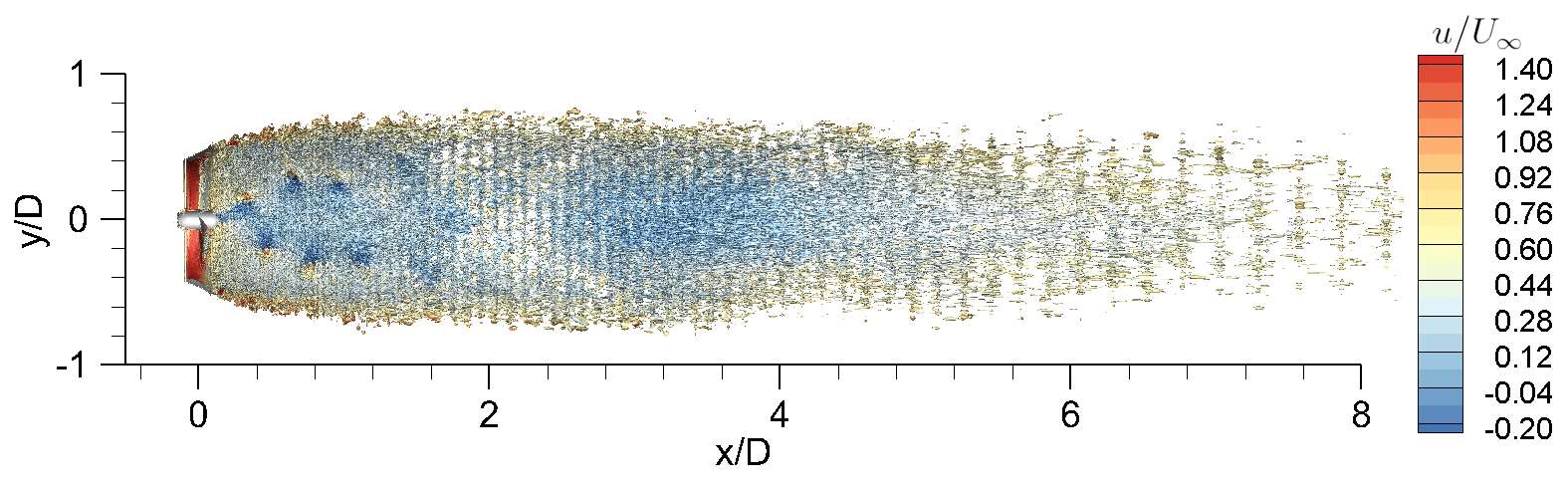}
\caption{Flow structures in the central x-y plane of the DWT at $\lambda=3.93$.}
\label{fig:dwt_Qcr_half}
\end{figure}

Isosurfaces of instantaneous Q-criterion, colored by the streamwise velocity, are plotted in Fig. \ref{fig:Qcr_ins} for both turbines. A view through the central plane of the DWT at $\lambda=3.93$ is shown in Fig. \ref{fig:dwt_Qcr_half}. Compared with the OWT's flow fields, there are dramatic changes in the DWT's due to the addition of the duct. Several important observations can be made. For example, 1) The DWT's flow fields are much wider in the radial direction. 2) The flow fields are more turbulent with a much broader range of flow structures. 3) Tip vortices have been dramatically weakened and can hardly be observed. 4) Hub vortices have almost been entirely suppressed. 5) The blades' trailing-edge vortices have been enhanced (see Fig. \ref{fig:dwt_Qcr_half}).

\subsection{Velocity Fields}
\begin{figure}[!htb]
\centering
\begin{subfigure}[b]{0.98\textwidth}
\centering
\includegraphics[width=3.0in]{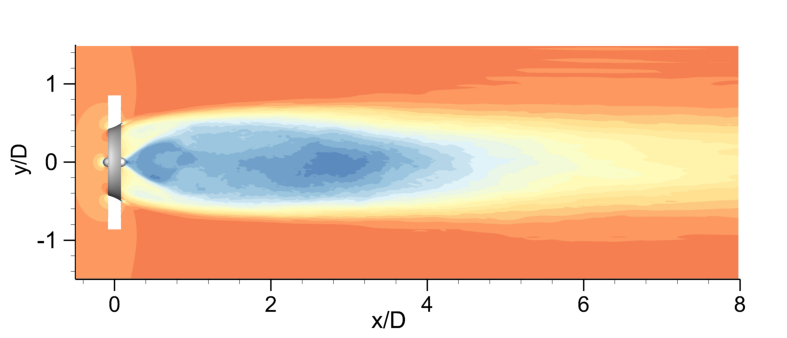}
\includegraphics[width=3.0in]{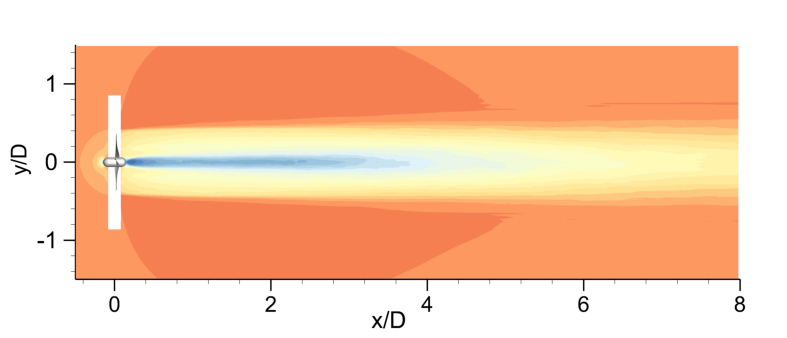}
\includegraphics[width=0.45in]{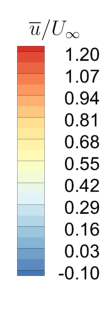}
\caption{$\lambda=3.11$}
\label{fig:ave_u_stream_l_3.7}
\end{subfigure}
\begin{subfigure}[b]{0.98\textwidth}
\centering
\includegraphics[width=3.0in]{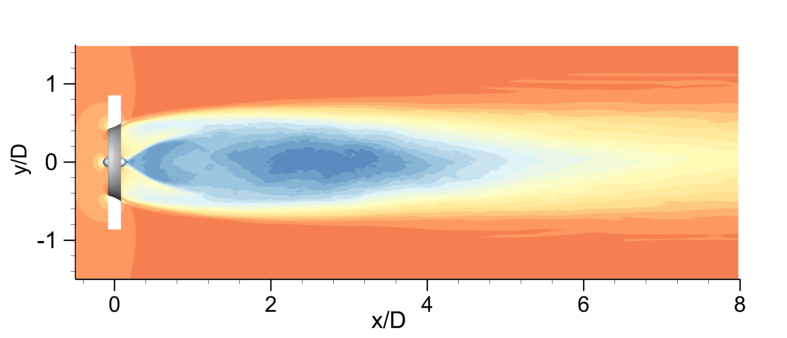}
\includegraphics[width=3.0in]{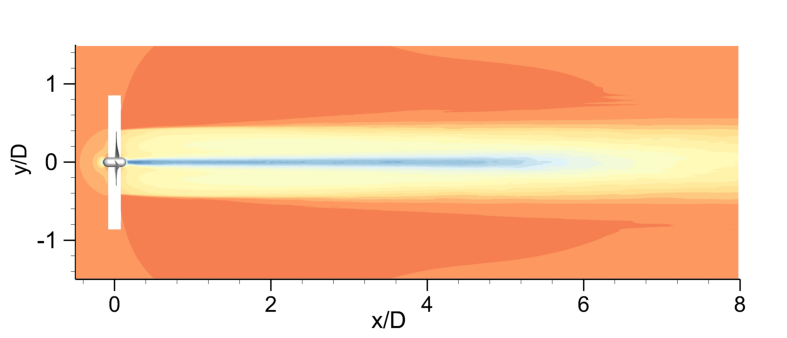}
\includegraphics[width=0.45in]{figures/Ch4/Velocity/DWT/u_stream_legend.png}
\caption{$\lambda=3.93$}
\label{fig:ave_u_stream_l_4.7}
\end{subfigure}
\begin{subfigure}[b]{0.98\textwidth}
\centering
\includegraphics[width=3.0in]{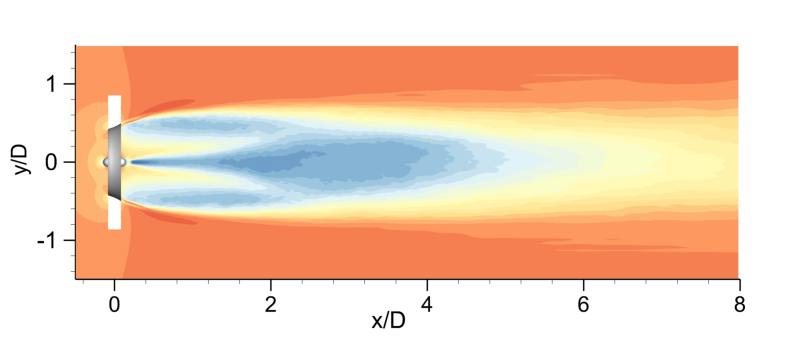}
\includegraphics[width=3.0in]{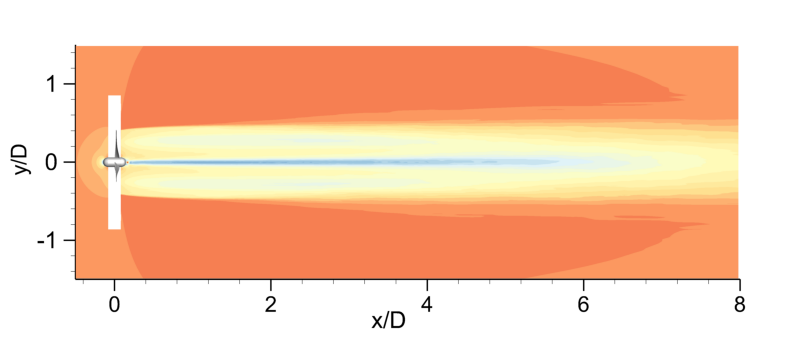}
\includegraphics[width=0.45in]{figures/Ch4/Velocity/DWT/u_stream_legend.png}
\caption{$\lambda=4.75$}
\label{fig:ave_u_stream_l_5.7}
\end{subfigure}
\caption{Contours of mean streamwise velocity  $\overline{u}$ in the central x-y plane: left, DWT; right, OWT.}
\label{fig:ave_u_stream}
\end{figure}

Contours of the mean streamwise velocity, $\overline{u}$, in the central x-y plane are shown in Fig. \ref{fig:ave_u_stream} for the three tip speed ratios. It is seen that the flows downstream of both turbines are at reduced speeds compared with the free-stream flow, which represents a fluid kinetic energy loss. The DWT's wakes have much larger low-speed regions than those of the OWT, indicating more fluid kinetic energy loss, which is consistent with the higher power coefficients of the DWT.

At $\lambda=3.11$ and $3.93$, the DWT's low-speed regions have similar diverging-converging shapes. However, at $\lambda=4.75$, the DWT's wake becomes bifurcated. It is conjectured that the bifurcation is caused by the blockage effects of the blades and the duct. From Fig. \ref{fig:dwt_geo}, it is observed that each blade is almost perpendicular to the incoming flow around the tip but more skewed (in other words, more aligned with the flow) towards the root. When a blade rotates faster (i.e., as $\lambda$ increases), the tip blockage effect increases more substantially, which, when combined with the presence of the duct, dramatically reduces the flow speed downstream of the duct. It is worth mentioning that the gap between the tips and the duct's inner surface is too small (about $0.03D$) to allow much fluid to go through. Meanwhile, the enhanced blockage effect forces more fluid to go through the root region of the blades, resulting in larger root flow speeds at $\lambda=4.75$ than at $3.11$ and $3.93$. The combination of the flows in these two regions (i.e., tips and roots) results in a bifurcation in the wake. The tip blockage effects actually also exist in the OWT's flow fields. As can be seen from Fig. \ref{fig:ave_u_stream}, the flow speeds downstream of the OWT's tips also decrease as $\lambda$ increases.

\begin{figure}[!htb]
\centering
\begin{subfigure}[b]{0.19\textwidth}
\centering
\includegraphics[width=1.2in]{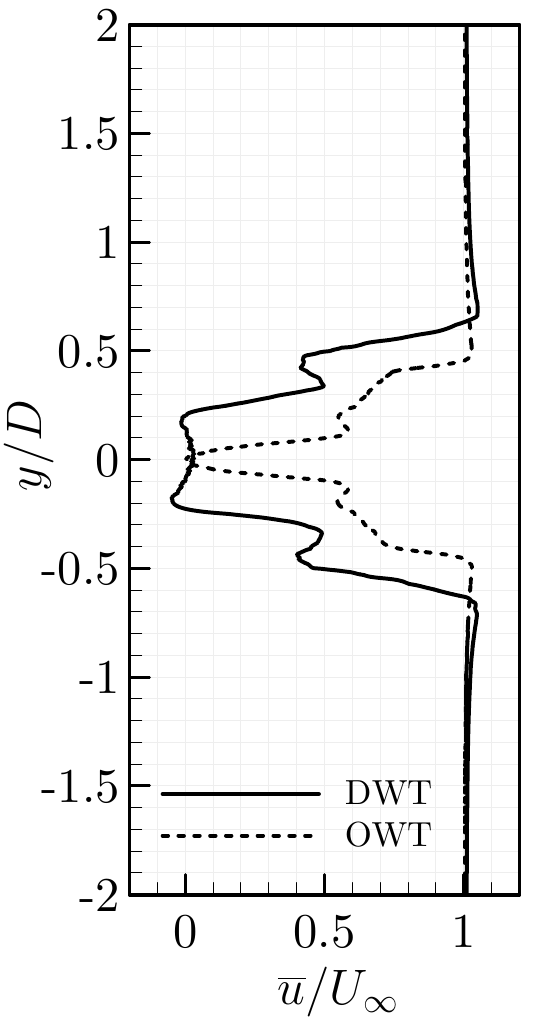}
\caption{$x/D=0.5$}
\label{fig:u_profile_l_3.7_0.5}
\end{subfigure}
\begin{subfigure}[b]{0.19\textwidth}
\centering
\includegraphics[width=1.2in]{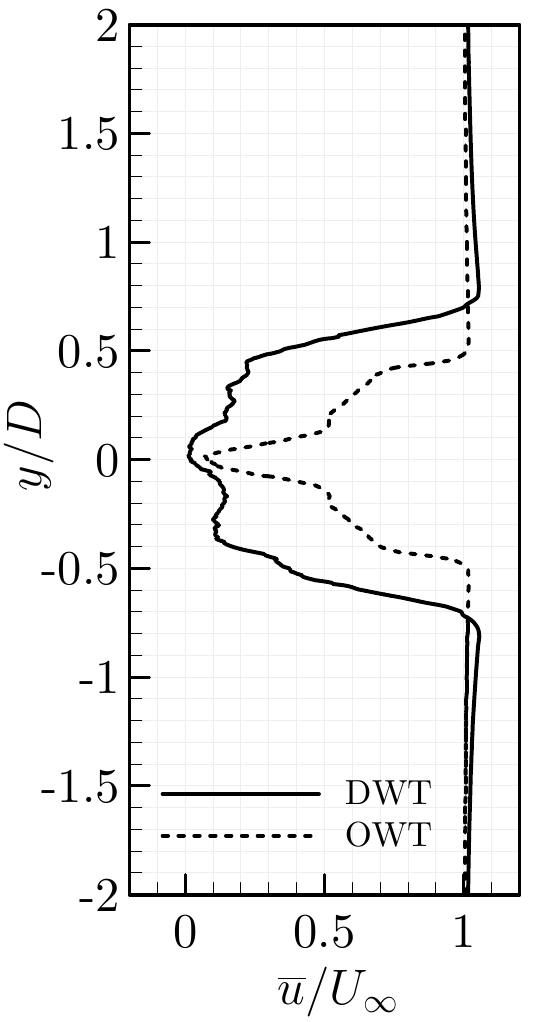}
\caption{$x/D=1$}
\label{fig:u_profile_l_3.7_1}
\end{subfigure}
\begin{subfigure}[b]{0.19\textwidth}
\centering
\includegraphics[width=1.2in]{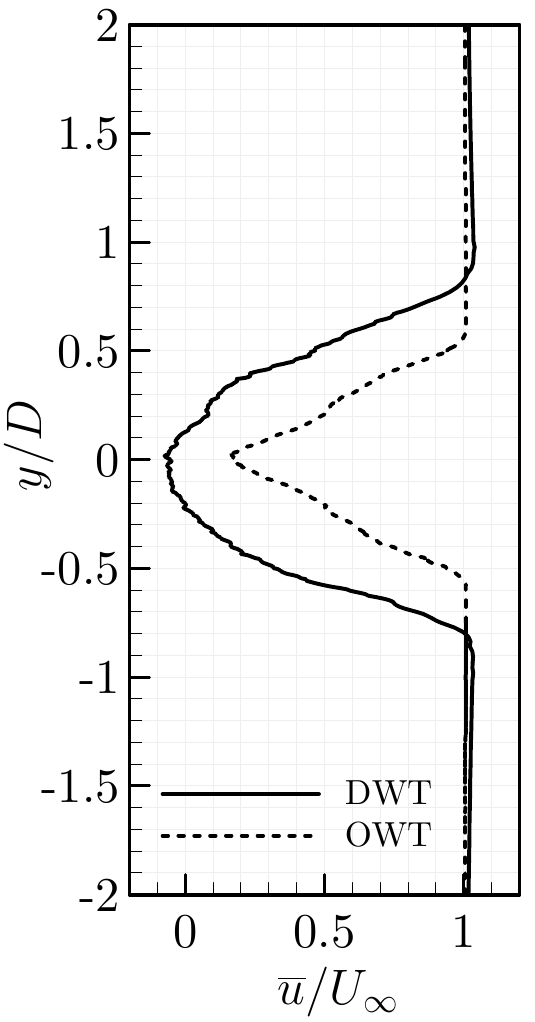}
\caption{$x/D=3$}
\label{fig:u_profile_l_3.7_3}
\end{subfigure}
\begin{subfigure}[b]{0.19\textwidth}
\centering
\includegraphics[width=1.2in]{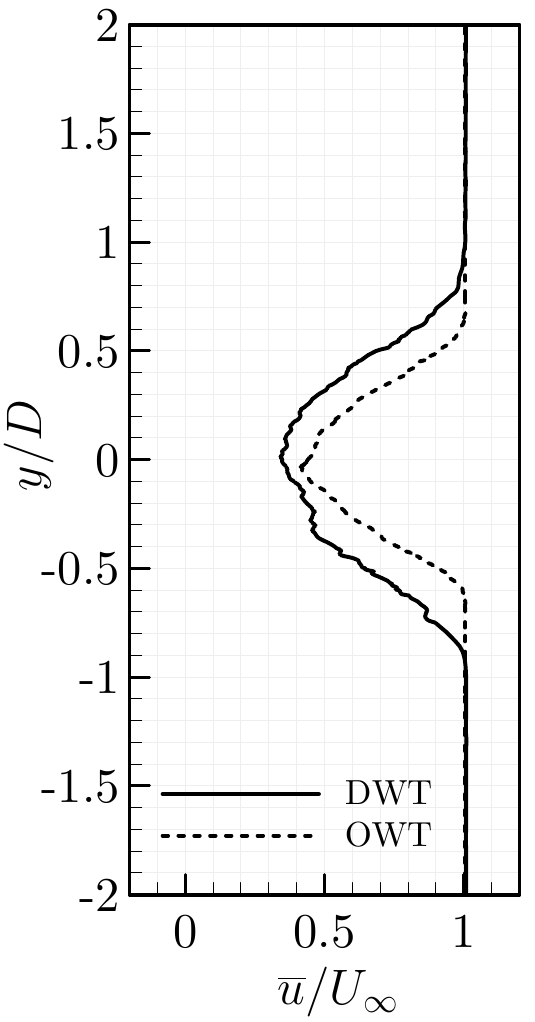}
\caption{$x/D=5$}
\label{fig:u_profile_l_3.7_5}
\end{subfigure}
\caption{Profiles of mean streamwise velocity for $\lambda=3.11$ at different streamwise locations in the central x-y plane.}
\label{fig:ave_u_profile_l_3.7}
\end{figure}

\begin{figure}[!htb]
\centering
\begin{subfigure}[b]{0.19\textwidth}
\centering
\includegraphics[width=1.2in]{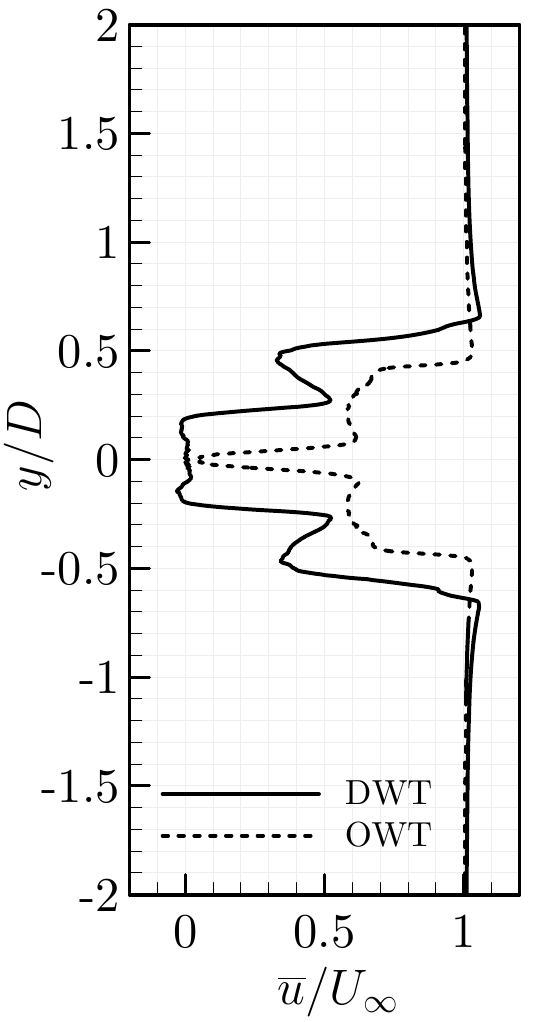}
\caption{$x/D=0.5$}
\label{fig:u_profile_l_4.7_0.5}
\end{subfigure}
\begin{subfigure}[b]{0.19\textwidth}
\centering
\includegraphics[width=1.2in]{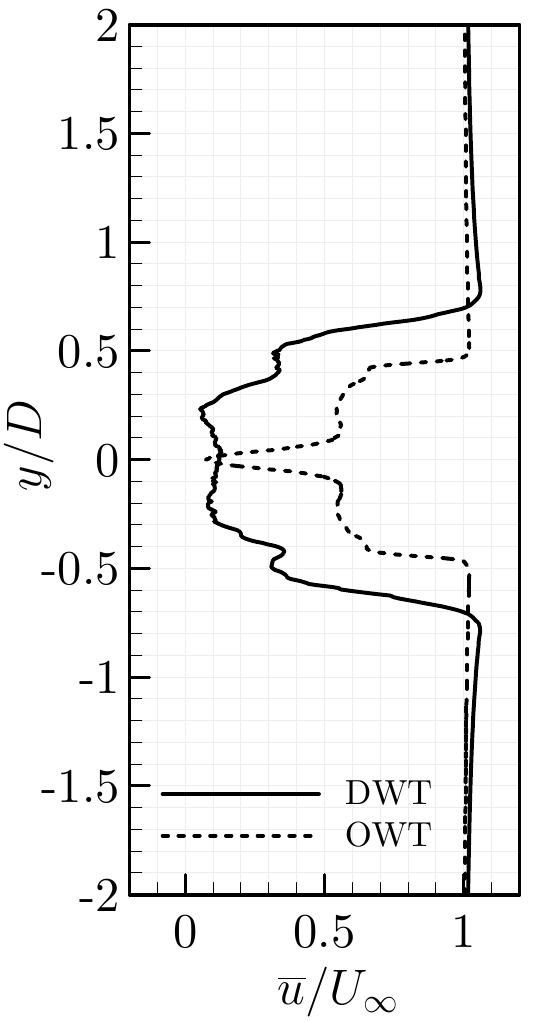}
\caption{$x/D=1$}
\label{fig:u_profile_l_4.7_1}
\end{subfigure}
\begin{subfigure}[b]{0.19\textwidth}
\centering
\includegraphics[width=1.2in]{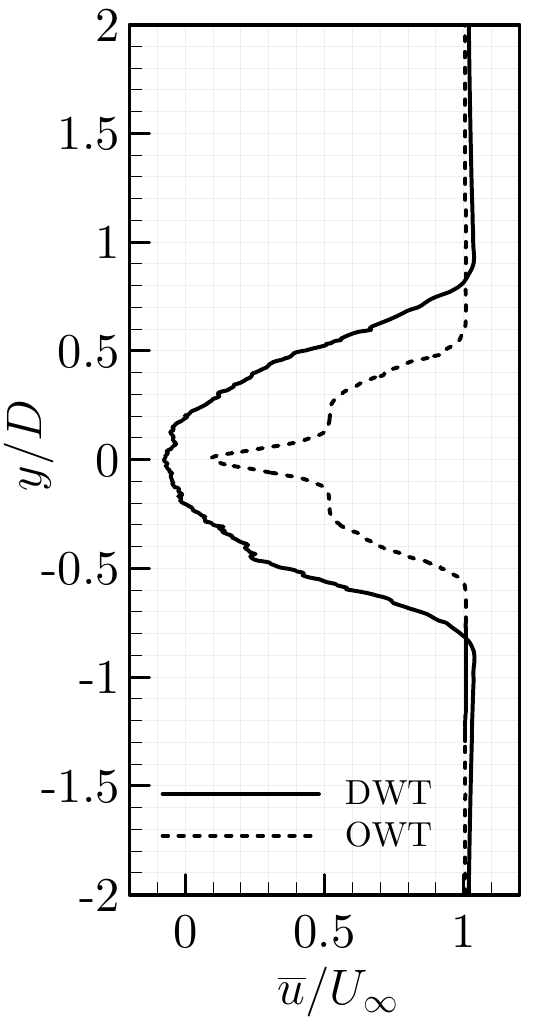}
\caption{$x/D=3$}
\label{fig:u_profile_l_4.7_3}
\end{subfigure}
\begin{subfigure}[b]{0.19\textwidth}
\centering
\includegraphics[width=1.2in]{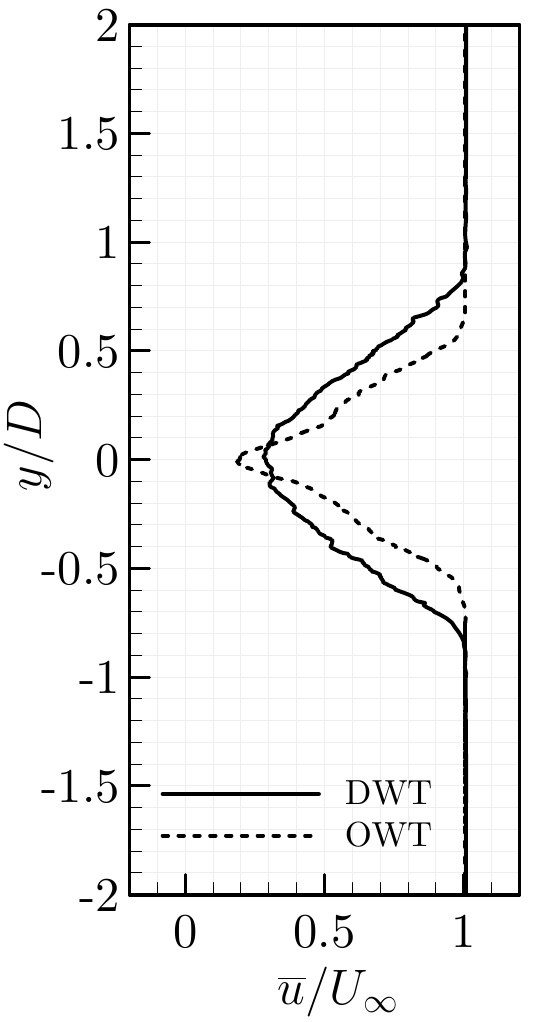}
\caption{$x/D=5$}
\label{fig:u_profile_l_4.7_5}
\end{subfigure}
\caption{Profiles of mean streamwise velocity for $\lambda=3.93$ at different streamwise locations in the central x-y plane.}
\label{fig:ave_u_profile_l_4.7}
\end{figure}

\begin{figure}[H]
\centering
\begin{subfigure}[b]{0.19\textwidth}
\centering
\includegraphics[width=1.2in]{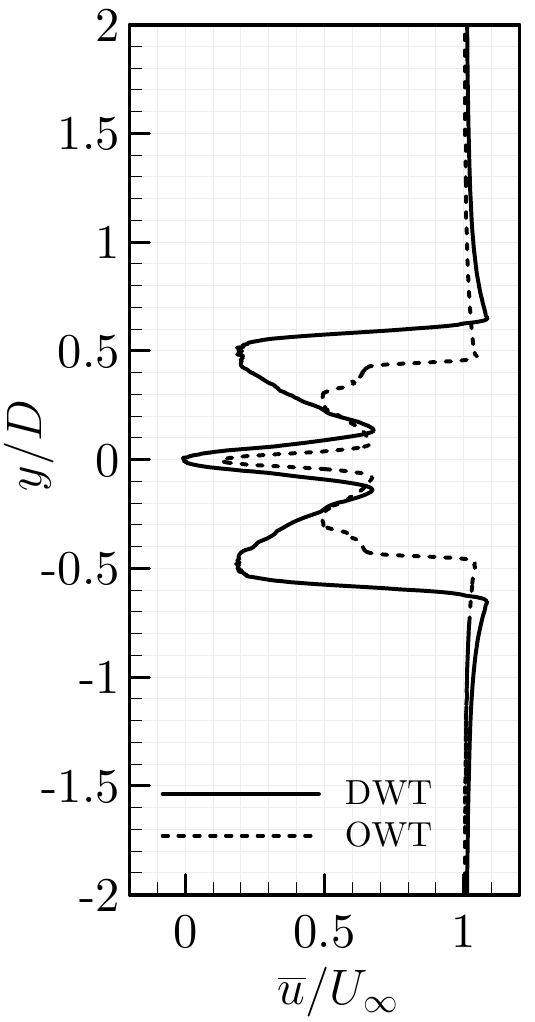}
\caption{$x/D=0.5$}
\label{fig:u_profile_l_5.7_0.5}
\end{subfigure}
\begin{subfigure}[b]{0.19\textwidth}
\centering
\includegraphics[width=1.2in]{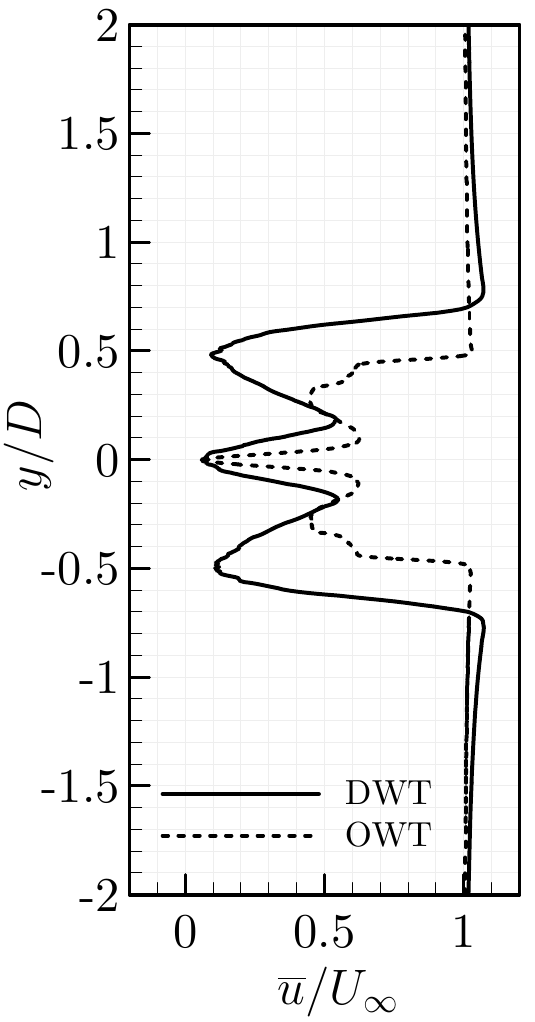}
\caption{$x/D=1$}
\label{fig:u_profile_l_5.7_1}
\end{subfigure}
\begin{subfigure}[b]{0.19\textwidth}
\centering
\includegraphics[width=1.2in]{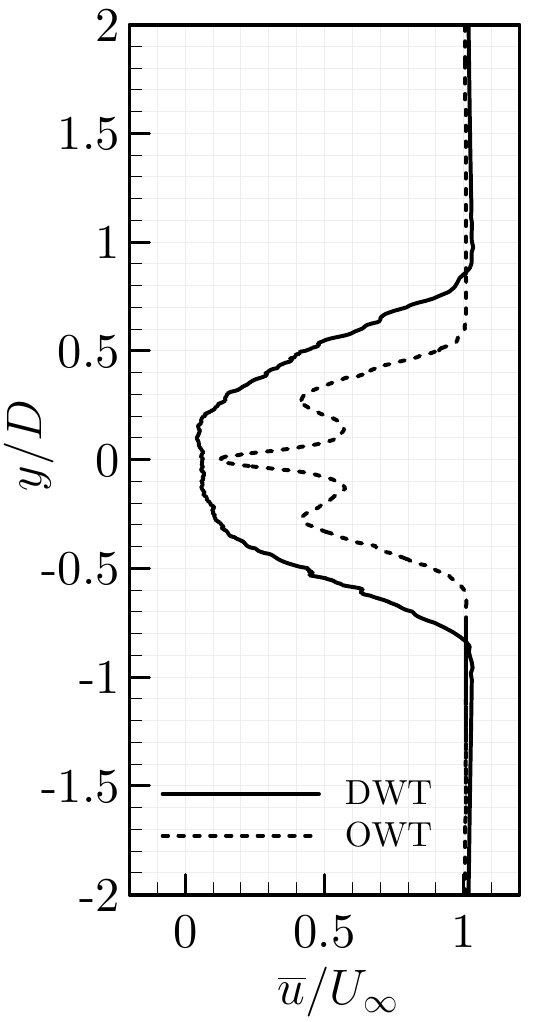}
\caption{$x/D=3$}
\label{fig:u_profile_l_5.7_3}
\end{subfigure}
\begin{subfigure}[b]{0.19\textwidth}
\centering
\includegraphics[width=1.2in]{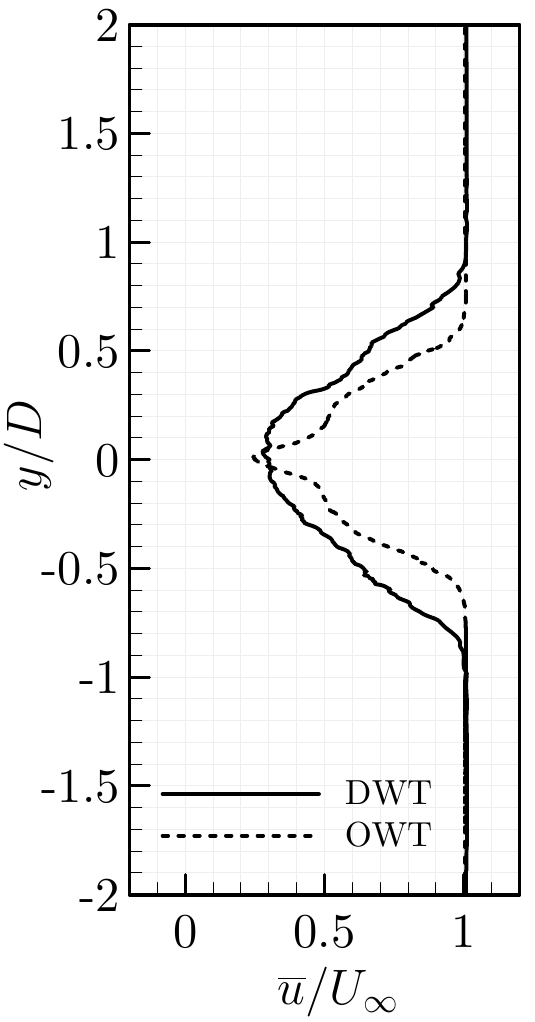}
\caption{$x/D=5$}
\label{fig:u_profile_l_5.7_5}
\end{subfigure}
\caption{Profiles of mean streamwise velocity for $\lambda=4.75$ at different streamwise locations in the central x-y plane.}
\label{fig:ave_u_profile_l_5.7}
\end{figure}

To quantitatively see how the wake flows develop, we plot the mean streamwise velocity profiles in Figs. \ref{fig:ave_u_profile_l_3.7}-\ref{fig:ave_u_profile_l_5.7}. Overall, the wakes show a consistent recovery trend as the flows travel downstream. At almost every location and for every $\lambda$, the DWT's wake has lower speeds than the OWT's. The notable exceptions are at $x/D=5$ and for the $\lambda=3.93$ and $4.75$ cases, where the OWT's wakes have lower speeds around $y/D=0$ because of the presence of strong hub vortices.

The effects of the blades and the duct on the wakes are most evident on the profiles at $x/D=0.5$. Since the mean flow is symmetric about $y/D=0$, we thus only focus on the $y/D>0$ part to examine the effects. For the DWT, there is always a local minimum of $\overline{u}$ around $y/D=0.5$, which is caused by the blockage effect of the duct. The value of this local minimum decreases as $\lambda$ increases (the values are approximately 0.41, 0.32, and 0.19 for $\lambda=3.11$, 3.93, and 4.75, respectively), suggesting an increasing blockage effect.  Right below this minima is a local speed maximum. As $\lambda$ increases, the maximum's location moves downward (toward the root), and its value increases. More specifically, the locations are $y/D=0.33$, 0.28, and 0.14, and the corresponding values are 0.50, 0.52, and 0.66, respectively, for $\lambda=3.11$, 3.93, and 4.75. These speed distributions agree with our conjecture that increasing the rotational speed will cause stronger tip and duct blockage effects and enhanced root flows, which are responsible for the bifurcation.

\section{Simulation of Turbines in Yawed Flows}
\label{sec:yaw}

\begin{figure}[!htb]
\centering
\includegraphics[width=2.0in]{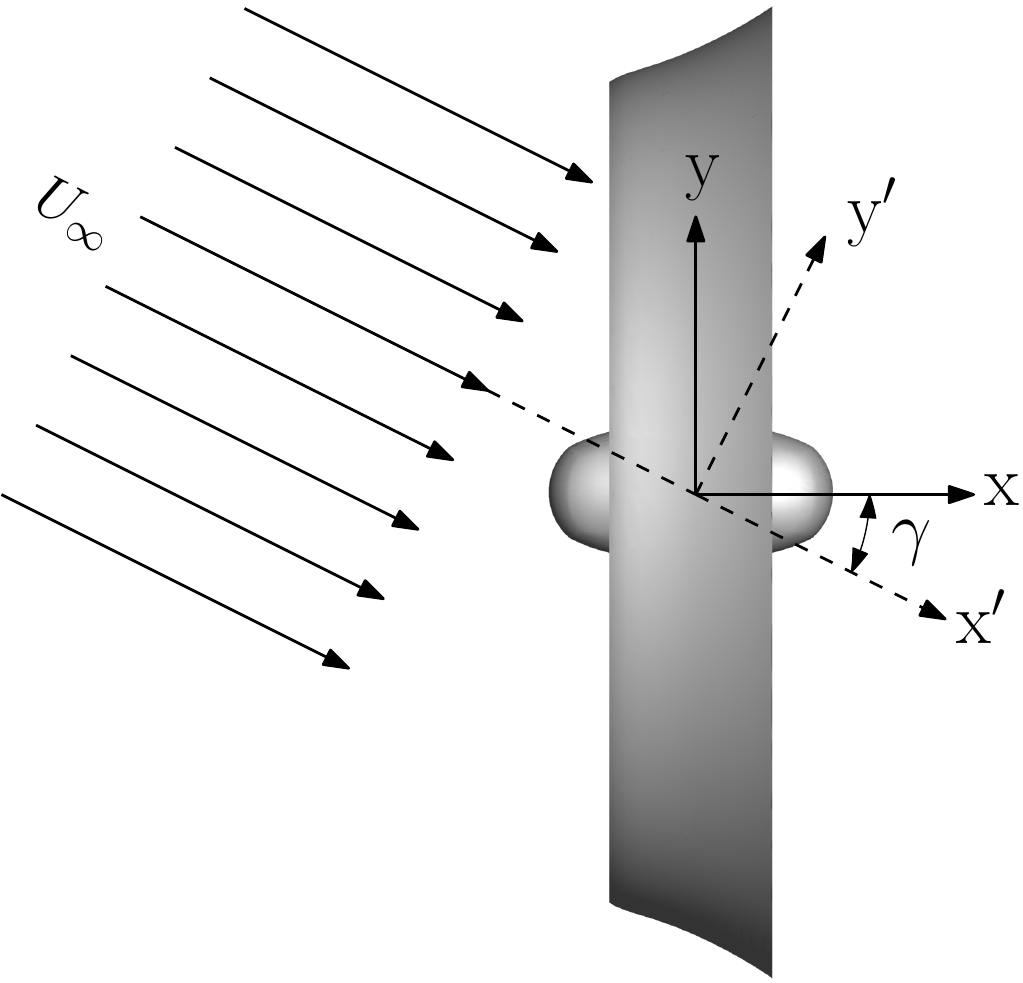}
\caption{Definitions of coordinates and yaw angle.}
\label{fig:dwt_geo_yaw}
\end{figure}

This section focuses on several yawed flows at $\lambda=3.93$. To facilitate the discussion, besides the original physical coordinate system, we introduce a flow coordinate system x$'$-y$'$ as shown in Fig. \ref{fig:dwt_geo_yaw}, where x$'$ is along the freestream flow direction, and y$'$ is perpendicular to the freestream flow direction. The yaw angle $\gamma$ is defined as the angle between the flow direction and the axial direction of the turbine, i.e., the angle between x$'$ and x. Four yaw angles: $\gamma=0^{\circ}$, $10^{\circ}$, $20^{\circ}$, and $30^{\circ}$, are considered for both the DWT and the OWT. It is worth noting that because of axial symmetry, we do not consider the sign of the yaw angle here.

\subsection{Load Analysis}
\begin{figure}[!htb]
\centering
\begin{subfigure}[b]{0.4\textwidth}
\includegraphics[width=2.4in]{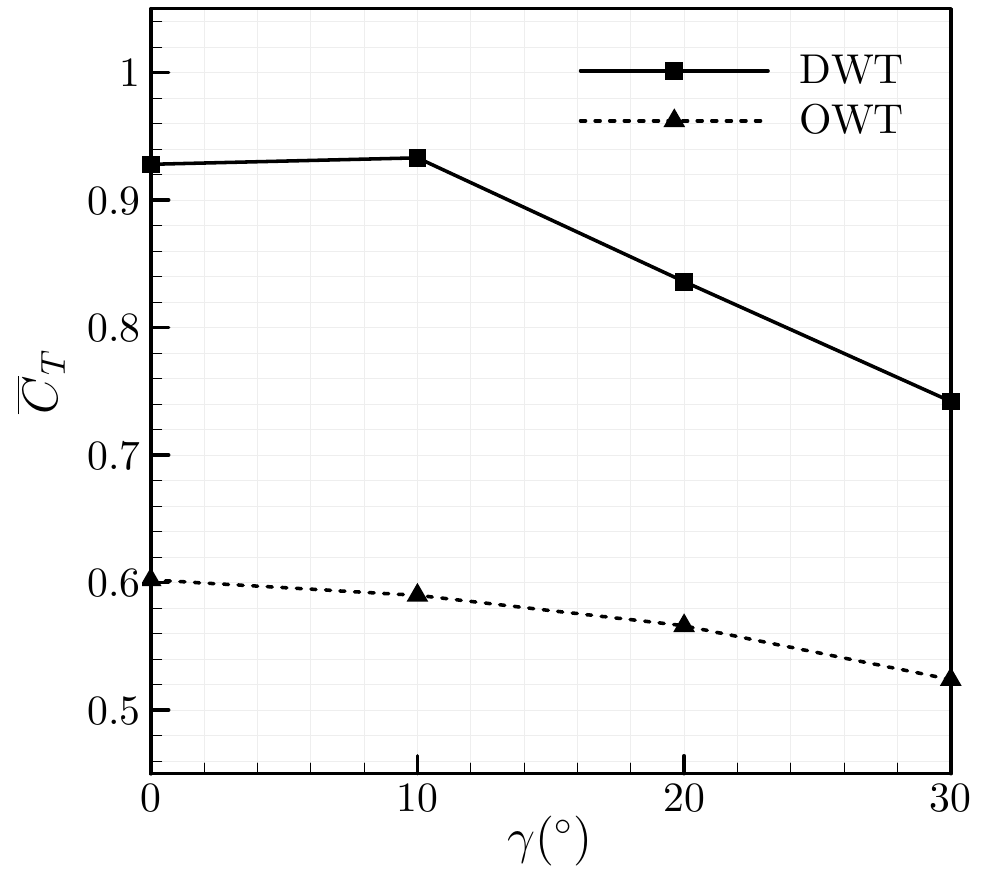}
\caption{Thrust}
\label{fig:yaw_ct}
\end{subfigure}
\begin{subfigure}[b]{0.4\textwidth}
\includegraphics[width=2.4in]{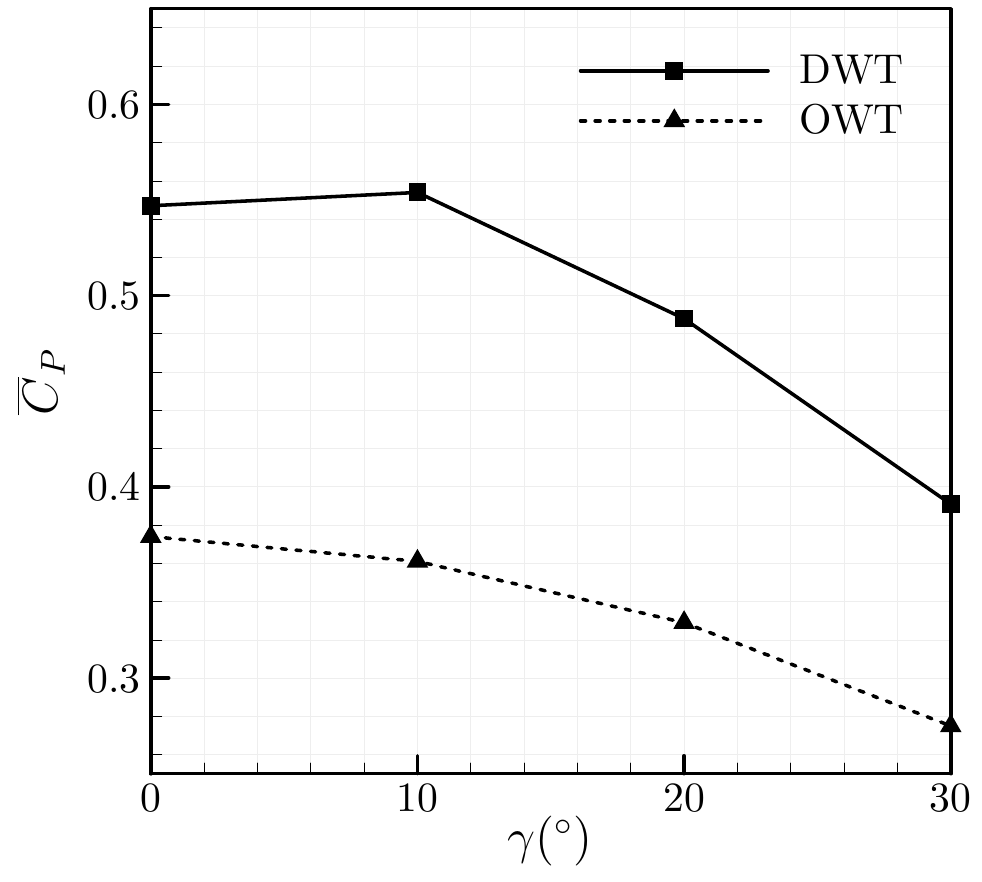}
\caption{Power}
\label{fig:yaw_cp}
\end{subfigure}
\caption{Mean thrust and power coefficients for yawed flows at $\lambda=3.93$.}
\label{fig:yaw_perform}
\end{figure}

The two turbines' mean thrust and power coefficients at the four yaw angles are plotted in Fig. \ref{fig:yaw_perform}. It is seen that the DWT experiences more drag and extracts more energy than the OWT at all yaw angles. The DWT's performance is found to be insensitive to small yaw angles (e.g., for $\gamma < 10^\circ$). Other than that, the two coefficients of both turbines decrease as the yaw angle increases, which is a direct result of the flow rate decrease across the turbines' swept areas.

\begin{figure}[H]
\centering
\begin{subfigure}[b]{0.4\textwidth}
\includegraphics[width=2.4in]{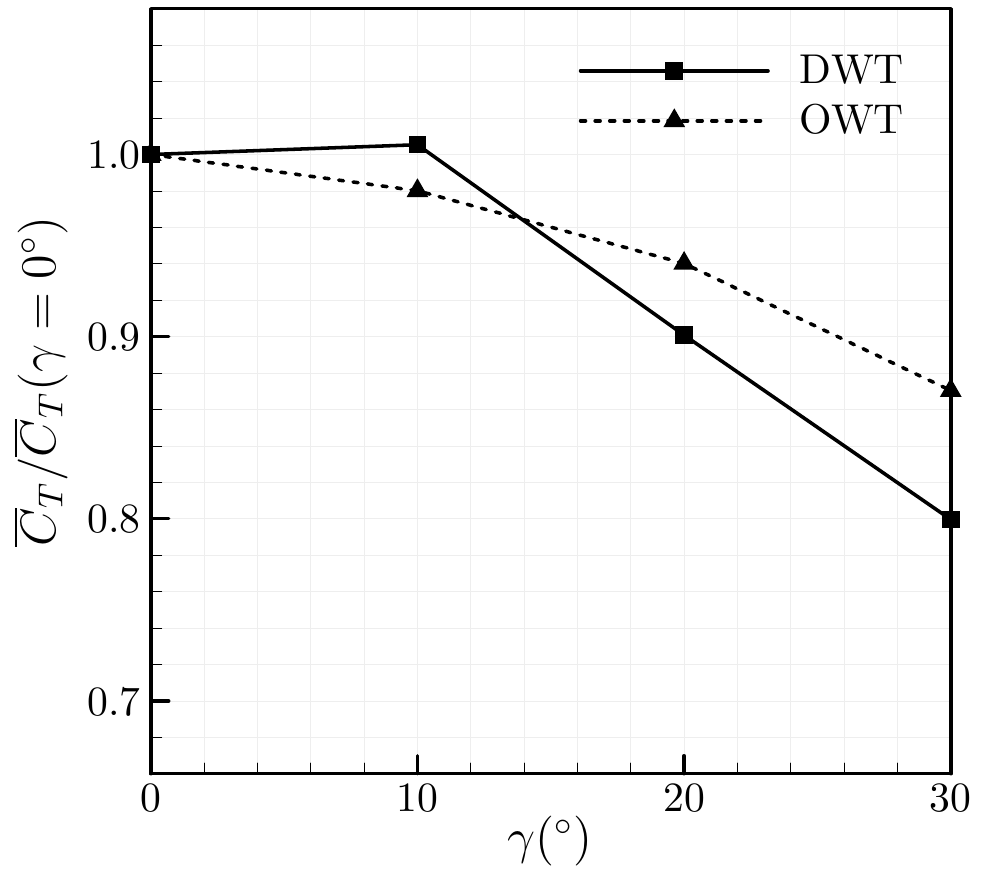}
\caption{Normalized $\overline{C}_T$}
\label{fig:yaw_ct_non}
\end{subfigure}
\begin{subfigure}[b]{0.4\textwidth}
\includegraphics[width=2.4in]{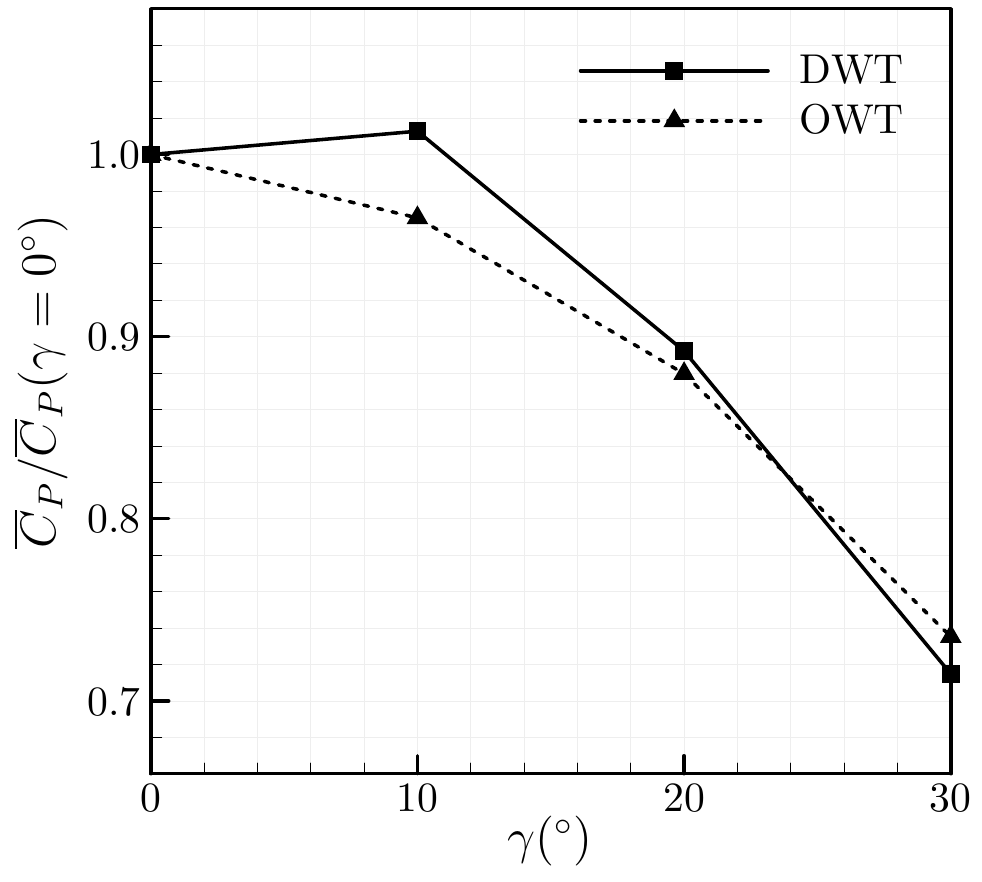}
\caption{Normalized $\overline{C}_P$}
\label{fig:yaw_cp_non}
\end{subfigure}
\caption{Normalized mean thrust and power coefficients for yawed flows at $\lambda=3.93$.}
\label{fig:yaw_non_perform}
\end{figure}

To see the relative performance change at different yaw angles, we normalize both coefficients with their values at $\gamma=0^\circ$. The normalized coefficients are plotted in Fig. \ref{fig:yaw_non_perform}. Except for small yaw angles for the DWT, the yaw angle is found to have more substantial effects on the power coefficients than on the thrust coefficients. For example, from $\gamma=0^\circ$ to $\gamma=20^\circ$, the OWT sees a $6\%$ relative drop in the thrust coefficient but a $12\%$ relative drop in the power coefficient. From $\gamma=0^\circ$ to $\gamma=30^\circ$, the OWT experiences a $13\%$ and $26\%$ relative drop on the $\overline{C}_T$ and $\overline{C}_P$, respectively, while the DWT sees a $20\%$ and $29\%$ relative drop in the two coefficients.

\subsection{Vortex Fields}
\begin{figure}[H]
\centering
\begin{subfigure}[b]{0.98\textwidth}
\centering
\includegraphics[width=3.1in]{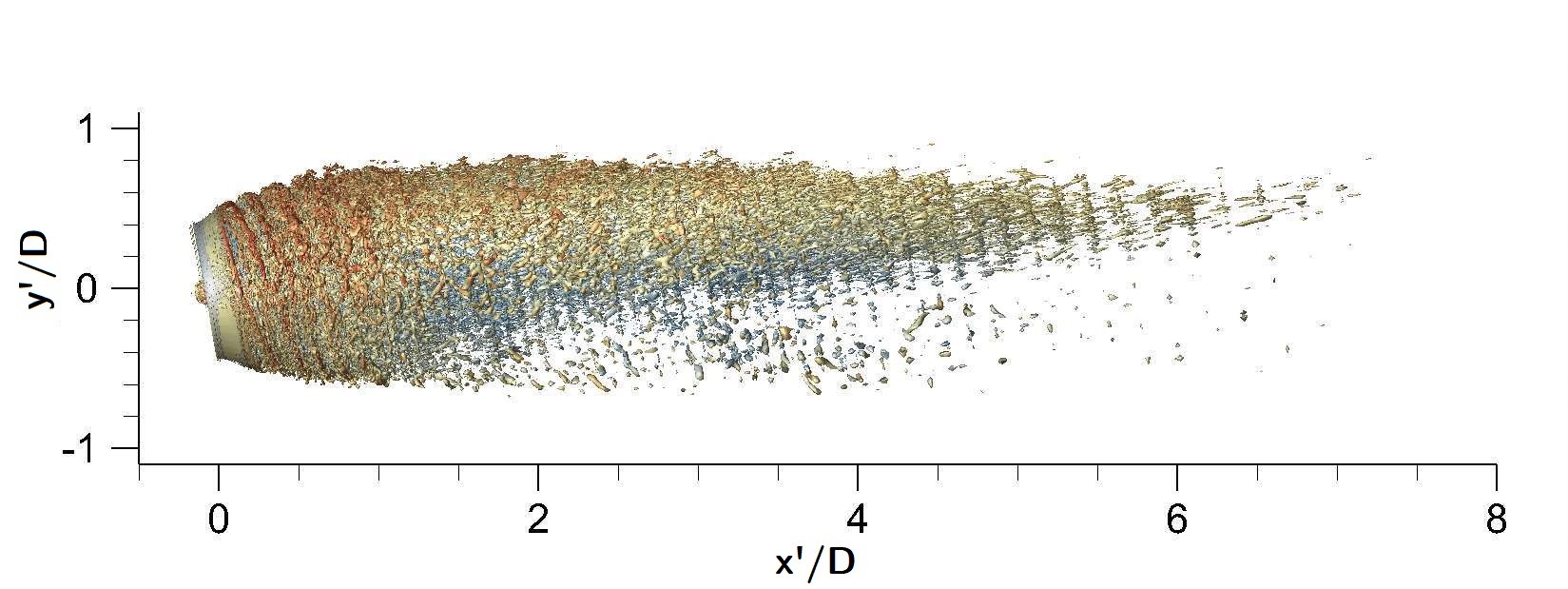}
\includegraphics[width=3.3in]{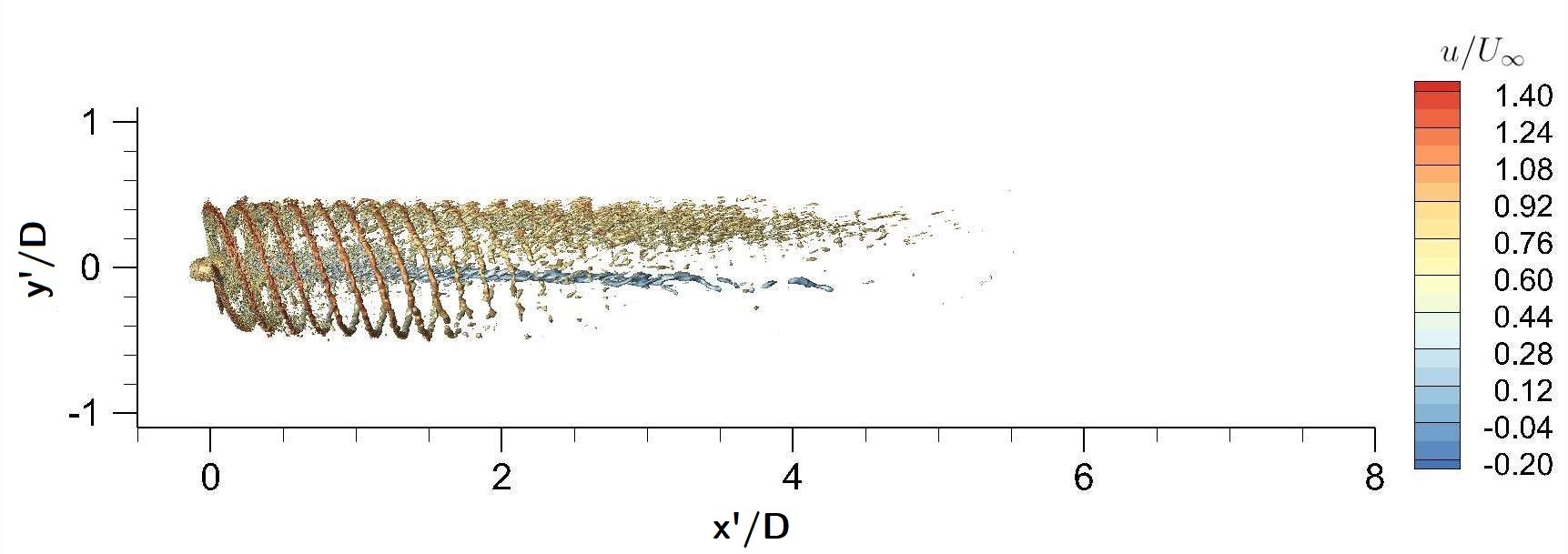}
\caption{$\gamma=10^{\circ}$}
\end{subfigure}
\begin{subfigure}[b]{0.98\textwidth}
\centering
\includegraphics[width=3.1in]{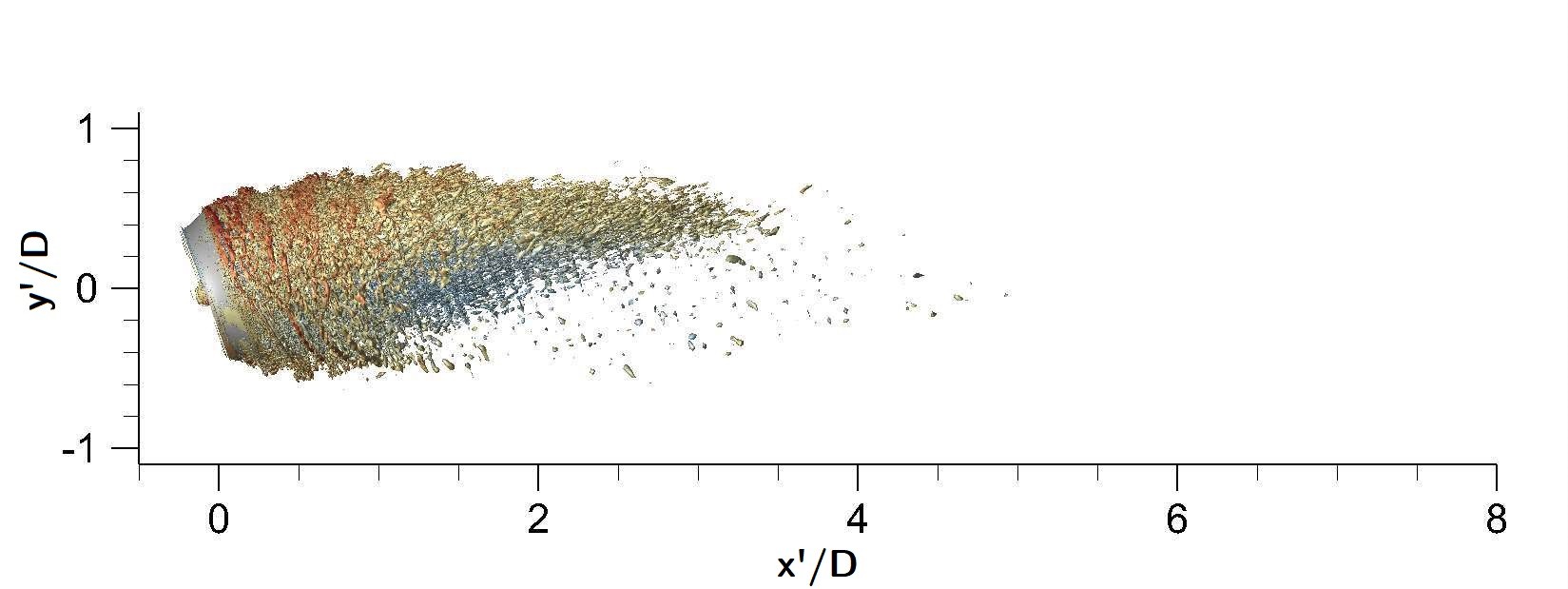}
\includegraphics[width=3.3in]{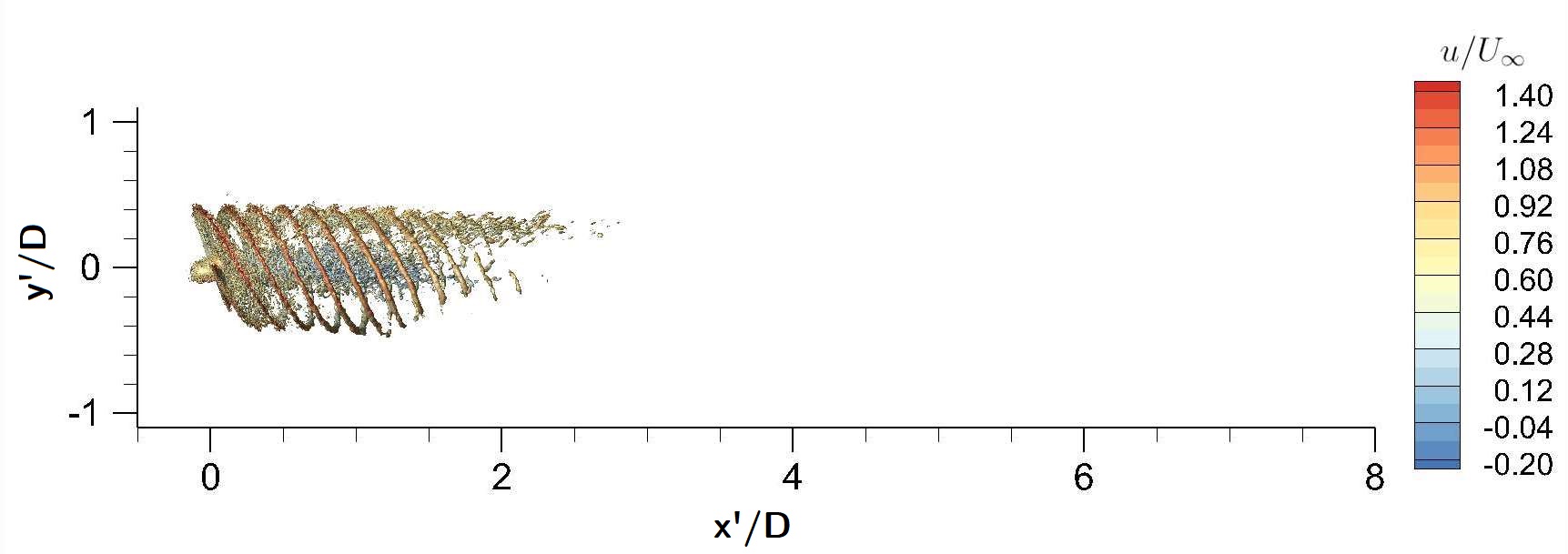}
\caption{$\gamma=20^{\circ}$}
\end{subfigure}
\begin{subfigure}[b]{0.98\textwidth}
\centering
\includegraphics[width=3.1in]{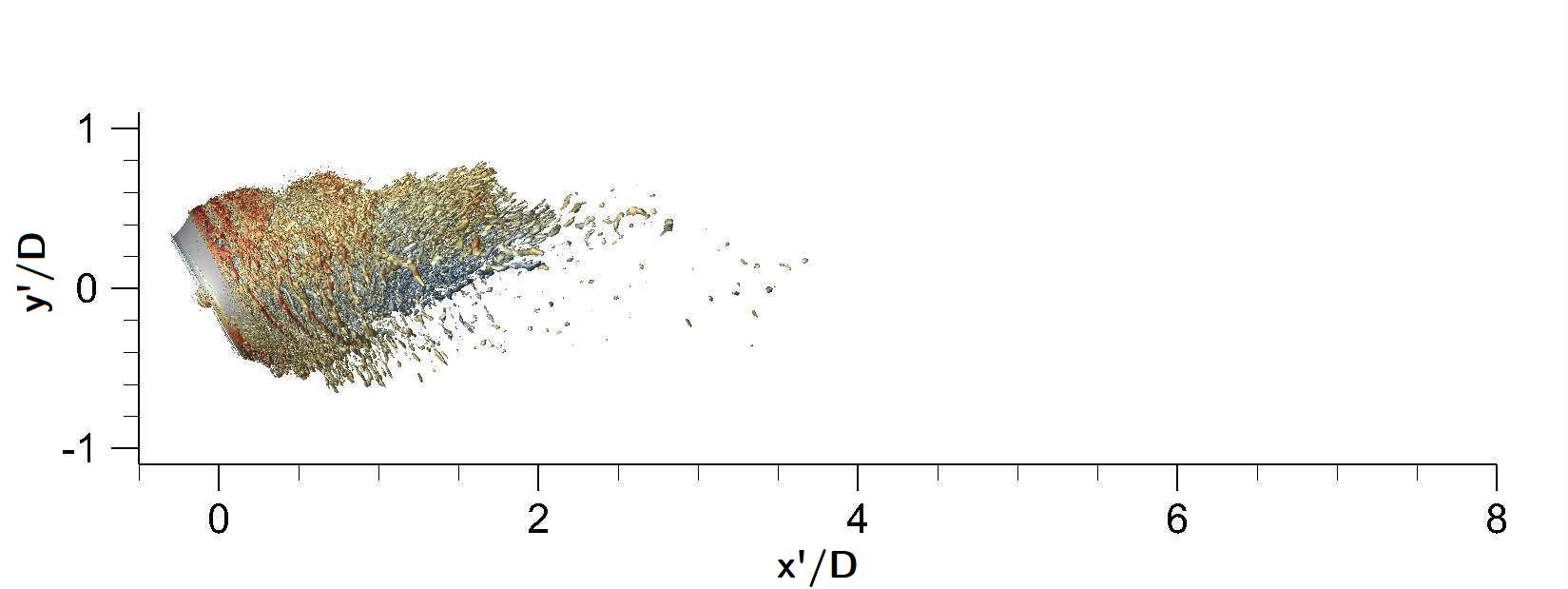}
\includegraphics[width=3.3in]{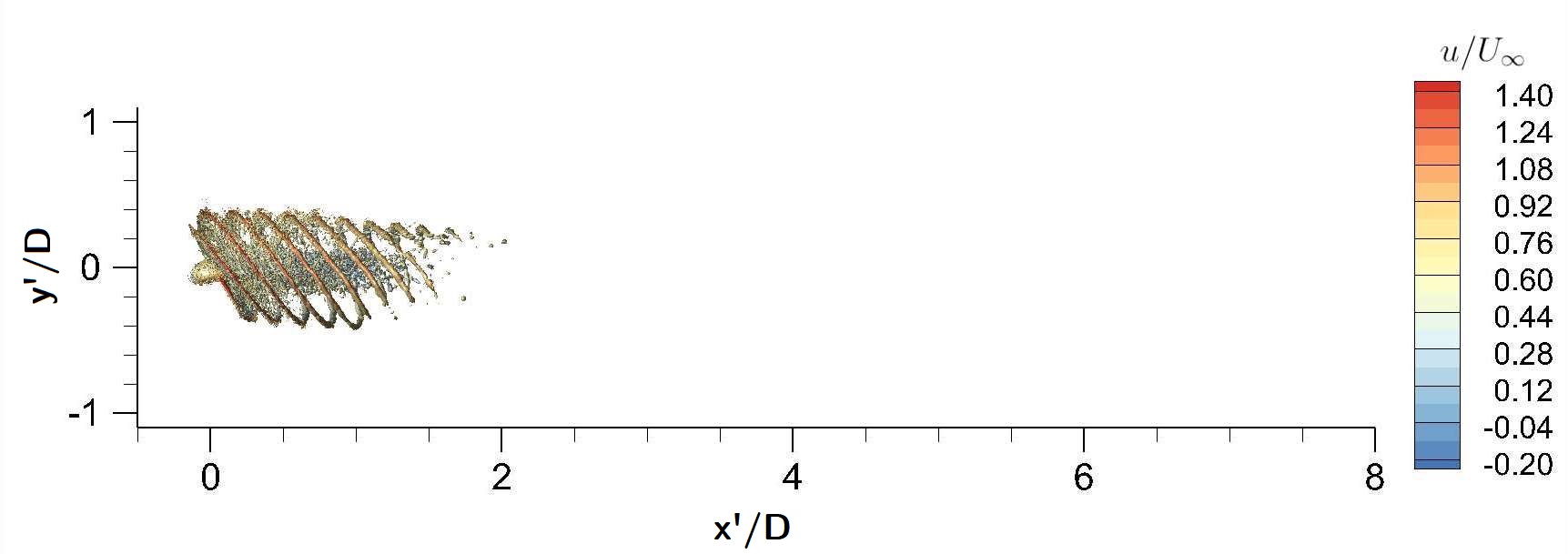}
\caption{$\gamma=30^{\circ}$}
\end{subfigure}
\caption{Isosurfaces of the instantaneous Q-criterion $Q_{cr}D^2/U_{\infty}^2=60$ for yawed flows.}
\label{fig:Qcr_ins_yaw}
\end{figure}

Isosurfaces of the instantaneous Q-criterion for the two turbines are shown in Fig. \ref{fig:Qcr_ins_yaw}. For simplicity and fair comparison, the plots are presented in the flow coordinates (i.e., the x$'$-y$'$ coordinates). Obviously, nonzero yaw angles have made the wakes asymmetric about the flow direction. These asymmetries can be qualitatively explained as follows. In the extreme situation of infinitely large blade rotation speed, the flows are completely blocked in the turbines' cross-sections, and the turbines behave like two disks. At a finite rotational speed, the turbines still roughly behave like disks with some cross-sectional flow rates. The upper leading edges of the ``disks'' cause much larger flow separations than the lower trailing edges, resulting in the present asymmetric shapes of the wakes. Furthermore, as the yaw angle increases, the relative angle of attack of the ``disks'' decreases, resulting in weaker wake vortices. Of course, the blades' rotation motions add more complexities to the flow fields. For example, for the OWT, the spiral tip vortices are still present even for large yaw angles, but the hub vortices gradually disappear as the yaw angle increases.

\subsection{Velocity Fields}
\begin{figure}[H]
\centering
\begin{subfigure}[b]{0.98\textwidth}
\centering
\includegraphics[width=3.1in]{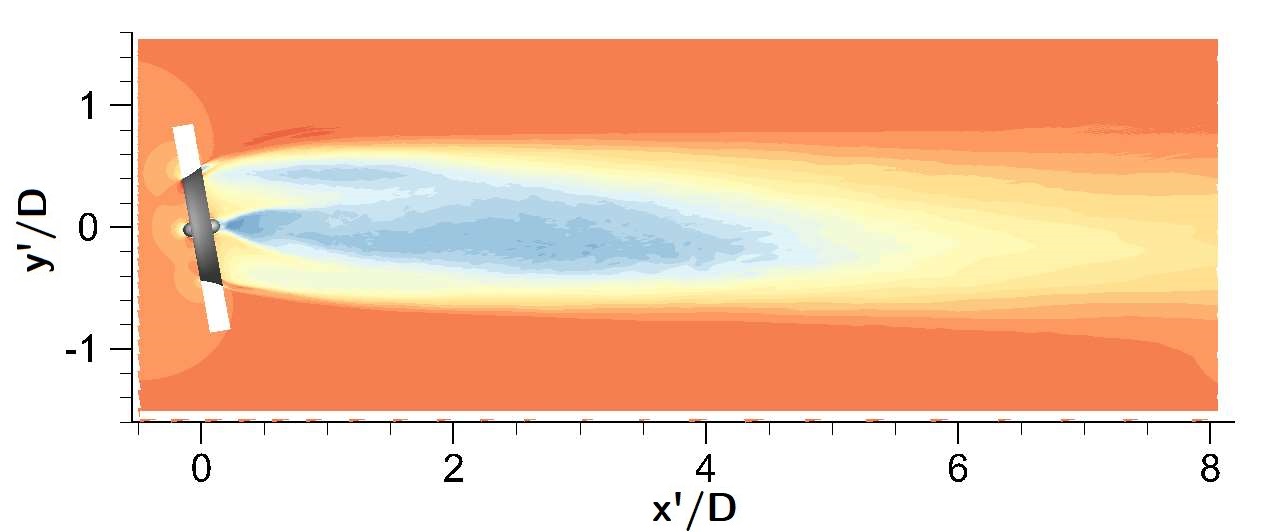}
\includegraphics[width=3.5in]{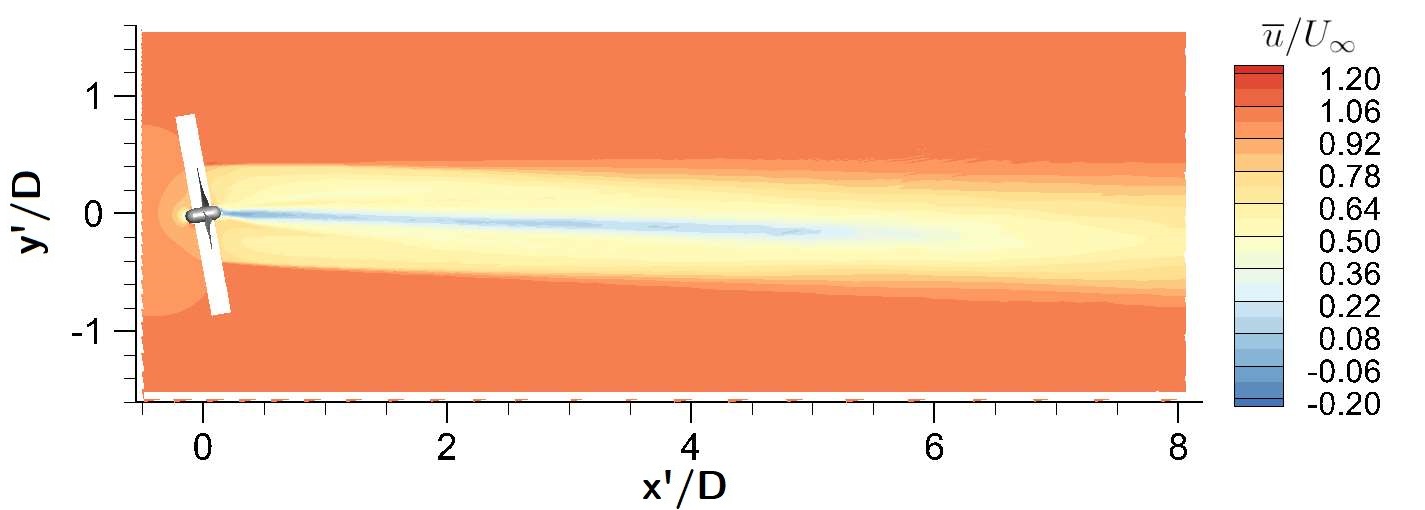}
\caption{$\gamma=10^{\circ}$}
\label{fig:vel_ave_yaw_10}
\end{subfigure}
\begin{subfigure}[b]{0.98\textwidth}
\centering
\includegraphics[width=3.1in]{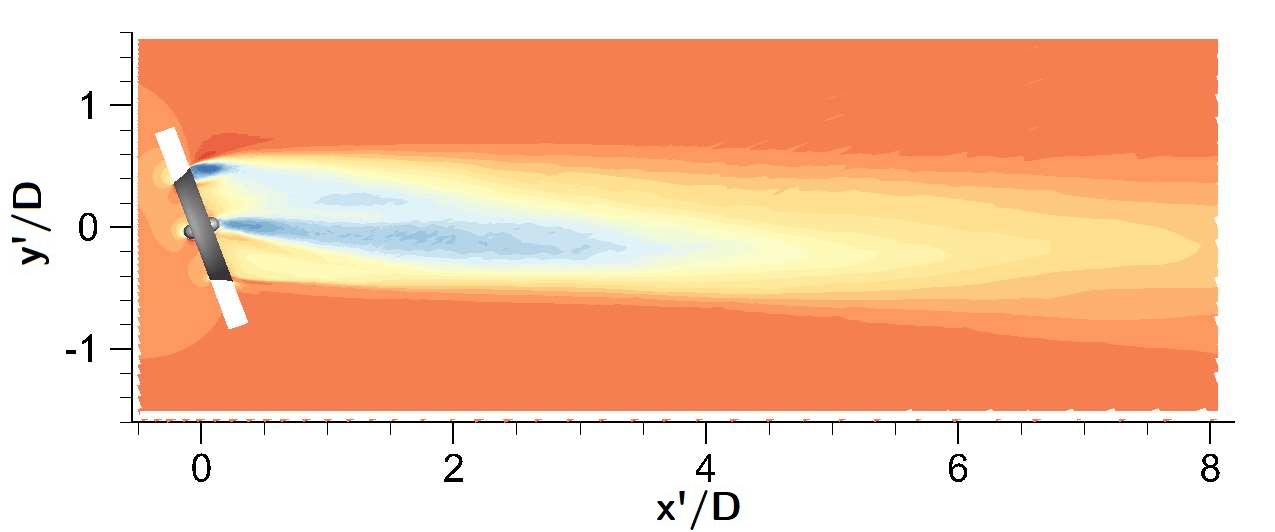}
\includegraphics[width=3.5in]{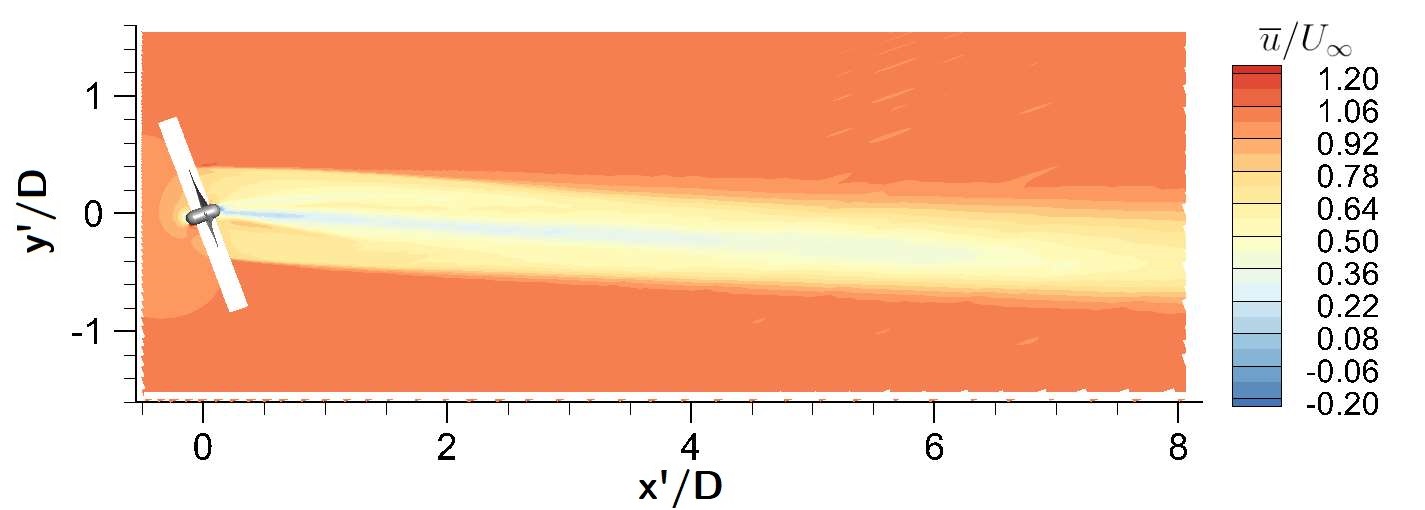}
\caption{$\gamma=20^{\circ}$}
\label{fig:vel_ave_yaw_20}
\end{subfigure}
\begin{subfigure}[b]{0.98\textwidth}
\centering
\includegraphics[width=3.1in]{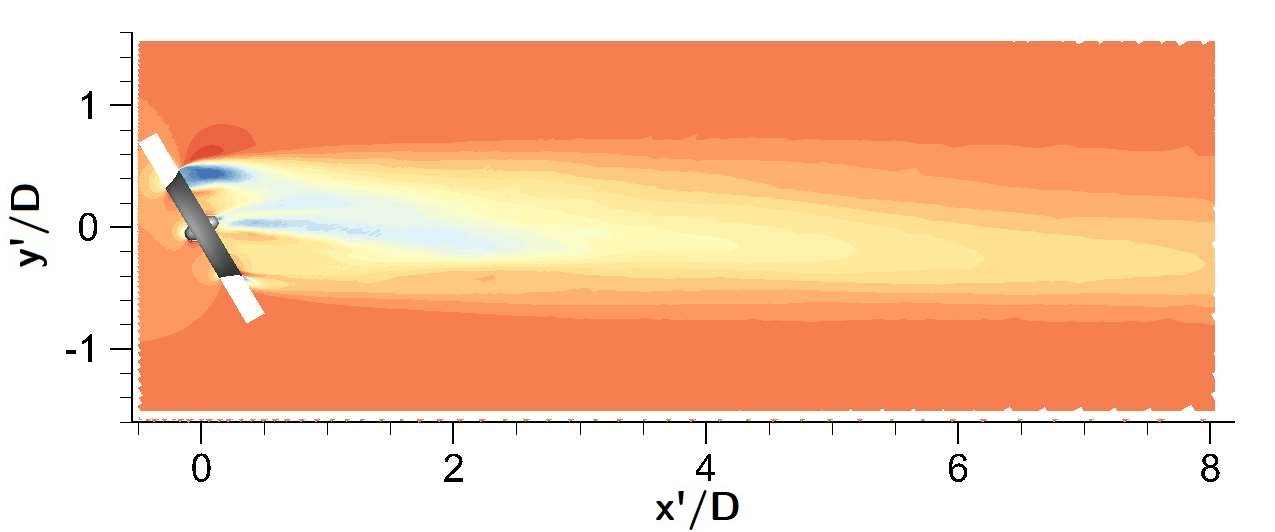}
\includegraphics[width=3.5in]{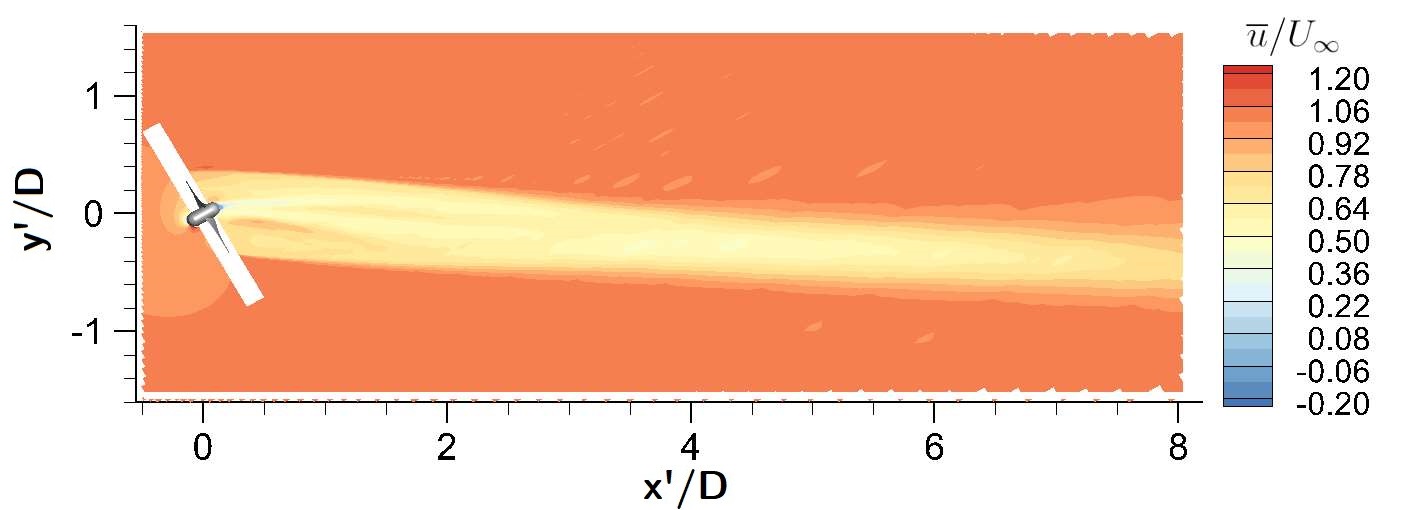}
\caption{$\gamma=30^{\circ}$}
\label{fig:vel_ave_yaw_30}
\end{subfigure}
\caption{Mean streamwise velocity contours at $\lambda=3.93$ in the central x$'$-y$'$ plane.}
\label{fig:vel_ave_yaw}
\end{figure}

Contours of the mean streamwise flow velocity in the central x$'$-y$'$ plane at different yaw angles are plotted in Fig. \ref{fig:vel_ave_yaw}. It is seen that the wakes are still at lower speeds compared with the freestream. However, the sizes of the low-speed regions decrease as the yaw angle increases, which agrees with the decreasing energy extraction performances revealed in Fig. \ref{fig:yaw_perform}. Although the results are presented in the flow coordinate, deflections of the wakes are still observed. Overall, the OWT shows more considerable wake deflections than the DWT. For the DWT, as the yaw angle increases, the angle of attack of the duct's upper portion increases, which has led to larger flow separations and blockage effects around the duct's upper trailing edge. Strong interactions are seen between the duct's upper wake and the DWT's main wake. For the duct's lower portion, the situation is just the opposite. The angle of attack decreases as the yaw angle increases, so no severe flow separation is observed in this region.

\begin{figure}[!htb]
\centering
\begin{subfigure}[b]{0.2\textwidth}
\centering
\includegraphics[width=1.2in]{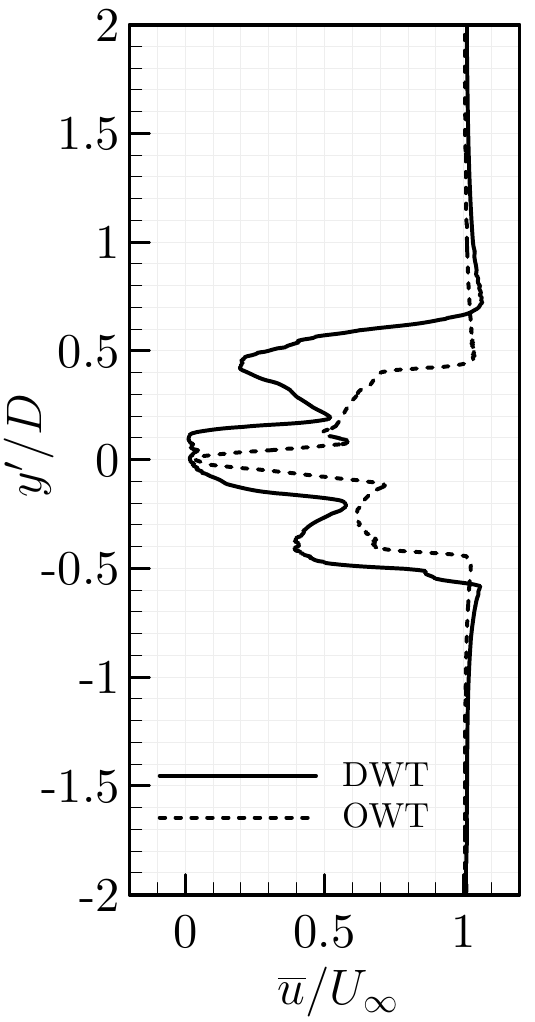}
\caption{$x'/D=0.5$}
\label{fig:yaw_0.5D_gamma_10}
\end{subfigure}
\begin{subfigure}[b]{0.2\textwidth}
\centering
\includegraphics[width=1.2in]{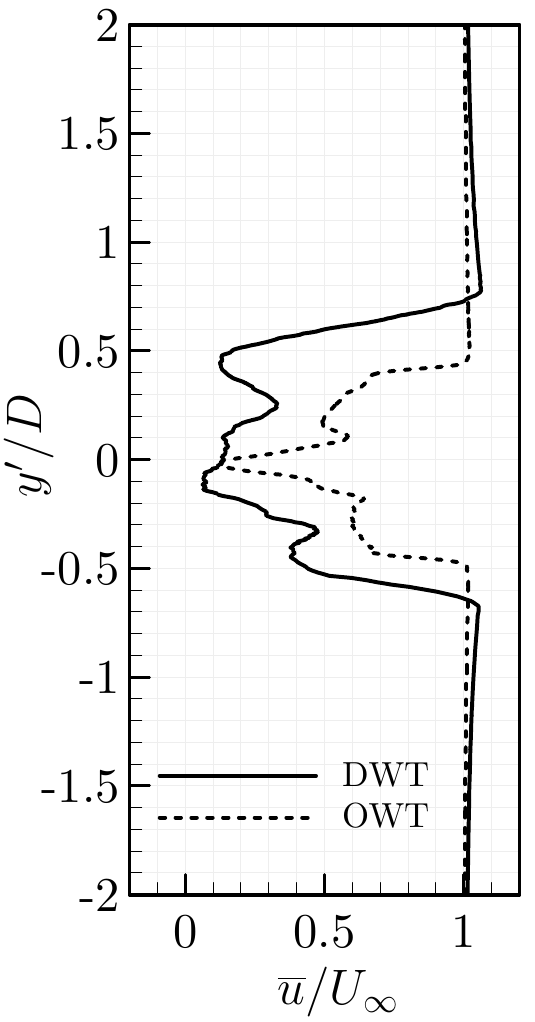}
\caption{$x'/D=1$}
\label{fig:yaw_1D_gamma_10}
\end{subfigure}
\begin{subfigure}[b]{0.2\textwidth}
\centering
\includegraphics[width=1.2in]{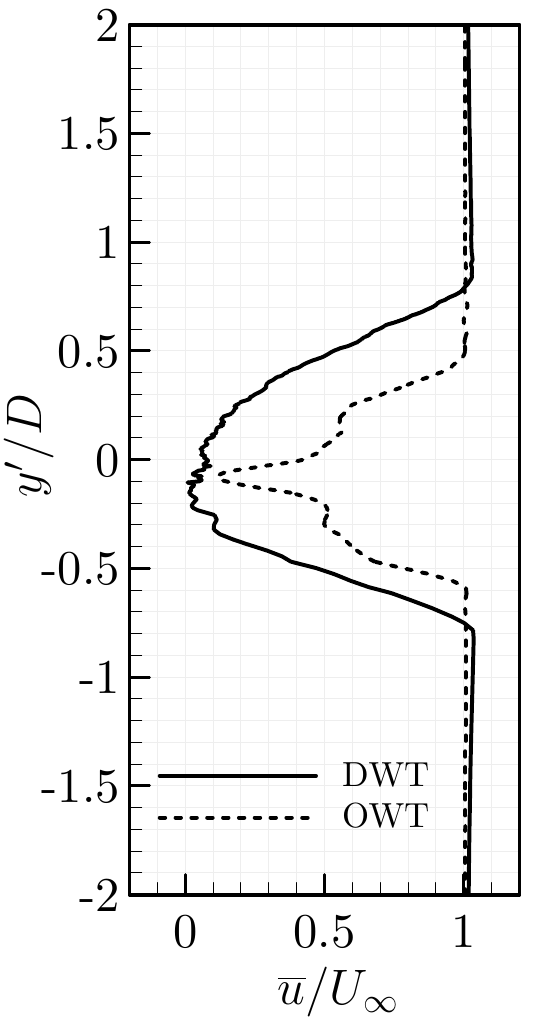}
\caption{$x'/D=3$}
\label{fig:yaw_3D_gamma_10}
\end{subfigure}
\begin{subfigure}[b]{0.2\textwidth}
\centering
\includegraphics[width=1.2in]{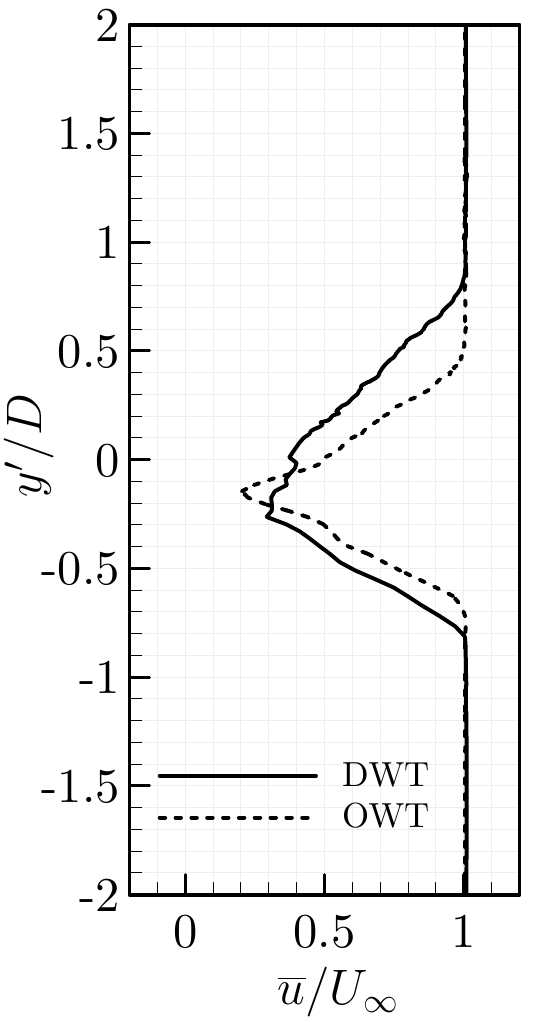}
\caption{$x'/D=5$}
\label{fig:yaw_5D_gamma_10}
\end{subfigure}
\caption{Mean streamwise velocity profiles for $\gamma=10^\circ$.}
\label{fig:yaw_profile_gamma10}
\end{figure}

\begin{figure}[!htb]
\centering
\begin{subfigure}[b]{0.2\textwidth}
\centering
\includegraphics[width=1.2in]{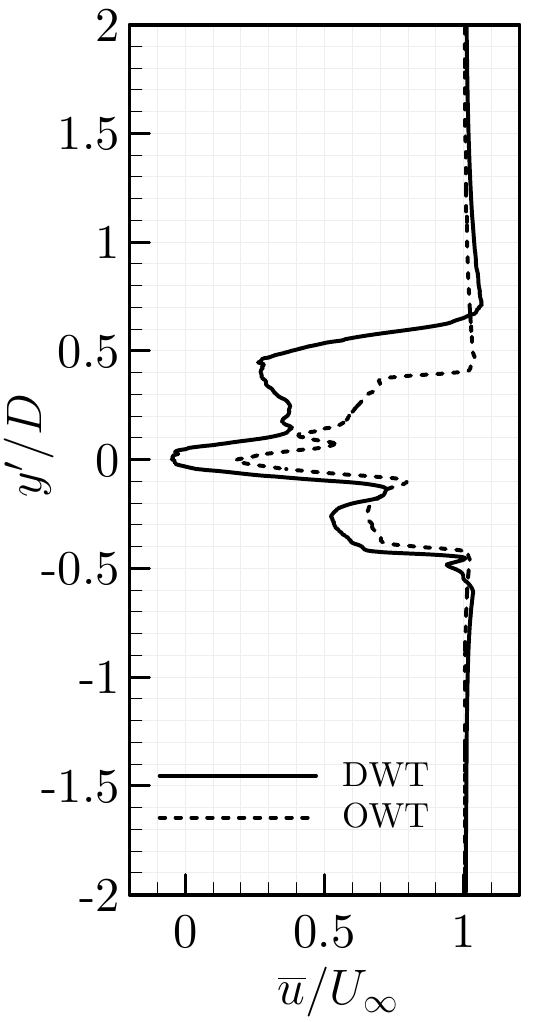}
\caption{$x'/D=0.5$}
\label{fig:yaw_0.5D_gamma_20}
\end{subfigure}
\begin{subfigure}[b]{0.2\textwidth}
\centering
\includegraphics[width=1.2in]{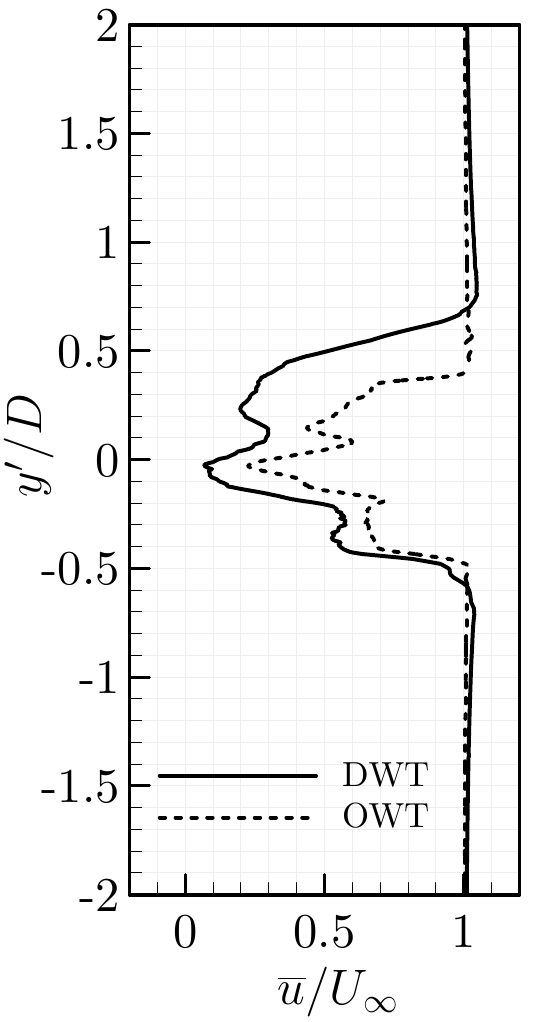}
\caption{$x'/D=1$}
\label{fig:yaw_1D_gamma_20}
\end{subfigure}
\begin{subfigure}[b]{0.2\textwidth}
\centering
\includegraphics[width=1.2in]{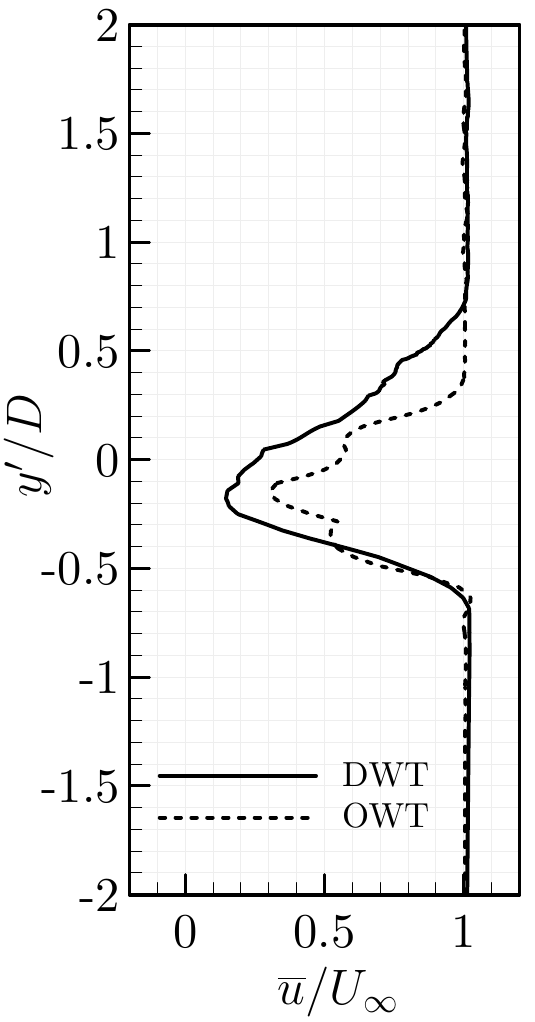}
\caption{$x'/D=3$}
\label{fig:yaw_3D_gamma_20}
\end{subfigure}
\begin{subfigure}[b]{0.2\textwidth}
\centering
\includegraphics[width=1.2in]{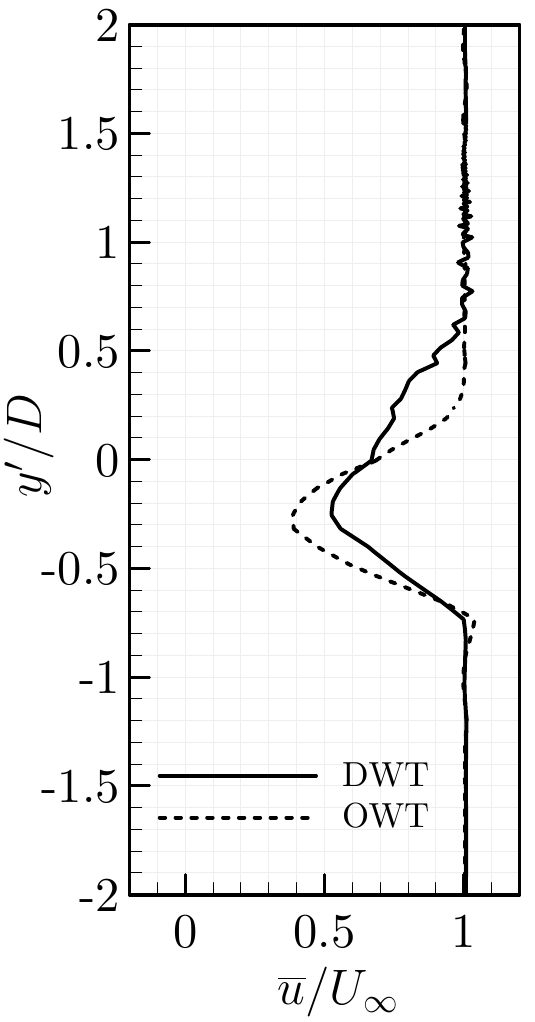}
\caption{$x'/D=5$}
\label{fig:yaw_5D_gamma_20}
\end{subfigure}
\caption{Mean streamwise velocity profiles for $\gamma=20^\circ$.}
\label{fig:yaw_profile_gamma20}
\end{figure}

\begin{figure}[H]
\centering
\begin{subfigure}[b]{0.2\textwidth}
\centering
\includegraphics[width=1.2in]{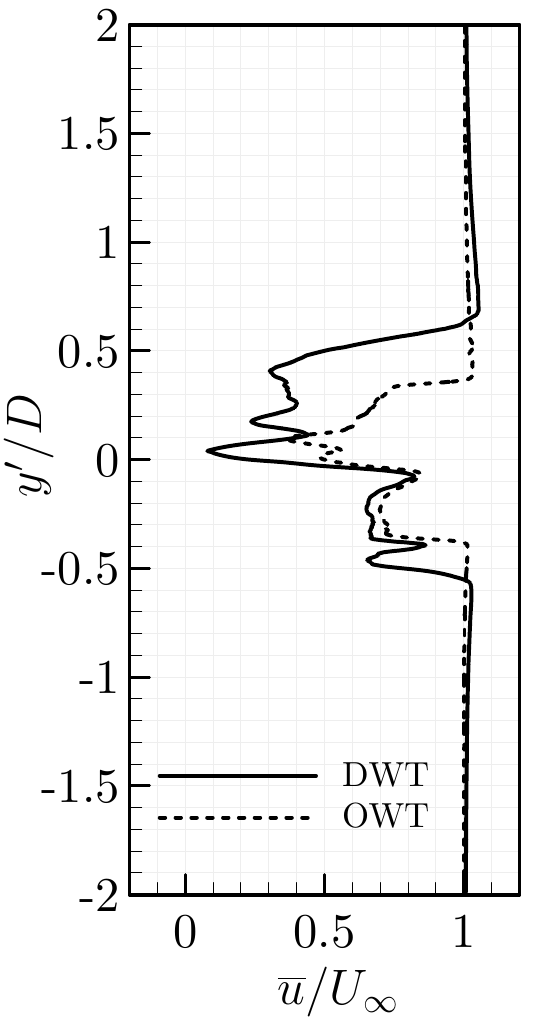}
\caption{$x'/D=0.5$}
\label{fig:yaw_0.5D_gamma_30}
\end{subfigure}
\begin{subfigure}[b]{0.2\textwidth}
\centering
\includegraphics[width=1.2in]{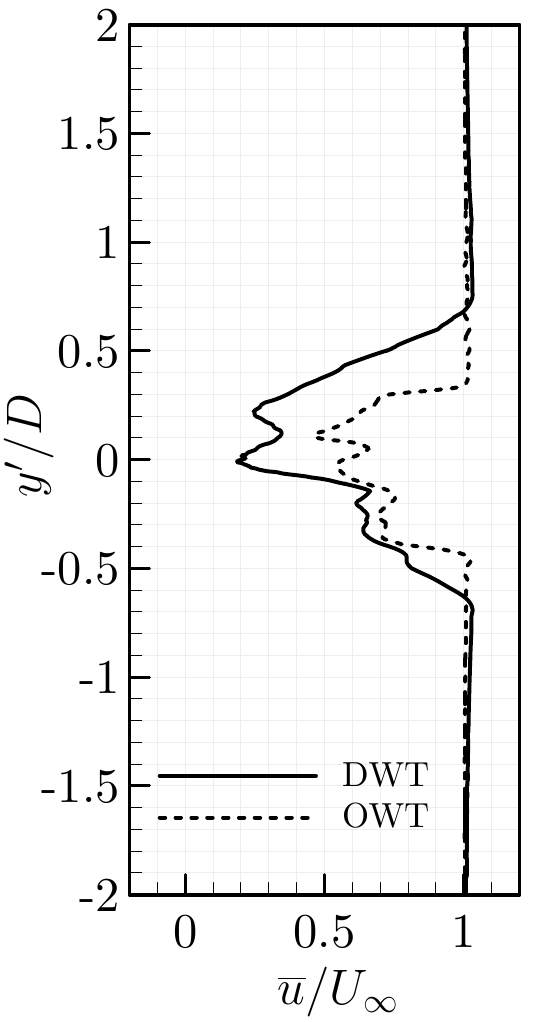}
\caption{$x'/D=1$}
\label{fig:yaw_1D_gamma_30}
\end{subfigure}
\begin{subfigure}[b]{0.2\textwidth}
\centering
\includegraphics[width=1.2in]{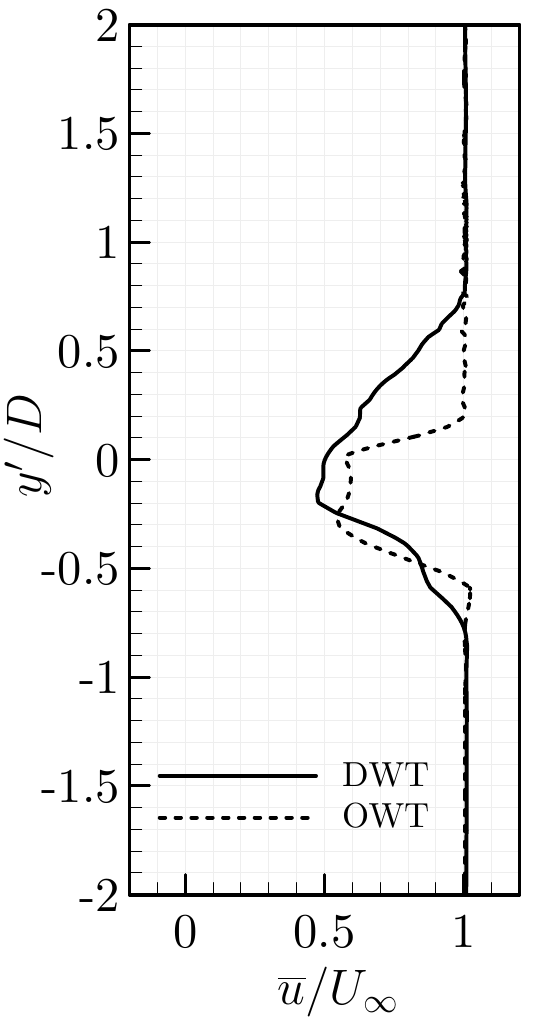}
\caption{$x'/D=3$}
\label{fig:yaw_3D_gamma_30}
\end{subfigure}
\begin{subfigure}[b]{0.2\textwidth}
\centering
\includegraphics[width=1.2in]{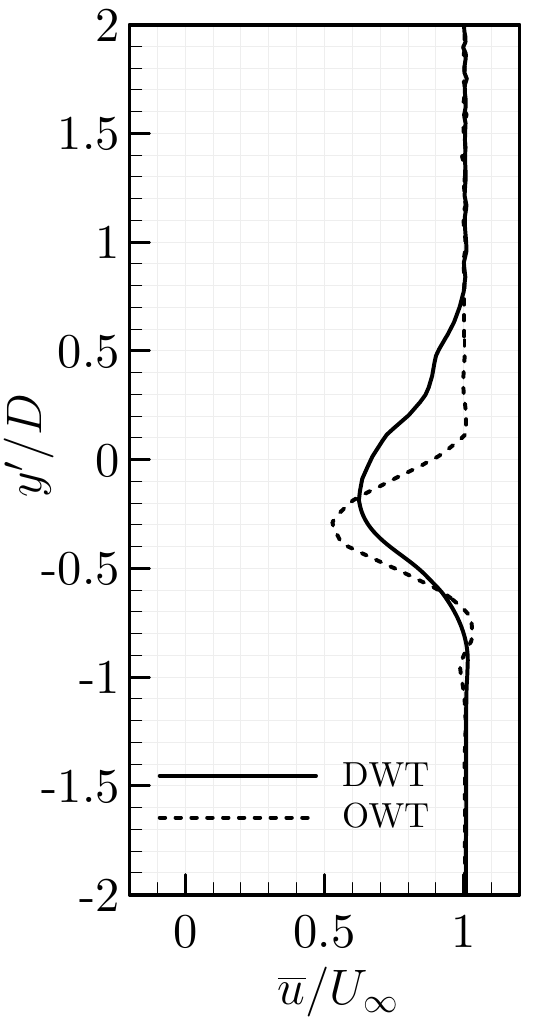}
\caption{$x'/D=5$}
\label{fig:yaw_5D_gamma_30}
\end{subfigure}
\caption{Mean streamwise velocity profiles for $\gamma=30^\circ$.}
\label{fig:yaw_profile_gamma30}
\end{figure}

The mean streamwise velocity profiles at several locations in the central x$'$-y$'$ plane are shown in Figs. \ref{fig:yaw_profile_gamma10}-\ref{fig:yaw_profile_gamma30} for the three nonzero yaw angles. To facilitate the discussion, we define the centerline of a wake as the line that connects points with minimum local streamwise flow speeds. This centerline divides a wake into an upper part and a lower part. From the profiles, it is seen that in the near field (e.g., $x'/D=0.5$ and $1.0$), the upper wakes overall travel slower than the lower wakes, indicating that the turbines may have extracted more energy via their upper parts. A comparison of the profiles further downstream, e.g., at $x'/D=3.0$ (Figs. \ref{fig:yaw_3D_gamma_10}, \ref{fig:yaw_3D_gamma_20}, and \ref{fig:yaw_3D_gamma_30}), and for different yaw angles reveals that the wakes of both turbines recover faster for larger yaw angles. Finally, the wake deflection is better judged from flows sufficiently downstream of the turbines, e.g., at $x'/D=5.0$. For $\gamma=10^\circ$ (see Fig. \ref{fig:yaw_5D_gamma_10}), the DWT's wake is slightly more deflected than the OWT's.  For $\gamma=20^\circ$ (see Fig. \ref{fig:yaw_5D_gamma_20}), the OWT's wake is slightly more deflected. For $\gamma=30^\circ$ (see Fig. \ref{fig:yaw_5D_gamma_30}), the OWT's wake is obviously more deflected.

\section{Summary}
\label{sec:concl}

High-order implicit large-eddy simulations of a ducted wind turbine and its open-rotor counterpart under different tip-speed ratios and yaw angles have been successfully performed. The simulations employed a fifth-order spatial scheme, a third-order temporal scheme, and about thirty million degrees of freedom. This is the first time that a high-order method being applied to a comprehensive study of a complete ducted wind turbine without any geometric simplification.

The simulation results reveal that the ducted turbine has much higher power outputs than its open-rotor counterpart for all the given tip-speed ratios. From pressure contours, it is identified that the loads concentrate more on the blades' outboard part, especially around the leading edges. From the isosurfaces of instantaneous Q-criterion, it is noticed that the presence of the duct has made the flow field more turbulent, weakened the tip and hub vortices, and enhanced the trailing edge vortices of the ducted turbine. As the tip-speed ratio increases, flow bifurcation is observed in the flow field of the ducted turbine. The contours and profiles of the mean streamwise velocity reveal that the bifurcation is caused by the increasing blockage effects of blade tips with increasing rotation speed.

Under yawed flow conditions, the ducted turbine is still found to have larger power outputs than the open configuration. It is confirmed that the ducted turbine's performance is insensitive to small yaw angles. Other than that, as the yaw angle increases, the thrust and the power coefficients both decrease, with the latter decreasing more than the former. At nonzero yaw angles, the vortex fields are highly asymmetric for both turbines, and an interpretation using disk analog and angle of attack has been established. The velocity fields and profiles show that a large yaw angle may cause obvious flow deflections in the wakes, especially in the open turbine's wakes. Finally, the wake flows are found to recover more quickly at larger yaw angles.

\section*{Acknowledgment}
The authors would like to express our acknowledgments to Clarkson University for financial support through the IGNITE Fellowship Program. C. Liang would also like to acknowledge the support by an Office of Naval Research grant (No. N00014-20-1-2007) monitored by Dr. Ki-Han Kim. The computing hours were granted through the DoD HPC Modernization Program.

\biboptions{sort&compress}
\bibliographystyle{elsarticle-num-names}
\bibliography{references}

\begin{thebibliography}{47}
\providecommand{\natexlab}[1]{#1}
\providecommand{\url}[1]{\texttt{#1}}
\providecommand{\urlprefix}{URL }
\expandafter\ifx\csname urlstyle\endcsname\relax
  \providecommand{\doi}[1]{doi:\discretionary{}{}{}#1}\else
  \providecommand{\doi}[1]{doi:\discretionary{}{}{}\begingroup
  \urlstyle{rm}\url{#1}\endgroup}\fi
\providecommand{\bibinfo}[2]{#2}

\bibitem[{Shepherd and Zhang(2017)}]{shepherd-2017}
\bibinfo{author}{W.~Shepherd}, \bibinfo{author}{L.~Zhang},
  \bibinfo{title}{Electricity generation using wind power},
  \bibinfo{publisher}{World Scientific}, \bibinfo{year}{2017}.

\bibitem[{Hau(2013)}]{hau-2013}
\bibinfo{author}{E.~Hau}, \bibinfo{title}{Wind turbines: fundamentals,
  technologies, application, economics}, \bibinfo{publisher}{Springer Science
  \& Business Media}, \bibinfo{year}{2013}.

\bibitem[{Burton et~al.(2011)Burton, Jenkins, Sharpe, and
  Bossanyi}]{burton-2011}
\bibinfo{author}{T.~Burton}, \bibinfo{author}{N.~Jenkins},
  \bibinfo{author}{D.~Sharpe}, \bibinfo{author}{E.~Bossanyi},
  \bibinfo{title}{Wind Energy Handbook}, \bibinfo{publisher}{John Wiley \&
  Sons}, \bibinfo{year}{2011}.

\bibitem[{Ochieng et~al.(2018)Ochieng, Hancock, Roberts, and
  Le~Kernec}]{ochieng-2018}
\bibinfo{author}{F.~X. Ochieng}, \bibinfo{author}{C.~M. Hancock},
  \bibinfo{author}{G.~W. Roberts}, \bibinfo{author}{J.~Le~Kernec},
  \bibinfo{title}{A review of ground-based radar as a noncontact sensor for
  structural health monitoring of in-field wind turbines blades},
  \bibinfo{journal}{Wind Energy} \bibinfo{volume}{21}~(\bibinfo{number}{12})
  (\bibinfo{year}{2018}) \bibinfo{pages}{1435--1449}.

\bibitem[{Dighe et~al.(2020)Dighe, Avallone, and Bussel}]{dighe-2020}
\bibinfo{author}{V.~Dighe}, \bibinfo{author}{F.~Avallone},
  \bibinfo{author}{G.~Bussel}, \bibinfo{title}{Effects of yawed inflow on the
  aerodynamic and aeroacoustic performance of ducted wind turbines},
  \bibinfo{journal}{Journal of Wind Engineering and Industrial Aerodynamics}
  \bibinfo{volume}{201} (\bibinfo{year}{2020}) \bibinfo{pages}{104174}.

\bibitem[{Bontempo and Manna(2020)}]{bontempo-2020}
\bibinfo{author}{R.~Bontempo}, \bibinfo{author}{M.~Manna}, \bibinfo{title}{On
  the potential of the ideal diffuser augmented wind turbine: an investigation
  by means of a momentum theory approach and of a free-wake ring-vortex
  actuator disk model}, \bibinfo{journal}{Energy Conversion and Management}
  \bibinfo{volume}{213} (\bibinfo{year}{2020}) \bibinfo{pages}{112794}.

\bibitem[{Lilley and Rainbird(1956)}]{lilley-1956}
\bibinfo{author}{G.~M. Lilley}, \bibinfo{author}{W.~J. Rainbird},
  \bibinfo{title}{A preliminary report on the design and performance of ducted
  windmills}, \bibinfo{publisher}{College of Aeronautics, Cranfield},
  \bibinfo{year}{1956}.

\bibitem[{Foreman et~al.(1978)Foreman, Gilbert, and Oman}]{foreman-1978}
\bibinfo{author}{K.~M. Foreman}, \bibinfo{author}{B.~Gilbert},
  \bibinfo{author}{R.~A. Oman}, \bibinfo{title}{Diffuser augmentation of wind
  turbines}, \bibinfo{journal}{Solar Energy}
  \bibinfo{volume}{20}~(\bibinfo{number}{4}) (\bibinfo{year}{1978})
  \bibinfo{pages}{305--311}.

\bibitem[{Gilbert and Foreman(1979)}]{gilbert-1979}
\bibinfo{author}{B.~L. Gilbert}, \bibinfo{author}{K.~M. Foreman},
  \bibinfo{title}{Experimental demonstration of the diffuser-augmented wind
  turbine concept}, \bibinfo{journal}{Journal of Energy}
  \bibinfo{volume}{3}~(\bibinfo{number}{4}) (\bibinfo{year}{1979})
  \bibinfo{pages}{235--240}.

\bibitem[{Igra(1981)}]{igra-1981}
\bibinfo{author}{O.~Igra}, \bibinfo{title}{Research and development for
  shrouded wind turbines}, \bibinfo{journal}{Energy Conversion and Management}
  \bibinfo{volume}{21}~(\bibinfo{number}{1}) (\bibinfo{year}{1981})
  \bibinfo{pages}{13--48}.

\bibitem[{Fletcher(1981)}]{fletcher-1981}
\bibinfo{author}{C.~Fletcher}, \bibinfo{title}{Computational analysis of
  diffuser-augmented wind turbines}, \bibinfo{journal}{Energy Conversion and
  Management} \bibinfo{volume}{21}~(\bibinfo{number}{3}) (\bibinfo{year}{1981})
  \bibinfo{pages}{175--183}, ISSN \bibinfo{issn}{0196-8904}.

\bibitem[{Vaz and Wood(2018)}]{vaz-2018}
\bibinfo{author}{J.~R. Vaz}, \bibinfo{author}{D.~H. Wood},
  \bibinfo{title}{Effect of the diffuser efficiency on wind turbine
  performance}, \bibinfo{journal}{Renewable Energy} \bibinfo{volume}{126}
  (\bibinfo{year}{2018}) \bibinfo{pages}{969--977}.

\bibitem[{Koras and Georgalas(1988)}]{koras-1988}
\bibinfo{author}{A.~D. Koras}, \bibinfo{author}{C.~G. Georgalas},
  \bibinfo{title}{Calculation of the Influence of Annular Augmentors on the
  Performance of a Wind Rotor}, \bibinfo{journal}{Wind Engineering}
  (\bibinfo{year}{1988}) \bibinfo{pages}{257--267}.

\bibitem[{Politis and Koras(1995)}]{politis-1995}
\bibinfo{author}{G.~K. Politis}, \bibinfo{author}{A.~D. Koras},
  \bibinfo{title}{A performance prediction method for ducted medium loaded
  horizontal axis windturbines}, \bibinfo{journal}{Wind Engineering}
  (\bibinfo{year}{1995}) \bibinfo{pages}{273--288}.

\bibitem[{Phillips et~al.(2002)Phillips, Richards, and Flay}]{phillips-2002}
\bibinfo{author}{D.~G. Phillips}, \bibinfo{author}{P.~J. Richards},
  \bibinfo{author}{R.~Flay}, \bibinfo{title}{{CFD} modelling and the
  development of the diffuser augmented wind turbine}, \bibinfo{journal}{Wind
  and Structures} \bibinfo{volume}{5}~(\bibinfo{number}{2\_3\_4})
  (\bibinfo{year}{2002}) \bibinfo{pages}{267--276}.

\bibitem[{Hansen et~al.(2000)Hansen, S{\o}rensen, and Flay}]{hansen-2000}
\bibinfo{author}{M.~Hansen}, \bibinfo{author}{N.~S{\o}rensen},
  \bibinfo{author}{R.~Flay}, \bibinfo{title}{Effect of placing a diffuser
  around a wind turbine}, \bibinfo{journal}{Wind Energy}
  \bibinfo{volume}{3}~(\bibinfo{number}{4}) (\bibinfo{year}{2000})
  \bibinfo{pages}{207--213}.

\bibitem[{Abe and Ohya(2004)}]{abe-2004}
\bibinfo{author}{K.~Abe}, \bibinfo{author}{Y.~Ohya}, \bibinfo{title}{An
  investigation of flow fields around flanged diffusers using {CFD}},
  \bibinfo{journal}{Journal of Wind Engineering and Industrial Aerodynamics}
  \bibinfo{volume}{92}~(\bibinfo{number}{3-4}) (\bibinfo{year}{2004})
  \bibinfo{pages}{315--330}.

\bibitem[{Venters et~al.(2018)Venters, Helenbrook, and Visser}]{venters-2018}
\bibinfo{author}{R.~Venters}, \bibinfo{author}{B.~Helenbrook},
  \bibinfo{author}{K.~Visser}, \bibinfo{title}{Ducted wind turbine
  optimization}, \bibinfo{journal}{Journal of Solar Energy Engineering}
  \bibinfo{volume}{140}~(\bibinfo{number}{1}).

\bibitem[{Sadeghi et~al.(2018)Sadeghi, Helenbrook, and Visser}]{bagheri-2018}
\bibinfo{author}{N.~B. Sadeghi}, \bibinfo{author}{B.~Helenbrook},
  \bibinfo{author}{K.~Visser}, \bibinfo{title}{Ducted wind turbine optimization
  and sensitivity to rotor position}, \bibinfo{journal}{Wind Energy Science}
  \bibinfo{volume}{3}~(\bibinfo{number}{1}) (\bibinfo{year}{2018})
  \bibinfo{pages}{221--229}.

\bibitem[{Hill(1973)}]{reed-1973}
\bibinfo{author}{W.~R.~T. Hill}, \bibinfo{title}{Triangular mesh methods for
  the neutron transport equation}, \bibinfo{type}{Tech. Rep.},
  \bibinfo{institution}{Los Alamos Scientific Lab., N. Mex.(USA)},
  \bibinfo{year}{1973}.

\bibitem[{Cockburn et~al.(2012)Cockburn, Karniadakis, and Shu}]{cockburn-2012}
\bibinfo{author}{B.~Cockburn}, \bibinfo{author}{G.~Karniadakis},
  \bibinfo{author}{C.~W. Shu}, \bibinfo{title}{Discontinuous Galerkin methods:
  theory, computation and applications}, vol.~\bibinfo{volume}{11},
  \bibinfo{publisher}{Springer Science \& Business Media},
  \bibinfo{year}{2012}.

\bibitem[{Patera(1984)}]{patera-1984}
\bibinfo{author}{A.~Patera}, \bibinfo{title}{A spectral element method for
  fluid dynamics: laminar flow in a channel expansion},
  \bibinfo{journal}{Journal of Computational Physics}
  \bibinfo{volume}{54}~(\bibinfo{number}{3}) (\bibinfo{year}{1984})
  \bibinfo{pages}{468--488}.

\bibitem[{Karniadakis and Sherwin(2013)}]{karniadakis-2013}
\bibinfo{author}{G.~Karniadakis}, \bibinfo{author}{S.~Sherwin},
  \bibinfo{title}{Spectral/hp element methods for computational fluid
  dynamics}, \bibinfo{publisher}{Oxford University Press},
  \bibinfo{year}{2013}.

\bibitem[{Kopriva and Kolias(1996)}]{kopriva-1996a}
\bibinfo{author}{D.~Kopriva}, \bibinfo{author}{J.~Kolias}, \bibinfo{title}{A
  conservative staggered-grid {Chebyshev} multidomain method for compressible
  flows}, \bibinfo{journal}{Journal of Computational Physics}
  \bibinfo{volume}{125}~(\bibinfo{number}{1}) (\bibinfo{year}{1996})
  \bibinfo{pages}{244--261}.

\bibitem[{Kopriva(1996)}]{kopriva-1996b}
\bibinfo{author}{D.~Kopriva}, \bibinfo{title}{A conservative staggered-grid
  {Chebyshev} multidomain method for compressible flows. II. A semi-structured
  method}, \bibinfo{journal}{Journal of Computational Physics}
  \bibinfo{volume}{128}~(\bibinfo{number}{2}) (\bibinfo{year}{1996})
  \bibinfo{pages}{475--488}.

\bibitem[{Kopriva(1998)}]{kopriva-1998}
\bibinfo{author}{D.~Kopriva}, \bibinfo{title}{A staggered-grid multidomain
  spectral method for the compressible {Navier-Stokes} equations},
  \bibinfo{journal}{Journal of Computational Physics}
  \bibinfo{volume}{143}~(\bibinfo{number}{1}) (\bibinfo{year}{1998})
  \bibinfo{pages}{125--158}.

\bibitem[{Liu et~al.(2006)Liu, Vinokur, and Wang}]{liu-2006}
\bibinfo{author}{Y.~Liu}, \bibinfo{author}{M.~Vinokur},
  \bibinfo{author}{Z.~Wang}, \bibinfo{title}{Spectral difference method for
  unstructured grids {I}: Basic formulation}, \bibinfo{journal}{Journal of
  Computational Physics} \bibinfo{volume}{216}~(\bibinfo{number}{2})
  (\bibinfo{year}{2006}) \bibinfo{pages}{780--801}.

\bibitem[{Huynh(2007)}]{huynh-2007}
\bibinfo{author}{H.~T. Huynh}, \bibinfo{title}{A flux reconstruction approach
  to high-order schemes including discontinuous {Galerkin} methods},
  \bibinfo{howpublished}{AIAA paper 2007-4079}, \bibinfo{year}{2007}.

\bibitem[{Huynh(2009)}]{huynh-2009}
\bibinfo{author}{H.~T. Huynh}, \bibinfo{title}{A reconstruction approach to
  high-order schemes including discontinuous {Galerkin} for diffusion},
  \bibinfo{howpublished}{AIAA paper 2009-403}, \bibinfo{year}{2009}.

\bibitem[{Wang and Gao(2009)}]{wang-2009}
\bibinfo{author}{Z.~J. Wang}, \bibinfo{author}{H.~Gao}, \bibinfo{title}{A
  unifying lifting collocation penalty formulation including the discontinuous
  {Galerkin}, spectral volume/difference methods for conservation laws on mixed
  grids}, \bibinfo{journal}{Journal of Computational Physics}
  \bibinfo{volume}{228} (\bibinfo{year}{2009}) \bibinfo{pages}{8161--8186}.

\bibitem[{Zhang and Liang(2015{\natexlab{a}})}]{zhang-2015a}
\bibinfo{author}{B.~Zhang}, \bibinfo{author}{C.~Liang}, \bibinfo{title}{A
  simple, efficient, high-order accurate sliding-mesh interface approach to
  {FR/CPR} method on coupled rotating and stationary domains},
  \bibinfo{howpublished}{AIAA paper 2015-1742},
  \bibinfo{year}{2015}{\natexlab{a}}.

\bibitem[{Zhang and Liang(2015{\natexlab{b}})}]{zhang-2015b}
\bibinfo{author}{B.~Zhang}, \bibinfo{author}{C.~Liang}, \bibinfo{title}{A
  simple, efficient, and high-order accurate curved sliding-mesh interface
  approach to spectral difference method on coupled rotating and stationary
  domains}, \bibinfo{journal}{Journal of Computational Physics}
  \bibinfo{volume}{295} (\bibinfo{year}{2015}{\natexlab{b}})
  \bibinfo{pages}{147--160}.

\bibitem[{Zhang et~al.(2016)Zhang, Liang, Yang, and Rong}]{zhang-2016a}
\bibinfo{author}{B.~Zhang}, \bibinfo{author}{C.~Liang},
  \bibinfo{author}{J.~Yang}, \bibinfo{author}{Y.~Rong}, \bibinfo{title}{A {2D}
  parallel high-order sliding and deforming spectral difference method},
  \bibinfo{journal}{Computers \& Fluids} \bibinfo{volume}{139}
  (\bibinfo{year}{2016}) \bibinfo{pages}{184--196}.

\bibitem[{Zhang and Liang(2016)}]{zhang-2016c}
\bibinfo{author}{B.~Zhang}, \bibinfo{author}{C.~Liang}, \bibinfo{title}{A
  high-order sliding-mesh spectral difference solver for simulating unsteady
  flows around rotating objects}, in: \bibinfo{booktitle}{31st Symposium on
  Naval Hydrodynamics}, \bibinfo{address}{Monterey, CA}, \bibinfo{year}{2016}.

\bibitem[{Zhang et~al.(2018)Zhang, Qiu, and Liang}]{zhang-2018}
\bibinfo{author}{B.~Zhang}, \bibinfo{author}{Z.~Qiu},
  \bibinfo{author}{C.~Liang}, \bibinfo{title}{A flux reconstruction method with
  nonuniform sliding-mesh interfaces for simulating rotating flows},
  \bibinfo{howpublished}{AIAA paper 2018-1094}, \bibinfo{year}{2018}.

\bibitem[{Zhang and Liang(2021)}]{zhang-2021b}
\bibinfo{author}{B.~Zhang}, \bibinfo{author}{C.~Liang}, \bibinfo{title}{A
  conservative high-order method utilizing dynamic transfinite mortar elements
  for flow simulation on curved sliding meshes}, \bibinfo{journal}{Journal of
  Computational Physics}  (\bibinfo{year}{2021}) \bibinfo{pages}{110522}.

\bibitem[{Zhang and Liang(2019{\natexlab{a}})}]{zhang-2019b}
\bibinfo{author}{B.~Zhang}, \bibinfo{author}{C.~Liang},
  \bibinfo{title}{High-order numerical simulation of flows over rotating
  cylinders of various cross-sectional shapes}, \bibinfo{howpublished}{AIAA
  paper 2019-3430}, \bibinfo{year}{2019}{\natexlab{a}}.

\bibitem[{Zhang and Liang(2019{\natexlab{b}})}]{zhang-2019a}
\bibinfo{author}{B.~Zhang}, \bibinfo{author}{C.~Liang},
  \bibinfo{title}{High-order numerical simulation of flapping wing for energy
  harvesting}, \bibinfo{howpublished}{AIAA paper 2019-3338},
  \bibinfo{year}{2019}{\natexlab{b}}.

\bibitem[{Zhang et~al.(2021)Zhang, Ding, and Liang}]{zhang-2021a}
\bibinfo{author}{B.~Zhang}, \bibinfo{author}{C.~Ding},
  \bibinfo{author}{C.~Liang}, \bibinfo{title}{High-Order Implicit Large-Eddy
  Simulation of Flow over a Marine Propeller}, \bibinfo{journal}{Computers \&
  Fluids}  (\bibinfo{year}{2021}) \bibinfo{pages}{104967}.

\bibitem[{Ding et~al.(2022)Ding, Zhang, Liang, Visser, and Yao}]{ding-2022a}
\bibinfo{author}{C.~Ding}, \bibinfo{author}{B.~Zhang},
  \bibinfo{author}{C.~Liang}, \bibinfo{author}{K.~D. Visser},
  \bibinfo{author}{G.~Yao}, \bibinfo{title}{High-order large-eddy simulations
  of a ducted wind turbine}, \bibinfo{howpublished}{AIAA paper 2022-1147},
  \bibinfo{year}{2022}.

\bibitem[{Thomas and Lombard(1979)}]{thomas-1979}
\bibinfo{author}{P.~D. Thomas}, \bibinfo{author}{C.~K. Lombard},
  \bibinfo{title}{Geometric conservation law and its application to flow
  computations on moving grids}, \bibinfo{journal}{AIAA Journal}
  \bibinfo{volume}{17} (\bibinfo{year}{1979}) \bibinfo{pages}{1030--1037}.

\bibitem[{Bathe(2006)}]{bathe-2006}
\bibinfo{author}{K.~J. Bathe}, \bibinfo{title}{Finite Element Procedures},
  \bibinfo{publisher}{Klaus-Jurgen Bathe}, \bibinfo{year}{2006}.

\bibitem[{Rusanov(1961)}]{rusanov-1961}
\bibinfo{author}{V.~V. Rusanov}, \bibinfo{title}{Calculation of interaction of
  non-steady shock waves with obstacles}, \bibinfo{journal}{Journal of
  Computational and Mathematical Physics USSR} \bibinfo{volume}{1}
  (\bibinfo{year}{1961}) \bibinfo{pages}{267--279}.

\bibitem[{Spiteri and Ruuth(2002)}]{spiteri-2002}
\bibinfo{author}{R.~J. Spiteri}, \bibinfo{author}{S.~J. Ruuth},
  \bibinfo{title}{A new class of optimal high-order strong-stability-preserving
  time discretization methods}, \bibinfo{journal}{SIAM Journal on Numerical
  Analysis} \bibinfo{volume}{40} (\bibinfo{year}{2002})
  \bibinfo{pages}{469--491}.

\bibitem[{Ruuth(2006)}]{ruuth-2006}
\bibinfo{author}{S.~Ruuth}, \bibinfo{title}{Global optimization of explicit
  strong-stability-preserving {Runge-Kutta} methods},
  \bibinfo{journal}{Mathematics of Computation}
  \bibinfo{volume}{75}~(\bibinfo{number}{253}) (\bibinfo{year}{2006})
  \bibinfo{pages}{183--207}.

\bibitem[{Zhang(2016)}]{zhang-2016b}
\bibinfo{author}{B.~Zhang}, \bibinfo{title}{A high-order computational
  framework for simulating flows around rotating and moving objects}, Ph.D.
  thesis, \bibinfo{school}{The George Washington University},
  \bibinfo{year}{2016}.

\bibitem[{Kanya and Visser(2018)}]{kanya-2018}
\bibinfo{author}{B.~Kanya}, \bibinfo{author}{K.~Visser},
  \bibinfo{title}{Experimental validation of a ducted wind turbine design
  strategy}, \bibinfo{journal}{Wind Energy Science}
  \bibinfo{volume}{3}~(\bibinfo{number}{2}) (\bibinfo{year}{2018})
  \bibinfo{pages}{919--928}.

\end{thebibliography}

\end{document}